\documentclass[fleqn,usenatbib]{mnras}

\usepackage{newtxtext,newtxmath}

\usepackage[T1]{fontenc}

\DeclareRobustCommand{\VAN}[3]{#2}
\let\VANthebibliography\thebibliography
\def\thebibliography{\DeclareRobustCommand{\VAN}[3]{##3}\VANthebibliography}

\usepackage{graphicx}	
\graphicspath{{Figures/}}
\usepackage{amsmath}	
\usepackage{color}
\usepackage[dvipsnames]{xcolor}
\usepackage{comment}
\usepackage{threeparttable}
\usepackage{subfig}

\newcommand{\dd}{\mathrm{d}}
\newcommand{\ee}{\mathrm{e}}
\newcommand{\ii}{\mathrm{i}}

\title[Warping in vertically oscillating discs]{Instability and warping in vertically oscillating accretion discs}

\author[L. E. Held and G. I. Ogilvie]{
Loren E. Held$^{1}$\thanks{E-mail: leh50@cam.ac.uk (LEH)} and Gordon I. Ogilvie$^{1}$
\\
$^{1}$Department of Applied Mathematics and Theoretical Physics, University of Cambridge, Centre for Mathematical Sciences, Wilberforce Road,\\ 
Cambridge CB3 0WA, United Kingdom\\}

\date{Accepted XXX. Received YYY; in original form ZZZ}

\pubyear{2024}

\begin{document}
\label{firstpage}
\pagerange{\pageref{firstpage}--\pageref{lastpage}}
\maketitle

\begin{abstract}
Many accretion discs have been found to be distorted: either warped due a misalignment in the system, or non-circular as a result of orbital eccentricity or tidal deformation by a binary companion. Warped, eccentric, and tidally distorted discs are not in vertical hydrostatic equilibrium, and thus exhibit vertical oscillations in the direction perpendicular to the disc, a phenomenon that is absent in circular and flat discs. In extreme cases, this vertical motion is manifested as a vertical `bouncing' of the gas, potentially leading to shocks and heating, as observed in recent global numerical simulations. In this paper we isolate the mechanics of vertical disc oscillations by means of quasi-2D and fully 3D hydrodynamic local (shearing-box) models. To determine the numerical and physical dissipation mechanisms at work during an oscillation we start by investigating unforced oscillations, examining the effect of initial oscillation amplitude, as well as resolution, boundary conditions, and vertical box size on the dissipation and energetics of the oscillations. We then drive the oscillations by introducing a time-dependent gravitational potential. A key result is that even a purely vertically oscillating disc is (parametrically) unstable to developing inertial waves, as we confirm through a linear stability analysis. The most important of these has the character of a bending wave, whose radial wavelength depends on the frequency of the vertical oscillation. The nonlinear phase of the instability exhibits shocks, which dampen the oscillations, although energy can also flow from the bending wave back to the vertical oscillation.
\end{abstract}




\section{Introduction}
\label{INTRO}
The classical picture of accretion discs assumes they are circular, flat (not warped), and not tilted or misaligned with respect to the rotation axis of their central object (or binary orbital plane in the case of discs in binaries). In fact, many accretion discs in nature have been observed or deduced to be distorted from this simple form, in that they are warped, eccentric or tidally deformed. ALMA and VLTI observations of discs around young stars have inferred warps from the shadows cast by a misaligned inner disc on the outer part of the disc \citep{kluska2020family, bohn2022probing}. Other examples include the misaligned (with the binary orbit) 
discs in the young binary HK Tau \citep{jensen2014misaligned,manara2019observational}, the triple star system GW Ori \citep{kraus2020triple, bi2020gw}, the tilted circumbinary disc of KH 15D \citep{chiang2004circumbinary,poon2021constraining} and the warped disc around the young star IRAS 04368+2557 \citep{sakai2019warped,villenave2024jwst}; the X-ray binaries Her X-1, SS 433, SMC MAXI J1820+070 (among others) are believed to have precessing tilted discs \citep{katz1973thirty, gerend1976optical, clarkson2003long, begelman2006nature, kotze2012characterizing, thomas2022large}, and warped discs have also been observed in the active galactic nuclei NGC 4258, NGC 1068, and the Circinus galaxy \citep{miyoshi1995evidence, herrnstein1996warp, greenhill2003warped}. Precessing eccentric discs have been deduced, for example, from observations of superhumps in close binary stars \citep{2005PASP..117.1204P} and from variable asymmetric emission lines from discs around Be stars \citep{2016ASPC..506....3O} and white dwarfs \citep{2021MNRAS.508.5657M}; eccentric discs have also been inferred from emission lines from some tidal disruption events \citep[e.g.][]{2022A&A...666A...6W}. Non-axisymmetric structure in tidally distorted discs in close binary stars has been measured using Doppler tomography \citep{2005Ap&SS.296..403M}.

Distorted discs exhibit a variety of dynamics not found in their idealized circular and planar counterparts. In particular, they are not in hydrostatic equilibrium in the direction perpendicular to the (local) orbital plane, resulting in periodic vertical contraction and expansion of the disc. This type of vertical oscillation, which we will refer to as \textit{bouncing}, can result in heating of the gas through adiabatic or dissipative effects, leading to non-axisymmetric emission \citep{ogilvie2002tidally, liska2021disc, kaaz2023nozzle}.

\subsection{Phenomena in distorted discs}
In addition to the aforementioned vertical oscillations (bouncing), distorted accretion discs can also exhibit behaviour such as nodal and/or apsidal precession of the disc \citep{tremaine2023dynamics, bardeen1975lense, deng2022non, lubow1991model, ogilvie2001}, disc breaking/tearing when the disc is strongly tilted or warped \citep{lodato2010diffusive, nixon2012broken, tremaine2014dynamics, raj2021disk, drewes2021dynamics, liska2021disc, kaaz2023nozzle, young2023conditions, nealon2022bardeen}, and hydrodynamic instability (known as \textit{parametric instability}). In this paper we will isolate the dynamics of vertical oscillations, though a new and interesting result is that these oscillations are parametrically unstable. We review both in turn below.

\subsubsection{Vertical oscillations or bouncing}
\label{SECTION_INTRO_VerticalOscillations}
Radial resonances (usually known as \textit{Lindblad} resonances) -- and, more generally, radial oscillations and waves -- in tidally forced discs have been studied in great detail \citep{goldreich1979excitation, 1987Icar...69..157M, artymowicz1994dynamics, nixon2015resonances, lubow2015tidal, armitage2022lecture, ju2016global}. However, \textit{vertical} oscillations and resonances, involving symmetric motion\footnote{For clarity, we are \textit{not} referring to vertical displacements of the mid-plane, such as those involved in bending-wave resonances in Saturn's rings, although these are also known as vertical resonances.} perpendicular to the plane of the disc, are also of interest, although they have been more sparsely studied in the literature \citep{lubow1981vertically, stehle1999hydrodynamics, stehle1999CVsimulations, ogilvie2002non, fairbairn2021non}. We will refer to the latter interchangeably as \textit{vertical oscillations}, \textit{vertical breathing modes}, or \textit{bouncing}, though some care is needed with the terminology, which we detail below.

\begin{figure}
\centering
\includegraphics[scale=0.3]{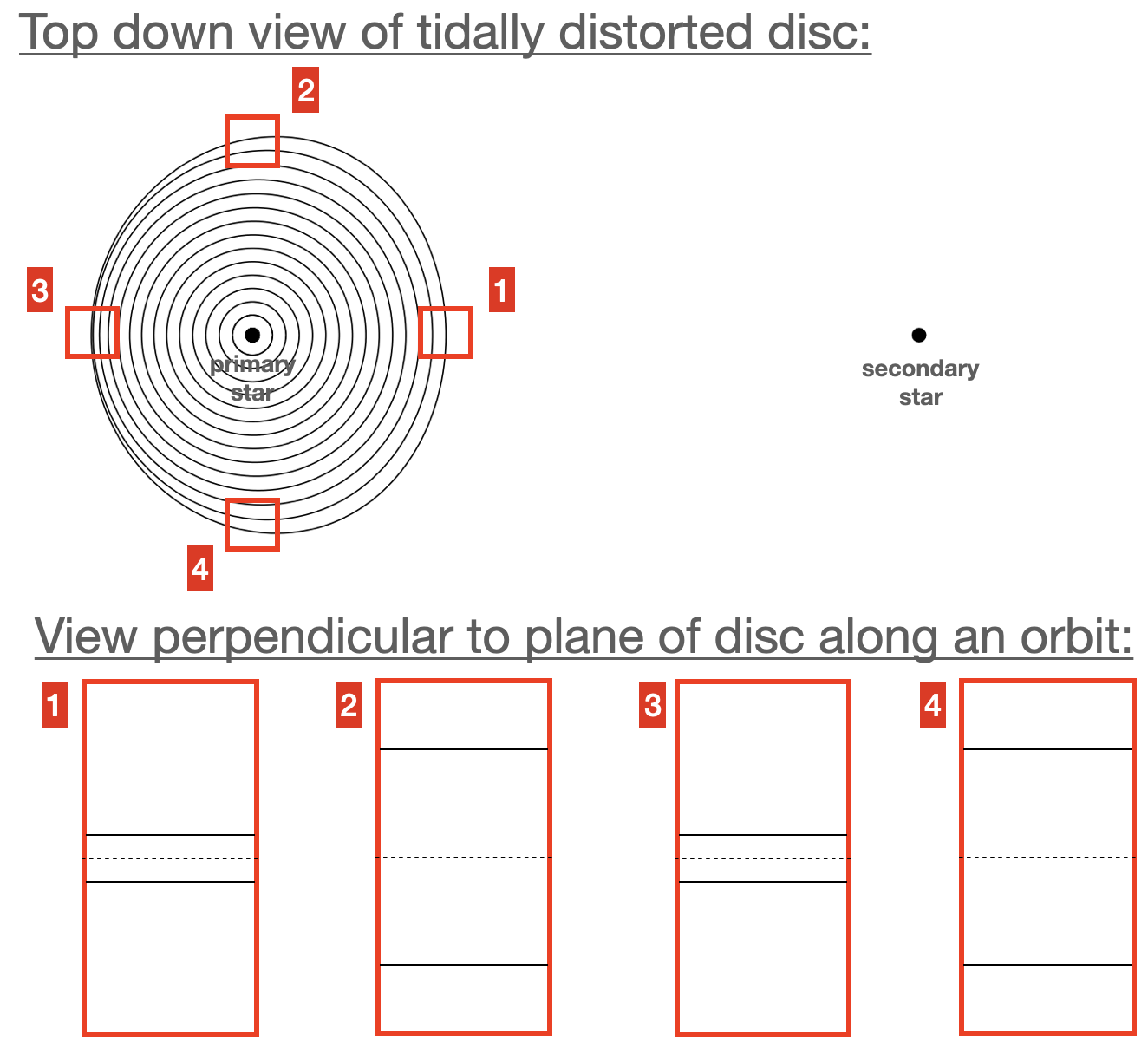}
\caption{Schematic of vertical oscillations in a tidally distorted single disc in a binary. Top: a tidally distorted disc viewed from above in the binary frame. Red boxes indicate locations along a relative orbit. Bottom: view perpendicular to the plane of the disc in the fluid frame. The disc is periodically compressed and stretched along a streamline due to the variation in the vertical component of gravity.}
\label{FIGURE_VerticalOscillationsSchematic}
\end{figure}

The specific vertical oscillations we are interested in here involves uniform expansion and contraction of the disc in the $z$-direction perpendicular to the local orbital plane, i.e.\ the displacement and velocity of the gas are proportional to $z$. See Fig.~\ref{FIGURE_VerticalOscillationsSchematic} for an illustrative example in a tidally distorted circumstellar disc in a circular binary. These oscillations can range from mild vertical breathing motion \citep{ogilvie2002non, ogilvie20083d, ryu2021impact} to a more extreme bouncing motion in which pressure support is unimportant most of the time and the disc expands and contracts in a ballistic manner in the vertical direction, with pressure acting impulsively to prevent the disc from collapsing \citep{lynch2021importance, fairbairn2021non}.\footnote{Technically, a bounce is precisely this second phenomenon (ballistic motion reversed by an impulsive force), such as when an elastic ball bounces off the ground, but we will not make this distinction in this paper.} It is worth emphasizing that these vertical oscillations of the gas occur around an orbit, in the sense that the disc (as viewed from a global reference frame) will appear `pinched' at certain locations. This can occur even in very thin discs and does not, in general, involve coupling between vertical and radial motion. It is therefore distinct from the homogeneous contraction and expansion of the entire disc in \textit{both} the radial and vertical directions. The latter oscillations can arise in slim or thick discs (tori), and are sometimes referred to as $+$- and breathing-modes \citep{blaes2006oscillation}.

Bouncing can be excited in all types of distorted disc. It has been studied in the context of warped discs \citep{ogilvie2013local, fairbairn2021non}, eccentric discs \citep{ogilvie2014local,ryu2021impact,chan2023magnetorotational}, and also in tidally distorted discs \citep{stehle1999hydrodynamics, ogilvie2002tidally}, though mechanisms driving the bouncing differ in each case. In an eccentric disc the vertical component of gravity due to the central star varies along each eccentric orbit \citep{ogilvie2001,barker2014hydrodynamic}, thus the vertical oscillation frequency is equal to the local orbital frequency. In a tidally distorted disc, the frequency at which the non-axisymmetric potential of the binary drives vertical oscillations range from about once to twice per orbit (in the frame of the fluid) depending on radial location in the disc (see Fig.~\ref{FIGURE_VerticalOscillationsSchematic}), and vertical resonances can occur at specific locations \citep{lubow1981vertically}. In a warped disc the mechanism driving vertical oscillations is more subtle: an observer moving along a fixed orbit sees neighbouring orbits tilt downwards and upwards along the trajectory. The combination of vertical motion (due to the tilting) and radial motion (due to radial pressure imbalances between neighboring vertical columns of fluid) induces vertical compression and rarefaction of the disc along an orbit at a frequency of twice per orbit, which can be enhanced by a nonlinear resonance \citep{ogilvie2013local,fairbairn2021non}.

\subsubsection{Parametric instability}
Distorted discs are unstable to a hydrodynamic instability involving the resonant amplification of inertial waves known as \textit{parametric instability}.\footnote{A classic example is the \textit{modulated pendulum} (a pendulum of variable length, or with a vertically oscillating pivot), such as when a person on a swing pumps their legs once per cycle, \citep{landau1960mechanics}. Another example (closer to what occurs in discs) is \textit{Faraday waves}, surface waves excited by vibrating a fluid container at twice the wave frequency \citep{miles1990parametrically}.} It was first discovered (theoretically) in the context of tidally distorted discs \citep{goodman1993local,1993ApJ...419..758L,ryu1994nonlinear}, and later also in warped discs \citep{papaloizou1995dynamics, gammie2000linear,ogilvie2013local} and eccentric discs \citep{papaloizou2005local, papaloizou2005global, barker2014hydrodynamic,ogilvie2014local,wienkers2018non}.

\subsection{Modelling distorted discs using local models}

Distorted discs have been modelled in global simulations using Lagrangian (smoothed-particle hydrodynamics [SPH]) \citep{lodato2010diffusive,nealon2022bardeen,deng2022non}, Eulerian (grid-based) \citep{kimmig2024warped}, and 
meshless approaches \citep{deng2021parametric,deng2022non}. However, global simulations struggle to capture some of the key physics of distorted discs, in particular the parametric instability, although this has been achieved with grid-based \citep{2020MNRAS.496.2849P} and meshless \citep{deng2021parametric} codes. For example, SPH simulations employ a smoothing length (typically some fraction of the disc thickness), which dampens the instability and also artificially damps the bouncing motion (confusing it with a shock-forming flow), while grid-based simulations introduce an advection error when the flow is not aligned with the grid.

These difficulties have motivated local (shearing-box) studies of distorted discs, which have led to warped \citep{ogilvie2013local, paardekooper2019local}, as well as eccentric \citep{ogilvie2014local, wienkers2018non} extensions to the standard shearing-box formalism. Perhaps surprisingly, much of the physics of distorted discs can actually also be captured in a \textit{standard} shearing box \citep{ogilvie2022hydrodynamics}, which is the approach we take in this work. There are limitations to this approach, however. First, the standard shearing box follows \textit{circular} streamlines in a disc. Thus we do not take into account any geometrical effects due to eccentric or tidally distorted orbits, in order to focus solely on the dynamics of the vertical oscillations. Second, the standard shearing box cannot pick up a large-scale warp, only a radially periodic warp (also known as a \textit{corrugation} or \textit{bending} wave/mode), although a non-periodic ring or torus with a large-scale warp can be studied in the local model \citep{fairbairn2021non}.

\subsection{Motivation and aims of paper}
Our main aim is to isolate the dynamics of vertical (i.e. perpendicular to the local plane of the disc) oscillations or bouncing expected to occur in tidally distorted, eccentric, and warped accretion discs. We do so by means of quasi-2D and fully 3D hydrodynamic shearing box simulations. For simplicity we employ an isothermal equation of state. Note that in astrophysical contexts we expect the oscillations to be forced (driven). However, to determine the main mechanisms damping the oscillations we first consider free (unforced) oscillations.

The structure of the paper is as follows. In Section \ref{METHODS_GoverningEquations} we discuss our set-up, initial conditions, parameters, and key diagnostics. In Section \ref{RESULTS_Theory} we present a 1D theory (including a stability analysis) of vertical disc oscillations. We present simulations of free oscillations in Section \ref{RESULTS_FreeBounceSimulations}. In Section \ref{RESULTS_ForcedBounceSimulations} we discuss forced oscillations (including astrophysically relevant forcing frequencies and amplitudes). We conclude in Section \ref{CONCLUSIONS}.

\section{Methods}
\label{METHODS}

\subsection{Governing equations}
\label{METHODS_GoverningEquations}
We work in the shearing-box approximation
\citep{goldreich1965,hawley1995,latter2017local},
which treats a local region of a disc as a Cartesian box located at some fiducial radius $r = r_0$ and orbiting with the angular frequency of the disc at that radius, $\Omega_0 \equiv \Omega(r_0)$. A point in the box has
Cartesian coordinates $(x, y, z)$ along the radial, azimuthal, and vertical directions, respectively. In this rotating frame, the equations of gas dynamics are:
\begin{align}
&\partial_t \rho + \nabla \cdot (\rho \mathbf{u}) = 0, \label{SB1}\\
&\partial_t \mathbf{u} + \mathbf{u}\cdot\nabla \mathbf{u} = -\frac{1}{\rho} \nabla P - 2\Omega_0 \mathbf{e}_z \times \mathbf{u} + \mathbf{g_\text{eff}}+\frac{1}{\rho}\nabla \cdot \mathbf{T}, \label{SB2}
\end{align}
with the symbols taking their usual meanings. We close the system with the equation of state for an isothermal gas, $P = c_\text{s}^2 \rho$, where $c_\text{s}$ is the constant sound speed.

All our simulations are stratified in the vertical direction. In our simulations of a freely vertically oscillating disc, the effective gravitational potential is embodied in the usual (constant-in-time) tidal acceleration $\mathbf{g_\text{eff}}=2q\Omega_0^2x\,\mathbf{e}_x-\Omega_0^2 z\,\mathbf{e}_z$ (third term on the right-hand side of Equation \ref{SB2}), where $q$ is the dimensionless shear parameter $q \equiv -\left.d\ln{\Omega}/d\ln{r}\right\vert_{r=r_0}$. For Keplerian discs $q=3/2$, a value we adopt throughout this paper.

In order to drive vertical oscillations/bouncing (Section~\ref{RESULTS_ForcedBounceSimulations}), we introduce a \textit{time-dependent} gravitational acceleration in the vertical direction, by modulating the usual gravity of the shearing box according to $g_{\text{eff},z} = -\Omega_0^2 z (1+a\cos{\omega t})$, where $a$ is the forcing amplitude and $\omega$ the forcing frequency.
With these two parameters we can mimic the driving of the vertical component of gravity found, e.g.\ in a binary system in which a companion external to the disc modulates the vertical gravity experienced by gas in the disc. We determine realistic values of these parameters to use in our forced bounce simulations in Section~\ref{SECTION_ForcedBounceAstrophysicalApplications}.

Finally, the viscous stress tensor is given by $\mathbf{T} \equiv 2\rho \nu \mathbf{S}$, where $\nu$ is the kinematic viscosity, and $\mathbf{S} \equiv (1/2)[\nabla \mathbf{u} + (\nabla \mathbf{u})^\text{T}] - (1/3)(\nabla\cdot\mathbf{u})\mathbf{I}$ is the traceless shear tensor \citep{landau1987}. We keep the viscosity fixed in space and time in any given simulation. We have also run select simulations without an explicit viscosity.

\subsection{Numerical set-up}
\label{METHODS_NumericalSetUp}

\subsubsection{Code}
\label{Methods_Codes}
For our simulations we use the conservative, finite-volume code \textsc{PLUTO} \citep{mignone2007}. We employ the HLLC Riemann solver, 2nd-order-in-space linear interpolation, and the 2nd-order-in-time Runge-Kutta algorithm. To allow for longer time-steps, we take advantage of the \textsc{FARGO} scheme \citep{mignone2012}. When explicit viscosity $\nu$ is included, we further reduce the computational time via the Super-Time-Stepping (STS) scheme \citep{alexiades1996super}. Ghost zones are used to implement the boundary conditions.

We use the built-in shearing box module in \textsc{PLUTO}
\citep{mignone2012}. Rather than solving Equations \eqref{SB1}--\eqref{SB2} (primitive
form), \textsc{PLUTO} solves the governing equations in conservative form.

\subsubsection{Units}
\label{METHODS_Units}
Note that from this point onwards, all quantities are given in terms of dimensionless (code) units. Time units are selected so that $\Omega_0 = 1$. The length unit is chosen so that the initial sound speed $c_{\text{s}0} = 1$, which in turn defines a reference \textit{equilibrium} scale-height $H_0\equiv c_{\text{s}0} / \Omega_0=1$, which can be thought of as the disc thickness in vertical hydrostatic equilibrium. Finally the mass unit is set by the initial mid-plane density $\rho_0 = 1$. Pressure, stresses, and energy densities are expressed in units of $c_{\text{s}0}^2 \rho_0$. From here onwards we will drop the subscripts on $\Omega_0$ and $c_{\text{s}0}$, since these quantities are constant. However, we will retain the subscript on $H_0$ (the equilibrium scale-height) to distinguish it from the the time-dependent \textit{dynamical} scale-height $H(t)$ (formally defined below).

\subsubsection{Initial conditions}
\label{METHODS_InitialConditions}
We employ different initial conditions depending on whether the vertical oscillations are free or forced. In the former case, we start from a disc that has been uniformly stretched in the vertical direction, with a Gaussian density profile of the form
\begin{equation}
\rho = \rho_0 \exp{\{-z^2/[2H^2(0)]\}},
\label{EQUN_densityprofile}
\end{equation}
where $H(0)$ is the dynamical scale-height (defined below) at initialization ($t=0$). In other words $H(0)<1$ ($H(0)>1$) corresponds to a disc that is compressed (stretched) in the vertical direction compared to its equilibrium thickness $H_0$. Note that this initial condition is \text{not} in equilibrium by design, in order to induce the disc to oscillate. In our freely bouncing disc simulations we select $H(0)$ to study bounces of small and large amplitude, respectively. In our forced bounce simulations we set $H(0)$ based on the theoretical predicted value for the disc thickness at a given forcing frequency (see Section \ref{SECTION_ForcedBounceTheory}).

The background velocity is given by $\mathbf{u}_0 = -q \Omega x \,\mathbf{e}_y$. At initialization we usually perturb all the velocity components with random noise exhibiting a flat power spectrum (different random noise is used for each component). The perturbations $\delta \mathbf{u}$ have maximum amplitude of about $5\times10^{-2}\,c_\text{s}$.

\subsubsection{Box size and resolution}
\label{METHODS_BoxSizeAndResolution}
The majority of our simulations are fully three-dimensional (3D) and run at a resolution of $32$ cells per scale-height $H_0$ (denoted $32/H_0$), in a box of size $L_y = 4H_0$ in the azimuthal direction and $L_z = 12H_0$ in the vertical direction. (We also repeated our fiducial 3D run of free [unforced] vertical oscillations at $64/H_0$). As the radial wavelength of the fastest growing mode of the parametric instability is sensitive to the forcing frequency $\omega$, we vary the radial box size at different forcing frequencies to ensure we always capture the fastest growing mode, from $L_x = 4H_0$ (at $\omega/\Omega = 1$) to $L_x = 30H_0$ (at $\omega/\Omega = 2$). To investigate the dependence on the vertical box size we have also repeated select unforced runs in boxes of size $[4,4,8]H_0$ and $[4,4,16]H_0$ (keeping the resolution per scale-height fixed at $32$ cells per $H_0$). Finally, we have also run a series of quasi-2D simulations in boxes of size $[8,0.008, 12]H_0$. These enable us to reach resolutions up to $512/H_0$. All the simulations described in this paper are listed in Tables \ref{TABLE_3DSimsFreelyBouncing_ComparisonBounceAmplitude}--\ref{TABLE_3DSimsFreelyBouncing_ComparisonExplicitNumericalViscosity} in Appendix \ref{APPENDIX_TablesOfSimulations}.

\subsubsection{Boundary conditions}
\label{METHODS_BoundaryConditions}
We use standard shear-periodic boundary conditions (BCs) in the $x$-direction \cite[see][]{hawley1995}, and periodic boundary conditions in the $y$-direction. In the vertical direction we mostly employ reflective boundary conditions ($\rho \rightarrow \rho$, $u_{x,y} \rightarrow u_{x,y}$, and $u_z \rightarrow -u_z$). However, we have also checked our fiducial 3D simulation with \textit{outflow} boundary conditions, for which we set $\partial_z u_i = 0$ ($i \in {x,y,z}$), described in greater detail in \cite{held2024mri}. Note that when we use outflow boundary conditions mass (and energy) can be advected across the vertical boundaries, and are not added back into the domain using a mass source term as was done in \cite{held2024mri}. 

\subsection{Diagnostics}
\label{METHODS_Diagnostics}

\subsubsection{Averaged quantities}
\label{METHODS_AveragedQuantities}
The volume-average of a quantity $X$ is denoted $\langle X \rangle$ and is defined as 
\begin{equation}
\langle X \rangle(t) \equiv \frac{1}{V} \int_V X(x, y, z, t) \, \dd V,
\end{equation}
where $V=L_x L_y L_z$ is the volume of the box.

We denote the horizontal average of a quantity $X$ as $\langle{X}\rangle_{xy}$, defined as
\begin{equation}
    \langle X \rangle_{xy}(z,t) \equiv \frac{1}{A} \int_A X(x,y,z,t) \, \dd x \, \dd y,
\end{equation}
where $A = L_x L_y$.

\subsubsection{Reynolds stress}
\label{METHODS_ReynoldsAndMagneticStressesAndAlpha}
In purely hydrodynamic accretion discs, the radial transport of angular momentum is related to the $xy$-component of the Reynolds stress $R_{xy} \equiv \rho u_x \delta u_y$, where $\delta u_y \equiv u_y + q\Omega x$ is the perturbation of the y-component of the total velocity $u_y$ about the background Keplerian flow $u_{0y} = -q \Omega x$.\footnote{Note that the Reynolds stress is related to the classical dimensionless angular momentum transport parameter by, e.g. $\alpha \equiv \langle R_{xy} \rangle / \langle P \rangle$,  though $\alpha$ is defined in slightly different ways by different authors, so caution is needed when comparing values from different sources \citep[e.g.,][]{pessah2008fundamental, heldlatter2018}.} (From now on we will drop the $\delta$ and simply refer to the perturbed velocity as $\textbf{u}$.)

\subsubsection{Dynamical scale-height}
A key diagnostic in our simulations is the \textit{dynamical scale-height} $H(t)$ which measures the time-variability of the disc thickness as it oscillates in the vertical direction. We define the square of this quantity as follows:
\begin{equation}
H^2(t) = \frac{1}{\Sigma} \int_{-L_z/2}^{L_z/2} \langle \rho \rangle_{xy} z^2 \, \dd z,
\label{EQUN_DynamicalScaleheight}
\end{equation}
where $\Sigma \equiv \int_{-L_z/2}^{L_z/2} \langle \rho \rangle_{xy} dz$ is the surface density of the disc. Thus $H$ is related to the second moment of the density and can be interpreted as the standard deviation of $z$ (assuming zero mean) of the mass elements of the disc. 

Finally, we define the \textit{amplitude} of the bounce as
\begin{equation}
\Delta H_n=H_{\text{max},n}-H_{\text{min},n},
\label{EQUN_OscillationAmplitude}
\end{equation}
where $H_{\text{max},n}$ and $H_{\text{min},n}$ correspond to the maximum and minimum scale-heights, respectively, in the $n$th oscillation cycle (we define a cycle as starting when the disc is at maximum thickness, so $H_{\text{min},n}$ occurs half a cycle later).

\subsubsection{Energetics}
Another useful diagnostic is the total energy density, which is given by
\begin{equation}
E_\text{total} = \frac{1}{2}\rho u^2 + \rho \Phi + c_\text{s}^2 \rho \ln{\rho},
\label{EQUN_TotalEnergy}
\end{equation}
where the first two terms on the right-hand side correspond to the kinetic $E_{\text{kin}}$ and gravitational potential energy density, respectively.  
The term $c_\text{s}^2 \rho \ln{\rho}$ embodies the variable part of the Helmholtz free energy, which replaces the internal energy in an isothermal gas.\footnote{For a perfect gas, the specific internal energy is $e=RT/(\gamma-1)+\text{constant}$, where $RT=p/\rho=c_\text{s}^2$, while the specific Helmholtz free energy is $f=e-Ts=RT(1-\ln T)/(\gamma-1)+RT\ln \rho+\text{constant}$. Under isothermal conditions, the relevant free energy is therefore $c_\text{s}^2\ln\rho+\text{constant}$.} In our freely bouncing disc simulations, $\Phi =-\frac{3}{2}\Omega^2x^2+\frac{1}{2}\Omega^2z^2$, which is the standard (time-independent) effective gravitational potential in the shearing-box approximation for a Keplerian disc. In our forced bouncing disc simulations, the vertical component of the gravitational potential is \textit{time-dependent} (to model the time-variation in vertical gravity along an orbit in a distorted disc). In this case, a term of the form $\rho \partial\Phi/\partial t$ appears as a source term on the right-hand side of the total energy equation.

\section{Theory}
\label{RESULTS_Theory}

In this section we derive some basic theoretical predictions related to disc bouncing in the absence of damping or forcing. In Section \ref{SECTION_1DTheoryofVerticallyBouncingdisc} we write down the governing equations for 1D vertical bouncing, and, by considering first hydrostatic solutions, use these to help derive a governing equation for the 
dynamical scale-height $H(t)$ in the non-hydrostatic case. In Section \ref{SECTION_ParametricInstabilityTheory} we examine the stability of the basic bouncing flow to radially dependent perturbations, and discover that the bouncing flow is parametrically unstable. We defer a discussion of \textit{forced} bouncing to Section~\ref{SECTION_ForcedBounceTheory}.

\begin{figure}
\centering
\includegraphics[scale=0.7]{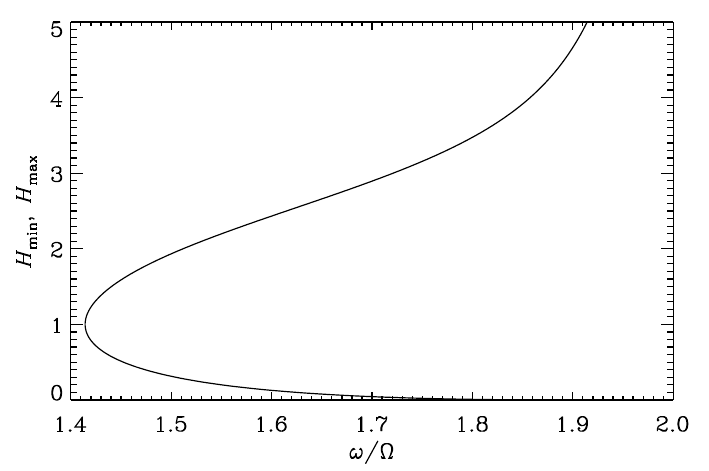}
\caption{Prediction of oscillation frequency $\omega$ (in units of the orbital frequency $\Omega$) of free (unforced) vertical oscillations as a function of oscillation amplitude. The top branch shows the maximum disc thickness $H_{\text{max}}$ (in units of equilibrium disc thickness $H_0$), and the bottom branch shows the minimum disc thickness $H_{\text{min}}$.}
\label{FIGURE_FreeBounceHmaxHminPrediction}
\end{figure}

\subsection{1D theory of a vertically bouncing disc}
\label{SECTION_1DTheoryofVerticallyBouncingdisc}
\subsubsection{Governing equations}

Using the shearing-box model of Section \ref{METHODS_GoverningEquations}, we consider a vertical motion of the disc, independent of $x$ and $y$, in which the vertical velocity $u_z(z,t)$ and the density $\rho(z,t)$ satisfy the continuity equation and the $z$-component of the momentum equation,
\begin{align}
    &\frac{D \ln{\rho}}{D t} = - \frac{\partial u_z}{\partial z},\label{EQUATION_1DTheoryContinuityEquation}\\
    &\frac{D u_z}{D t} = -\Omega^2 z - \frac{1}{\rho}\frac{\partial p}{\partial z},\label{EQUATION_1DTheoryMomentumEquation}
\end{align}
together with $p=c_\text{s}^2\rho$, where the material derivative is given by $D/Dt \equiv \partial_t + u_z \partial_z$.
As for the horizontal components of the momentum equation, the orbital shear flow is unaffected by the vertical motion.

\subsubsection{Hydrostatic equilibrium solution}
In hydrostatic equilibrium we take $u_z = 0$, $\rho = \rho(z)$ and $p=p(z)$ and the governing equations reduce to the structural equation
\begin{equation}
    \frac{\dd p}{\dd z} = - \rho \Omega^2 z.
    \label{EQUATION_1DTheoryHydrostaticEquilibrium}
\end{equation}
Together with the isothermal condition $p=c_\text{s}^2\rho$, this equation determines the usual Gaussian structure of a hydrostatic disc, which we write in the form
\begin{equation}
  \rho=\frac{\Sigma}{H}F_{\rho}(\zeta),
\label{rho_dimensionless}
\end{equation}
where
\begin{equation}
  F_{\rho}(\zeta)=\frac{1}{\sqrt{2\pi}}\exp\left(-\frac{1}{2}\zeta^2\right)\label{EQUATION_GaussianDimensionlessDensity}
\end{equation}
is a normalized Gaussian function and $\zeta=z/H$ is a dimensionless vertical coordinate. For hydrostatic equilibrium, the scale-height is given as usual by $H=c_\text{s}/\Omega$.

\begin{figure*}
\centering
\includegraphics[scale=0.49]{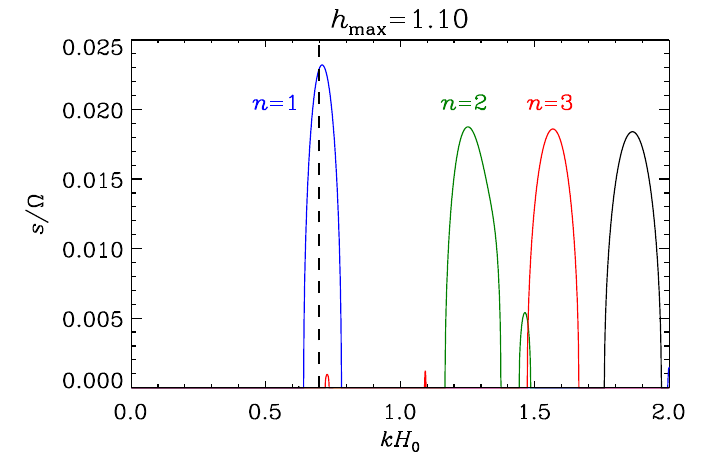}
\includegraphics[scale=0.49]{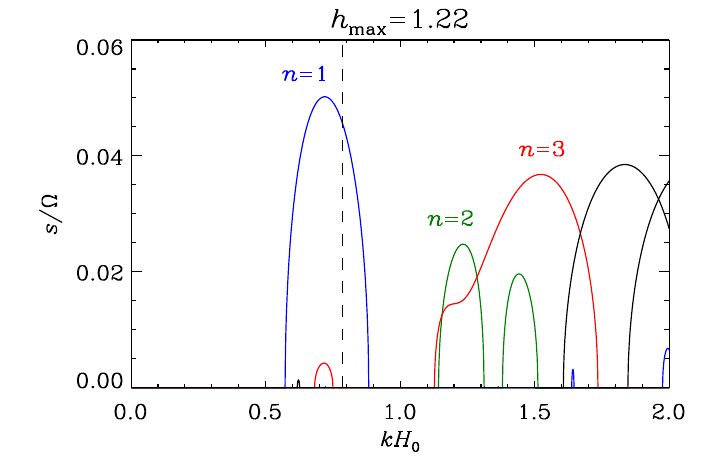}
\includegraphics[scale=0.49]{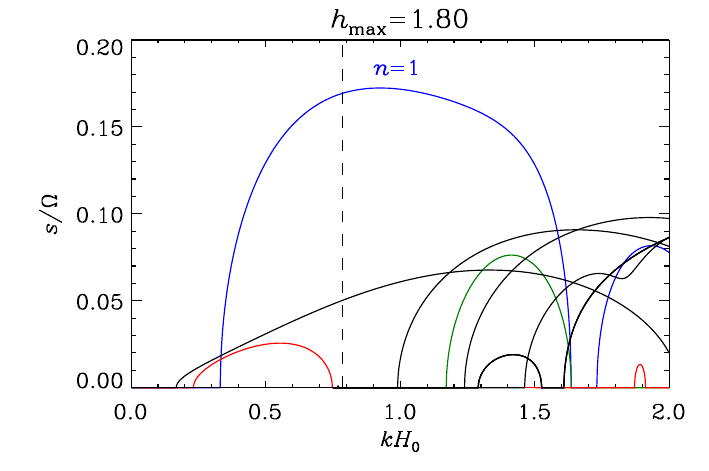}
\caption{Growth rate $s$ (in units of disc angular frequency $\Omega$) as a function of radial wavenumber $k$ of parametric instability for free (unforced) vertical oscillations with different disc thicknesses $h_\text{max}$ (this can be though of as a measure of the oscillation amplitude). The blue curve corresponds to the mode with vertical wavenumber $n=1$; other curves show higher order modes. The vertical dashed line marks the wavenumber $2\pi/L_x$ corresponding to the radial box size ($L_x=9H_0$ or $L_x=8H_0$) used in the simulations.}
\label{FIGURE_FreeBounceGrowthRates}
\end{figure*}

\subsubsection{Non-hydrostatic solutions}
\label{SECTION_1DTheoryofVerticallyBouncingdisc_NonHydrostaticSolutions}
To study the bouncing motion, we now aim to derive an equation governing the time-evolution of the disc thickness, or scale-height. We look for non-hydrostatic solutions in which the scale-height $H(t)$ used in the non-dimensionalization above depends on time. The solutions we are interested in have the same dimensionless density (and pressure) profile as the hydrostatic solution; they are achieved by a uniform, time-dependent expansion or contraction of the model. The dimensionless variable $\zeta$ becomes a Lagrangian coordinate, which means that the vertical velocity is given by $u_z=(\dd H/\dd t)\zeta=(\dd H/\dd t)(z/H)$. Mass conservation implies that the surface density $\Sigma$ is independent of time. Substituting the density~(\ref{rho_dimensionless}), with a time-dependent scale-height, into the momentum equation \ref{EQUATION_1DTheoryMomentumEquation}, we find that all terms are proportional to $\zeta$, and thereby obtain an evolution equation for the disc thickness or dynamical scale-height $H(t)$:
\begin{equation}
    \frac{\dd^2 H}{\dd t^2} = -\Omega^2 H + \frac{c_\text{s}^2}{H}.
    \label{EQUATION_1DTheoryDynamicalHEvolution2}
\end{equation}
The hydrostatic disc corresponds to the equilibrium solution $H=H_0=c_\text{s}/\Omega$.

In the units defined in Section~\ref{METHODS_Units}, we have
\begin{equation}
    \ddot H = -H + \frac{1}{H},
    \label{EQUATION_1DTheoryDynamicalHEvolution3}
\end{equation}
where a dot denotes a time-derivative.

We note the derivation above can be extended to allow for adiabatic thermodynamics and different vertical structure profiles, in which case the dimensionless scale-height satisfies
$\ddot H = -H + H^{-\gamma}$, where $\gamma$ is the adiabatic exponent \citep[cf.][]{ogilvie2014local}.

\subsubsection{Periodic solutions}
We now derive predictions for the oscillation period in the small- and large-oscillation amplitude limits. Let us first consider the case of small-amplitude oscillations $H=1+\epsilon(t)$. Substituting into Equation \ref{EQUATION_1DTheoryDynamicalHEvolution3} and linearizing we obtain $\ddot\epsilon=-2\epsilon$,
which describes small-amplitude oscillations of angular frequency $\sqrt{2}$ (or $\sqrt{\gamma+1}$ when adiabatic thermodynamics is taken into account).

To consider large-amplitude oscillations we first rewrite the right-hand side of Equation \ref{EQUATION_1DTheoryDynamicalHEvolution3} in terms of the gradient of a potential $V$,
\begin{equation}
    \frac{\dd^2 H}{\dd t^2} = -\frac{\dd V}{\dd H},
    \label{EQUATION_1DTheoryDynamicalHEvolution3Linearized}
\end{equation}
where $V(H) = (1/2)H^2 - \ln H$. Integrating, we obtain
\begin{equation}
\frac{1}{2}\left(\frac{\dd H}{\dd t}\right)^2 + V \equiv E,
\end{equation}
where the sum of kinetic and potential energies $E$ is a constant. Rearranging this equation and integrating between the minimum and maximum disc thickness over half an oscillation, we obtain an expression for the period of the oscillation for arbitrary oscillation amplitude:
\begin{equation}
    T = 2\int_{H_\text{min}}^{H_\text{max}} \frac{\dd H}{\sqrt{2(E-V)}}.
    \label{EQUATION_1DTheoryOscillationPeriod}
\end{equation}
In the limit of large amplitude we find $2\pi / T \rightarrow 2$ \citep{fairbairn2021non}.

The angular frequency increases from $\sqrt{2}$ to $2$ as the amplitude $H_\text{max}-H_\text{min}$ increases from $0$ to $\infty$ (see Fig.~\ref{FIGURE_FreeBounceHmaxHminPrediction}).

\subsection{Parametric instability of solutions}
\label{SECTION_ParametricInstabilityTheory}
Motivated by the simulation results to be described below, we analyse the stability of a periodically bouncing disc to axisymmetric perturbations that are independent of $y$ but depend on $x$ through a common factor $\ee^{\ii kx}$, where $k$ is the radial wavenumber. The linearized equations for small perturbations can be formulated in terms of the Lagrangian displacement $\boldsymbol{\xi}$. We find
\begin{align}
  &\frac{D^2\xi_x}{D t^2}=-\Omega^2\xi_x-\ii k\frac{p'}{\rho},\\
  &\frac{D^2\xi_z}{D t^2}=\frac{\ddot H}{H}\xi_z-\frac{\partial}{\partial z}\left(\frac{p'}{\rho}\right),\\
  &p'=c_\text{s}^2\rho',\\
  &\frac{\rho'}{\rho}=-\ii k\xi_x-\frac{\partial\xi_z}{\partial z}-\xi_z\frac{1}{\rho}\frac{\partial\rho}{\partial z}.
\end{align}
Transforming from Eulerian $(z,t)$ to Lagrangian $(\zeta,t)$ variables [recalling (Section \ref{SECTION_1DTheoryofVerticallyBouncingdisc_NonHydrostaticSolutions}) that $D/Dt \equiv \partial_t + u_z \partial_z$ and $u_z = (dH/dt)(z/H) \equiv \dot{H} \zeta$, and using the Gaussian dimensionless density profile (Equation \ref{EQUATION_GaussianDimensionlessDensity}) to evaluate $(1/\rho)\partial_z \rho$ in the last equation], we find
\begin{align}
  &\frac{\partial^2\xi_x}{\partial t^2}=-\Omega^2\xi_x-\ii k\frac{p'}{\rho},\\
  &\frac{\partial^2\xi_z}{\partial t^2}=\frac{\ddot H}{H}\xi_z-\frac{1}{H}\frac{\partial}{\partial\zeta}\left(\frac{p'}{\rho}\right),\\
  &\frac{p'}{\rho}=c_\text{s}^2\left(-\ii k\xi_x-\frac{1}{H}\frac{\partial\xi_z}{\partial\zeta}+\xi_z\frac{\zeta}{H}\right).
\end{align}
By analogy with the linear modes of a disc in hydrostatic equilibrium \citep{okazaki1987gloval}, we seek a solution in Hermite polynomials:
\begin{align}
  &\xi_x=X(t)\,\text{He}_n(\zeta),\\
  &\xi_z=Z(t)\,\text{He}_n'(\zeta),\\
  &\frac{p'}{\rho}=W(t)\,\text{He}_n(\zeta),
\end{align}
where the vertical mode number $n$ is a non-negative integer. The linearized equations are satisfied provided that
\begin{align}
  &\ddot X=-\Omega^2X-\ii kW,\label{floquet1}\\
  &\ddot Z=\frac{\ddot H}{H}Z-\frac{W}{H},\label{floquet2}\\
  &W=-c_\text{s}^2\left(\ii kX-\frac{nZ}{H}\right).\label{floquet3}
\end{align}
For an equilibrium disc ($H=H_0$), these equations have constant coefficients and therefore have harmonic solutions with angular frequency $\omega$ given by the dispersion relation
\begin{equation}
  (\omega^2-\Omega^2)(\omega^2-n\Omega^2)=\omega^2\Omega^2k^2H^2,
\end{equation}
in agreement with \citet{okazaki1987gloval}. For a bouncing disc, we have instead a fourth-order system of ordinary differential equations with time-periodic coefficients [because $H(t)$ is periodic] and there is the possibility of instability. Floquet's method can be used to analyse the solutions and their growth rates. 
We solve the dimensionless version of equations (\ref{floquet1})--(\ref{floquet3}), in which the factors of $\Omega$ and $c_\text{s}$ drop out and $H(t)$ satisfies $\ddot H=-H+H^{-1}$ (see Equation \ref{EQUATION_1DTheoryDynamicalHEvolution3}).
Starting at $t=0$ with an initial value $H(0)>1$ and with $\dot H(0)=0$, meaning that the disc is released from rest above its equilibrium scale-height. By integrating four different solutions of the system (\ref{floquet1})--(\ref{floquet3}) with four linearly independent initial conditions such as $(X,Z,\dot X,\dot Z)=(1,0,0,0)$, $(0,1,0,0)$, $(0,0,1,0)$ and $(0,0,0,1)$ and integrating for one oscillation period $T$ of $H(t)$, we construct the monodromy matrix that maps the initial values to the final values one period later \citep{teschl2024ordinary}. We calculate its eigenvalues, which are the Floquet multipliers $\lambda$, and convert these into exponential growth rates $s$ via $\exp(sT)=|\lambda|$. 

\begin{figure*}
\centering
\includegraphics[scale=0.39]{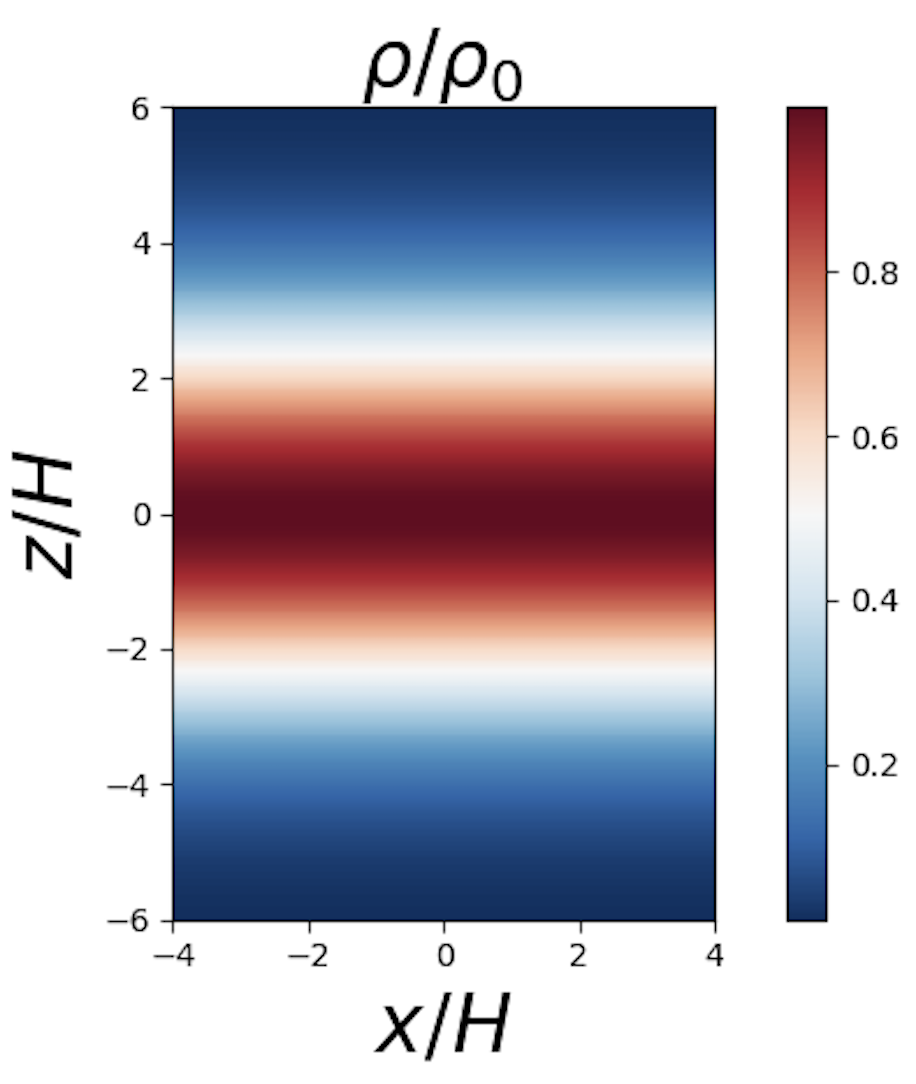}
\includegraphics[scale=0.39]{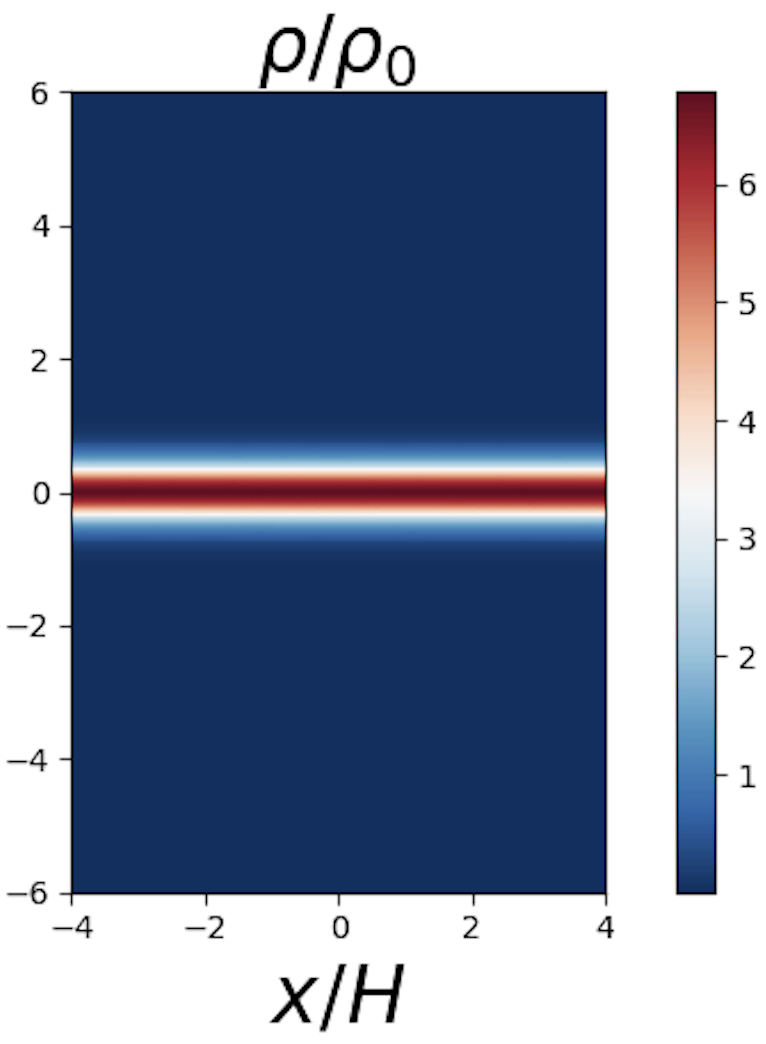}
\includegraphics[scale=0.39]{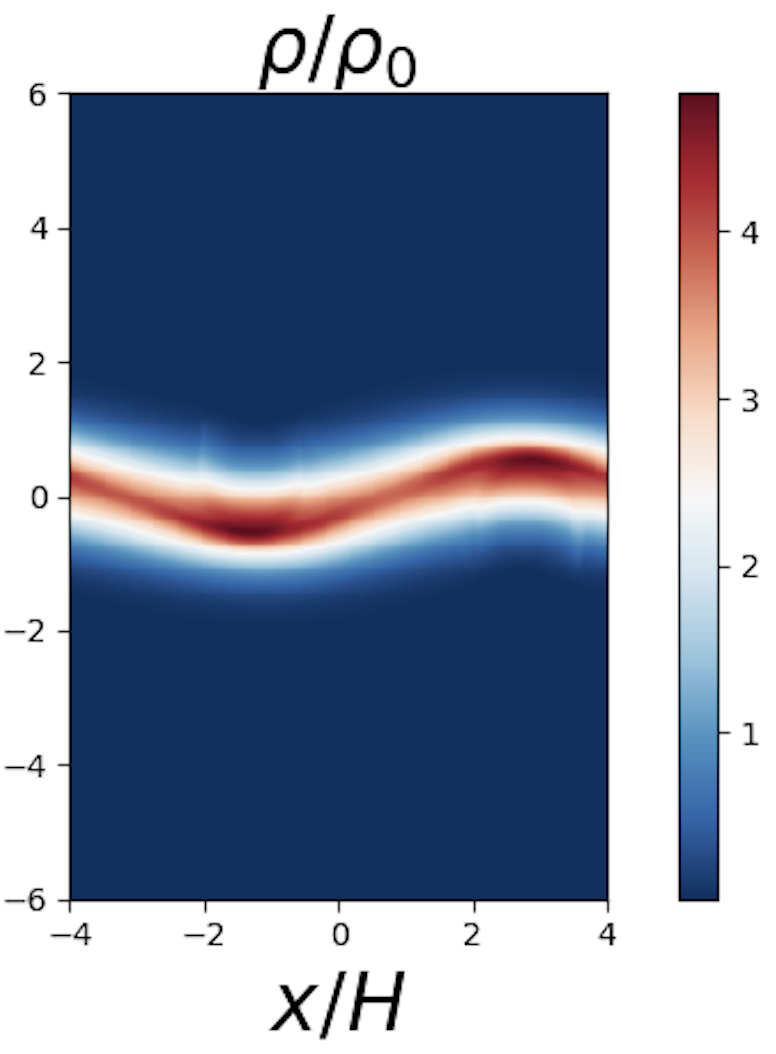}
\includegraphics[scale=0.26]{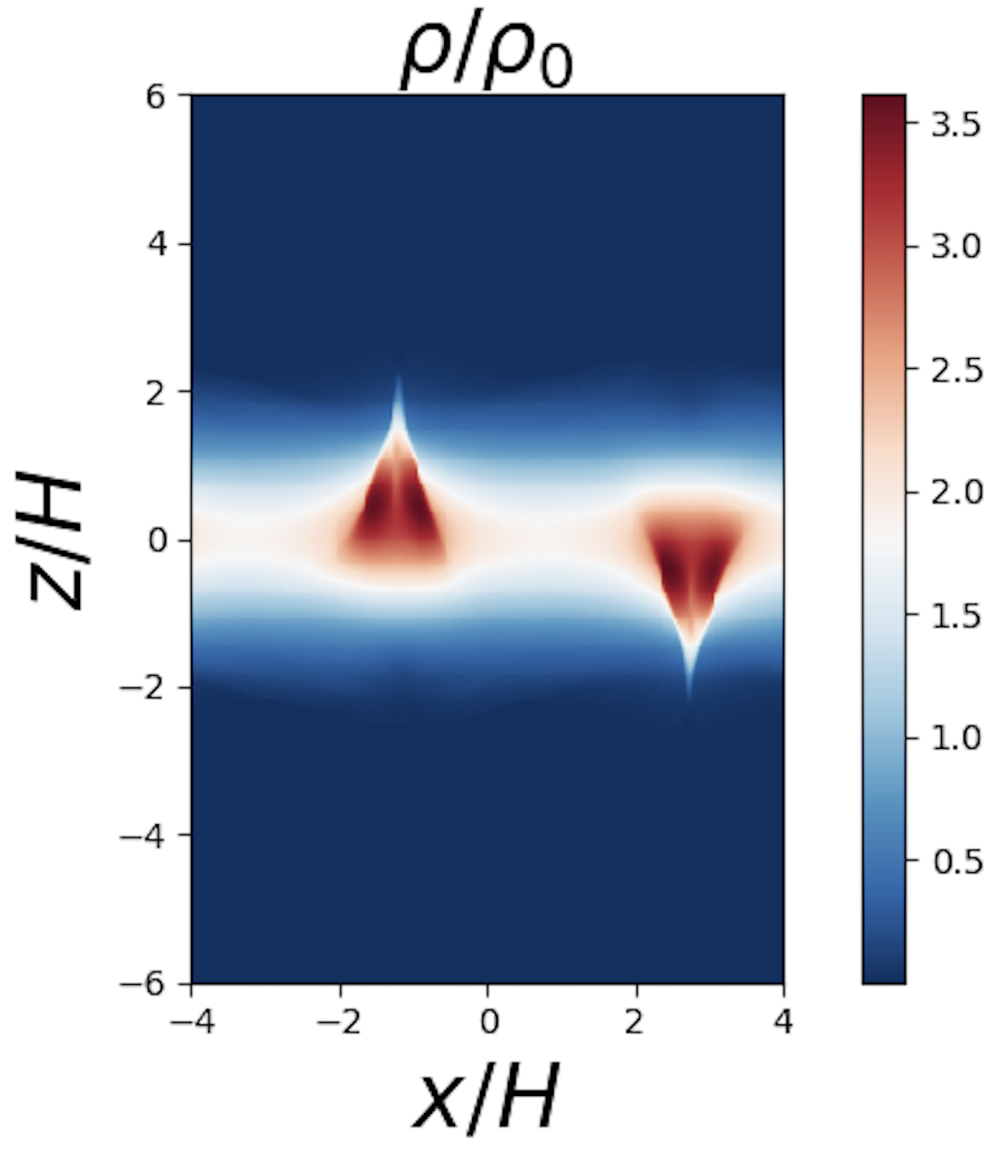}
\includegraphics[scale=0.31]{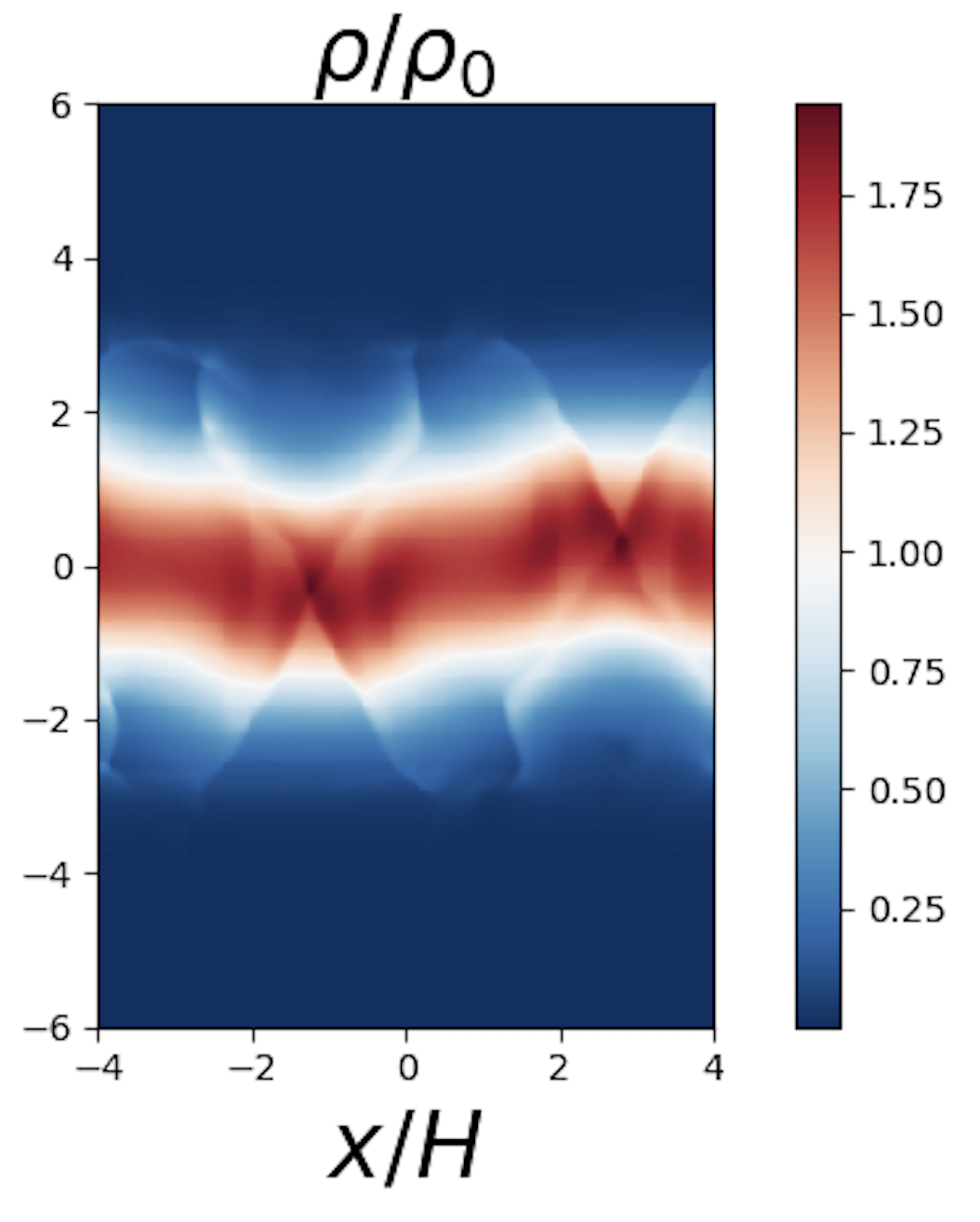}
\includegraphics[scale=0.31]{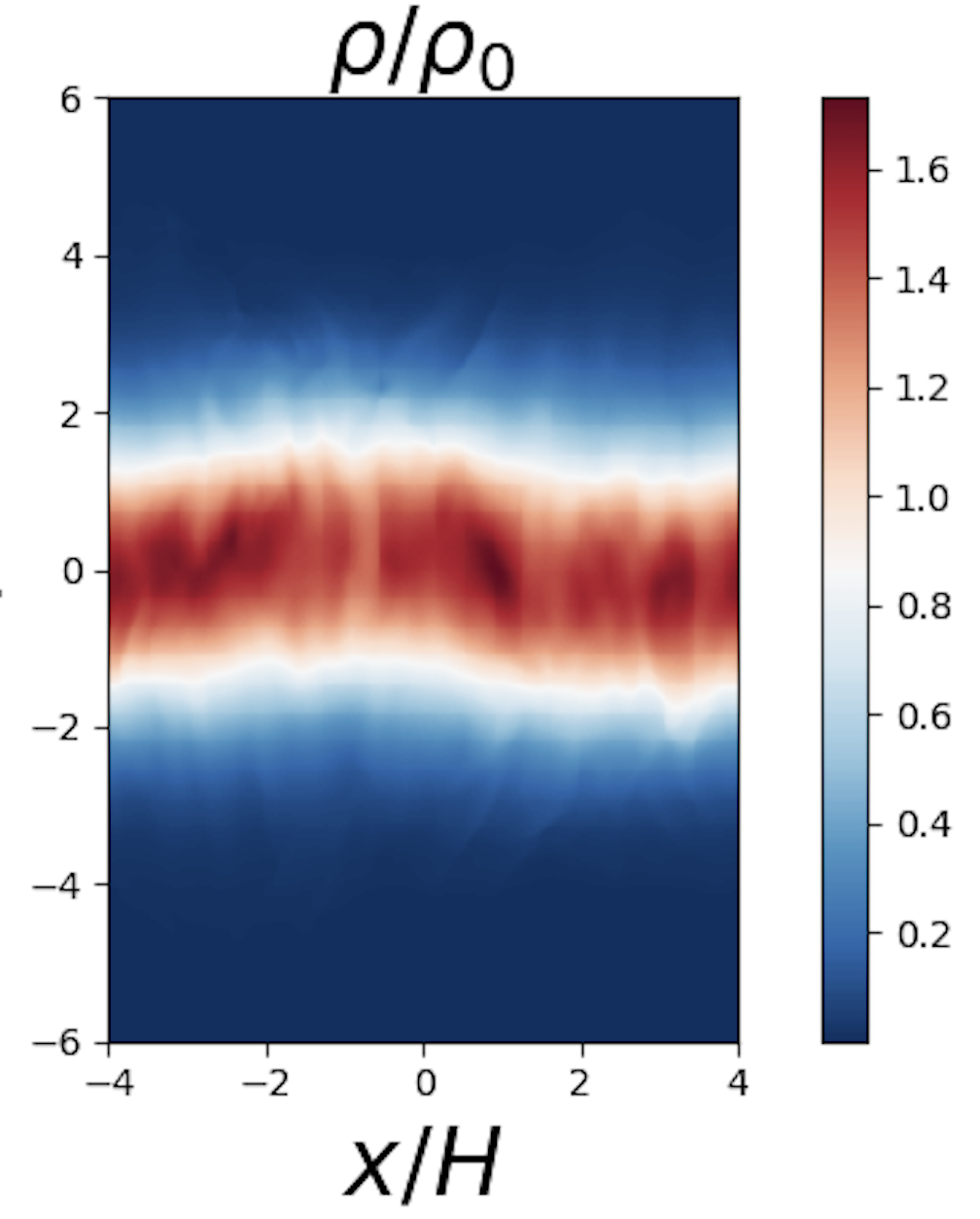}
\caption{Density snapshots ($xz$-plane) in the fiducial simulation of free (unforced) vertical oscillations. Top left: initial condition at orbit 0 (the initial disc thickness is $H(0)\sim 1.97_0$). Top-middle: The disc oscillates 1.5 times per orbit (orbit 0.32). Top-right: after several orbits a warp/bending wave develops in the disc due to the parametric instability (orbit 9). Bottom-left: the radial pressure mismatch in the warped disc drives counter-propagating radial flows that result in the disc becoming highly compressed where these waves meet, shown here shortly after nonlinear saturation (orbit 11). After nonlinear saturation, waves and shocks dampen the oscillation (bottom-middle and bottom-right panels at orbit 12 and orbit 19, respectively).}
\label{FIGURE_FiducialSimulationFlowField}
\end{figure*}

The linearized equations in the form written above are also valid when the bouncing motion is forced, provided that $H(t)$ now satisfies the appropriate forced equation such as $\ddot H=-(1+a\cos\omega t)H+H^{-1}$.

Typical results for the instability of freely bouncing discs are shown in Fig.~\ref{FIGURE_FreeBounceGrowthRates}. The nature of the instability is very similar to that analysed for eccentric discs analysed by \citet{barker2014hydrodynamic}. An inertial wave with a natural frequency close to half the bounce frequency is destabilized by the oscillating geometry of the disc. The instability is an example of three-wave coupling in which a parent mode (the bouncing motion) couples to two daughter modes (inertial waves travelling radially inwards and outwards). For the coupling to be exactly resonant, the frequencies and wavenumbers of the three modes must sum to zero for some choice of sign combinations. Since the parent mode has radial wavenumber $k=0$, the two daughter modes have the same value of $k$. Selection rules regarding the vertical mode number $n$ imply that the daughter modes are identical except for the direction of propagation; the combined growing mode has the character of a standing wave. For each $n$ there is a single $k$ at which the inertial-wave frequency is half the bounce frequency and the resonance condition is satisfied. The growth rate at exact resonance increases with the bounce amplitude. Each resonant band has a width that also increases with the bounce amplitude, because parametric resonance can tolerate some detuning.

Analytical expressions can be obtained for the radial wavenumber and growth rate of the parametric instability in the limit of small bounce amplitude. We omit the details of this calculation. If $\omega$ is the angular frequency of the bouncing motion (whether free or forced), then for vertical mode $n$ the resonance is centred at
\begin{equation}
  k=\frac{\sqrt{4-\omega^2}\sqrt{4n-\omega^2}}{2\omega}
\end{equation}
and has a maximum growth rate of
\begin{equation}
  \frac{n\omega(4-\omega^2)}{16n-\omega^4}(H_\text{max}-1),
\end{equation}
which is proportional to the semi-amplitude $H_\text{max}-1$. The fastest-growing mode is $n=1$, with growth rate
\begin{equation}
  \frac{\omega}{4+\omega^2}(H_\text{max}-1).
\end{equation}
In the case of free bouncing ($\omega=\sqrt{2}$) this simplifies to $(\sqrt{2}/6)(H_\text{max}-1)\approx0.236(H_\text{max}-1)$ and the resonant wavenumber is $k=1/\sqrt{2}$, corresponding to a wavelength of nearly~$9$ scale-heights.

It is interesting to compare these results with the case of eccentric discs. \cite{papaloizou2005local} first analysed the parametric instability of inertial waves in eccentric discs, using a model that neglects vertical gravity. In this case the instability is due to the oscillatory horizontal geometry of the disc associated with the non-circular orbital motion. \cite{barker2014hydrodynamic} revisited the problem and found that the growth rate was significantly enhanced by the vertical bouncing of the disc driven by the variation of vertical gravity around an elliptical orbit. In the present work we are finding a related instability that derives solely from the bouncing motion and not from any non-circular orbital motion.

\section{Simulations of unforced oscillations}
\label{RESULTS_FreeBounceSimulations}

Having presented our theoretical model in the previous section, we now carry out simulations of free (unforced) vertical oscillations. We initialize each simulation with an initial condition in which the Gaussian disc profile (see Equation \ref{EQUN_densityprofile}) has been stretched in the vertical direction. We perturb each velocity component at initialization with random perturbations. We begin with a description of our fiducial simulation in Section \ref{SECTION_FreeBounceFiducialSimulation}, and identify the main dissipative mechanisms at work during a bounce by means of an energy analysis. We then discuss a series of quasi-2D simulations, which enable us to check the key results at much higher resolutions (Section \ref{SECTION_FreeBounceQuasi2DSimulations}). Finally, in Section \ref{SECTION_FreeBounceParameterStudy} we systematically investigate the effects of changing various physical and numerical parameters such as the initial oscillation amplitude, vertical box size, vertical boundary conditions, and explicit versus numerical viscosity. 

\subsection{Fiducial simulation}
\label{SECTION_FreeBounceFiducialSimulation}
In this section we describe our fiducial simulation. The disc has been stretched in the vertical direction at initialization, so that the initial density profile is $\rho(z,t=0) = \rho_0 \exp(-z^2/2H(0)^2)$ where we take $H(0) = 2H_0$.\footnote{Note that due to the finite extent of the box in the vertical direction, the actual dynamical scale-height at initialization is around $H(0) \sim 1.97H_0$.} At initialization we perturb all three velocity components with (different) random noise of amplitude about $0.05 c_\text{s}$. The box size is $[L_x,L_y,L_z] = [8,4,12]H_0$ at a resolution of $32$ cells per $H_0$ in all three directions. We use reflective boundary conditions in the vertical direction, and run the simulation for 50 orbits ($\sim 314\,\Omega^{-1}$). Note that we repeated this fiducial simulation at double the resolution ($64/H_0$) and obtained very similar results. We also carried out a much more comprehensive resolution study, but in 2D, which we will discuss in Section \ref{SECTION_FreeBounceQuasi2DSimulations}. 

\begin{figure}
\centering
\includegraphics[scale=0.30]{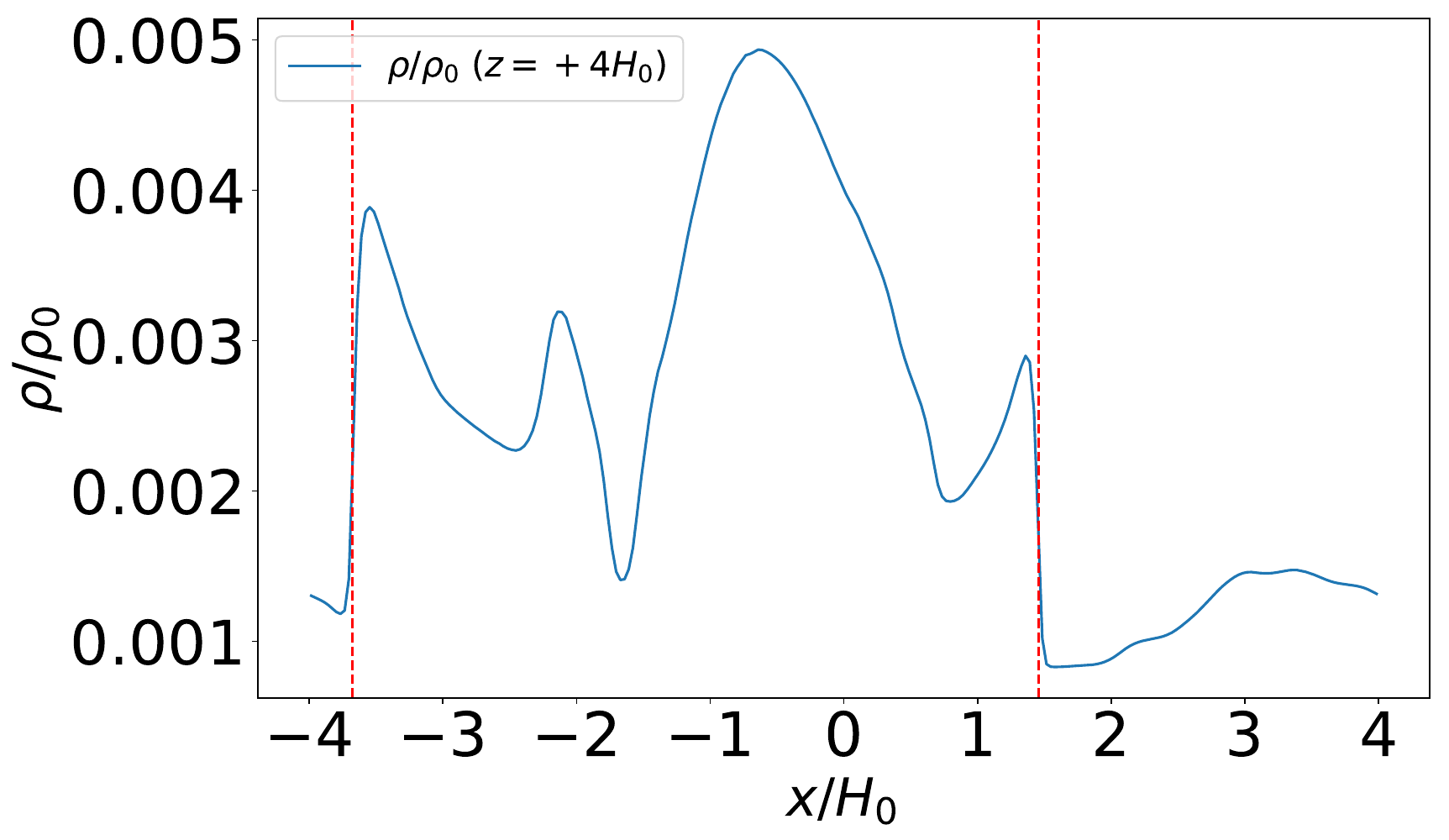}
\includegraphics[scale=0.30]{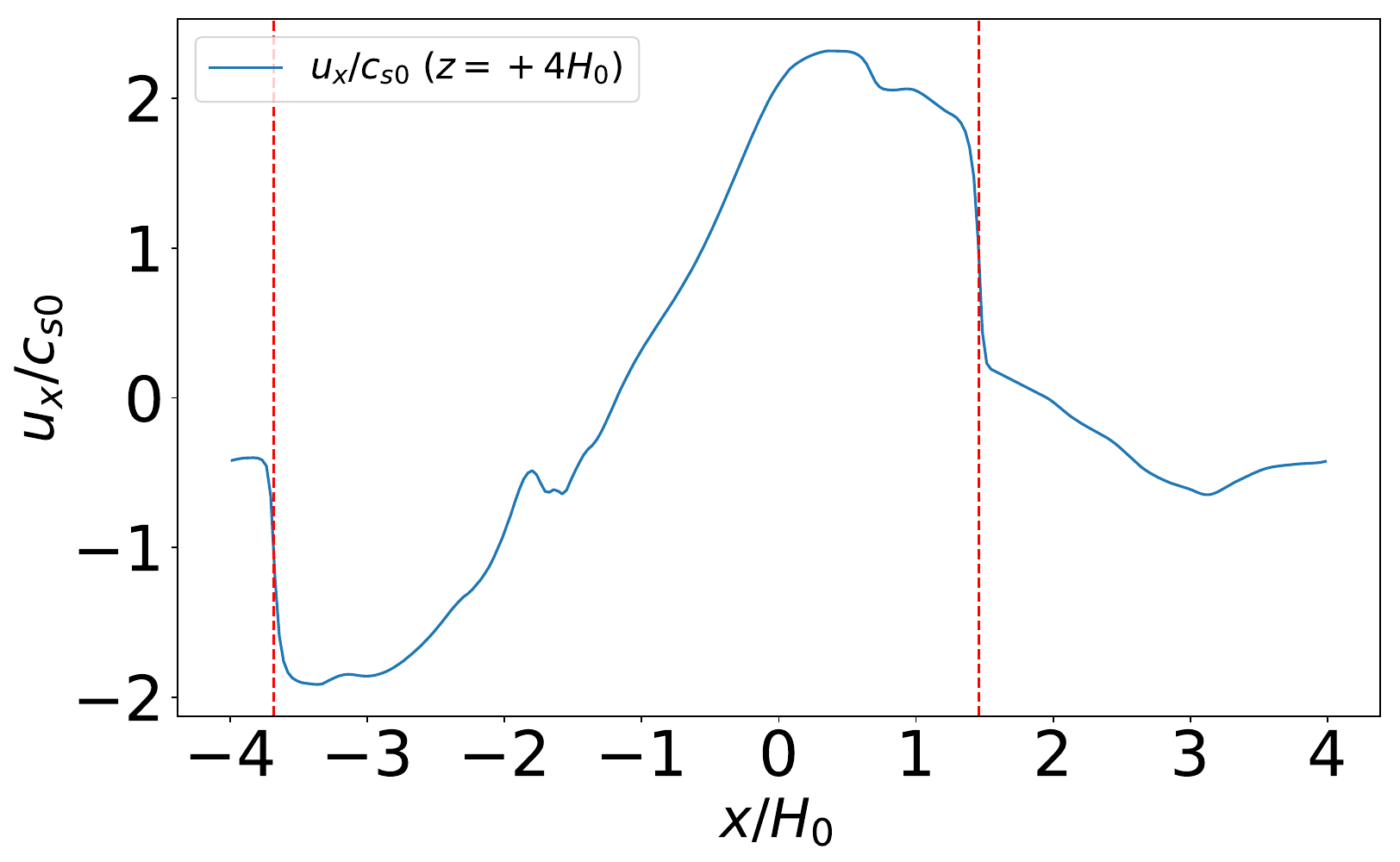}
\includegraphics[scale=0.30]{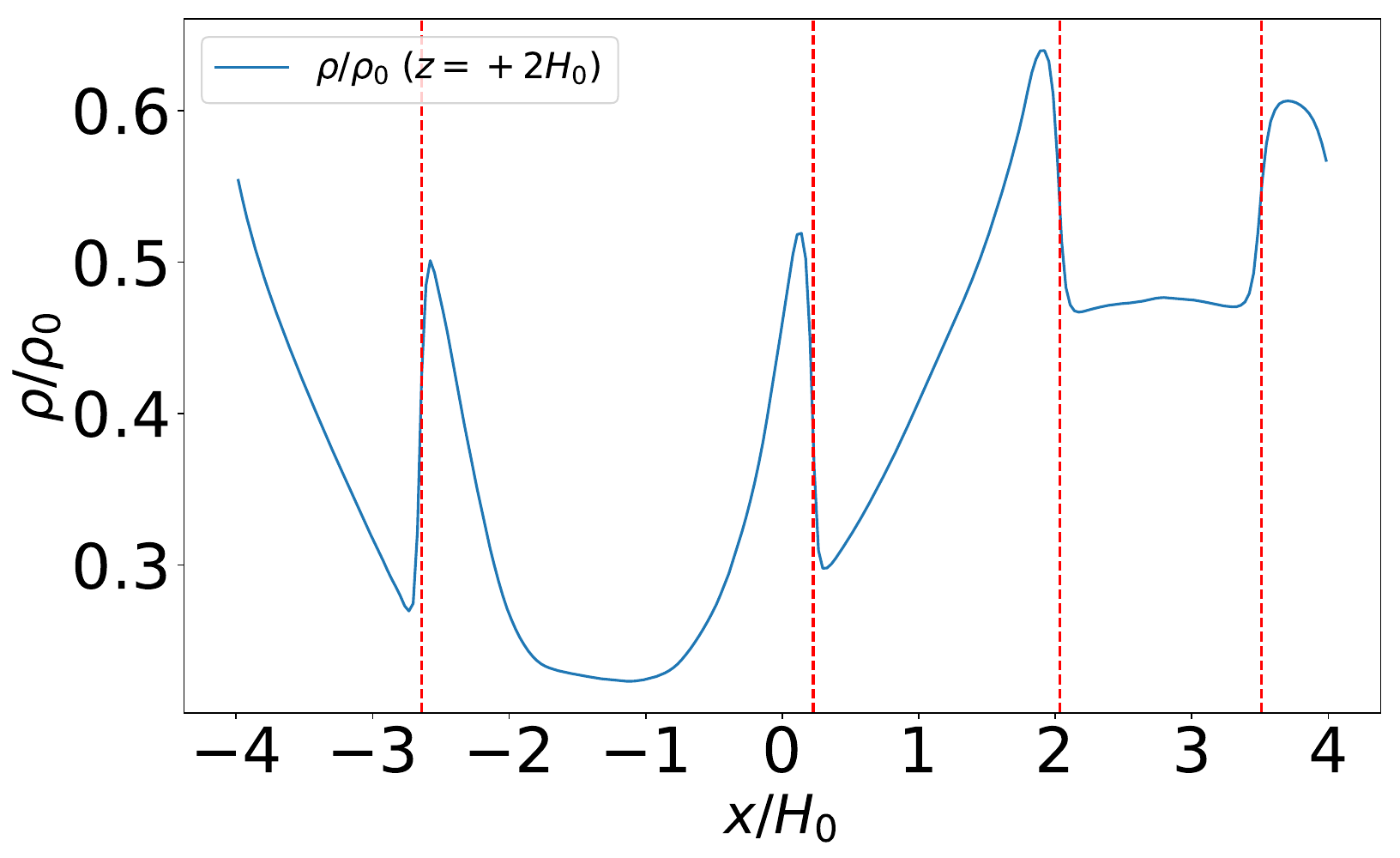}
\includegraphics[scale=0.30]{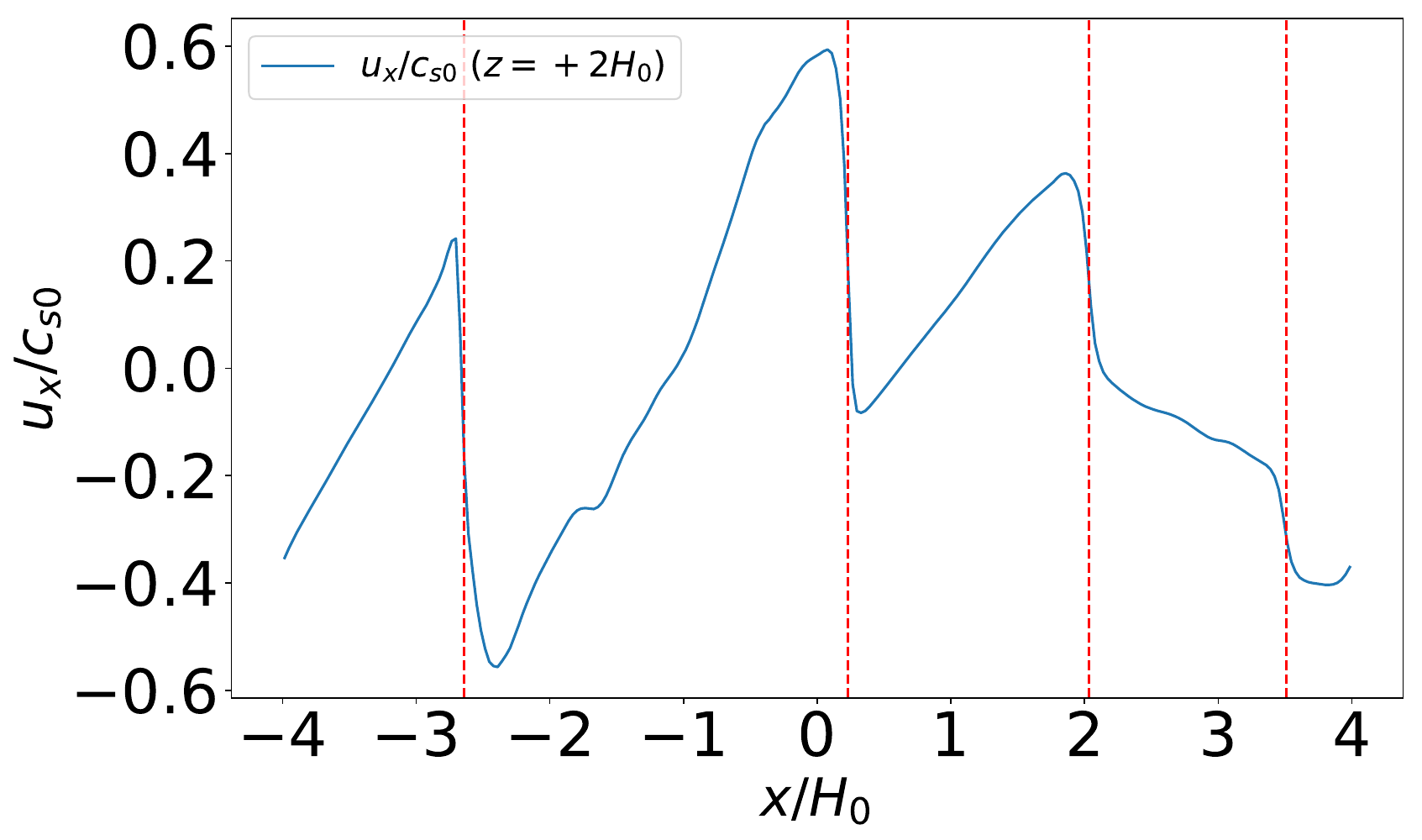}
\caption{Radial disc structure at $z=+4H_0$ (top two panels, for density $\rho$ and radial velocity $u_x$, respectively), and at $z=+2H_0$ (bottom two panels), taken from a snapshot around orbit 12 (see bottom-middle panel of Fig.~\ref{FIGURE_FiducialSimulationFlowField}) from our fiducial simulation of a freely bouncing disc. Red dashed lines mark locations of discontinuities in the density.}
\label{FIGURE_FiducialFreeBounceSimRadialdiscstructure}
\end{figure}

\begin{figure}
\centering
\includegraphics[scale=0.22]{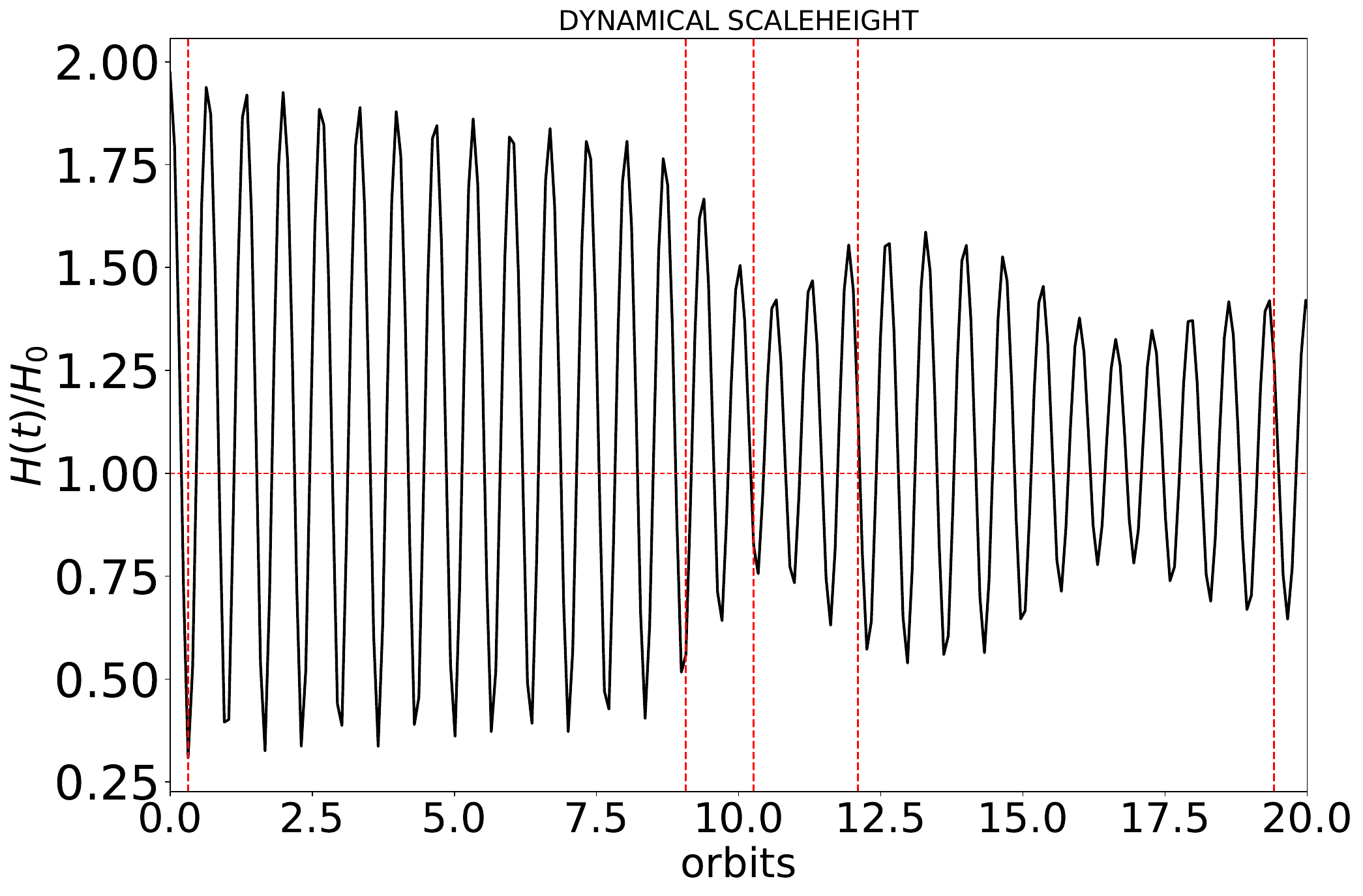}
\includegraphics[scale=0.22]{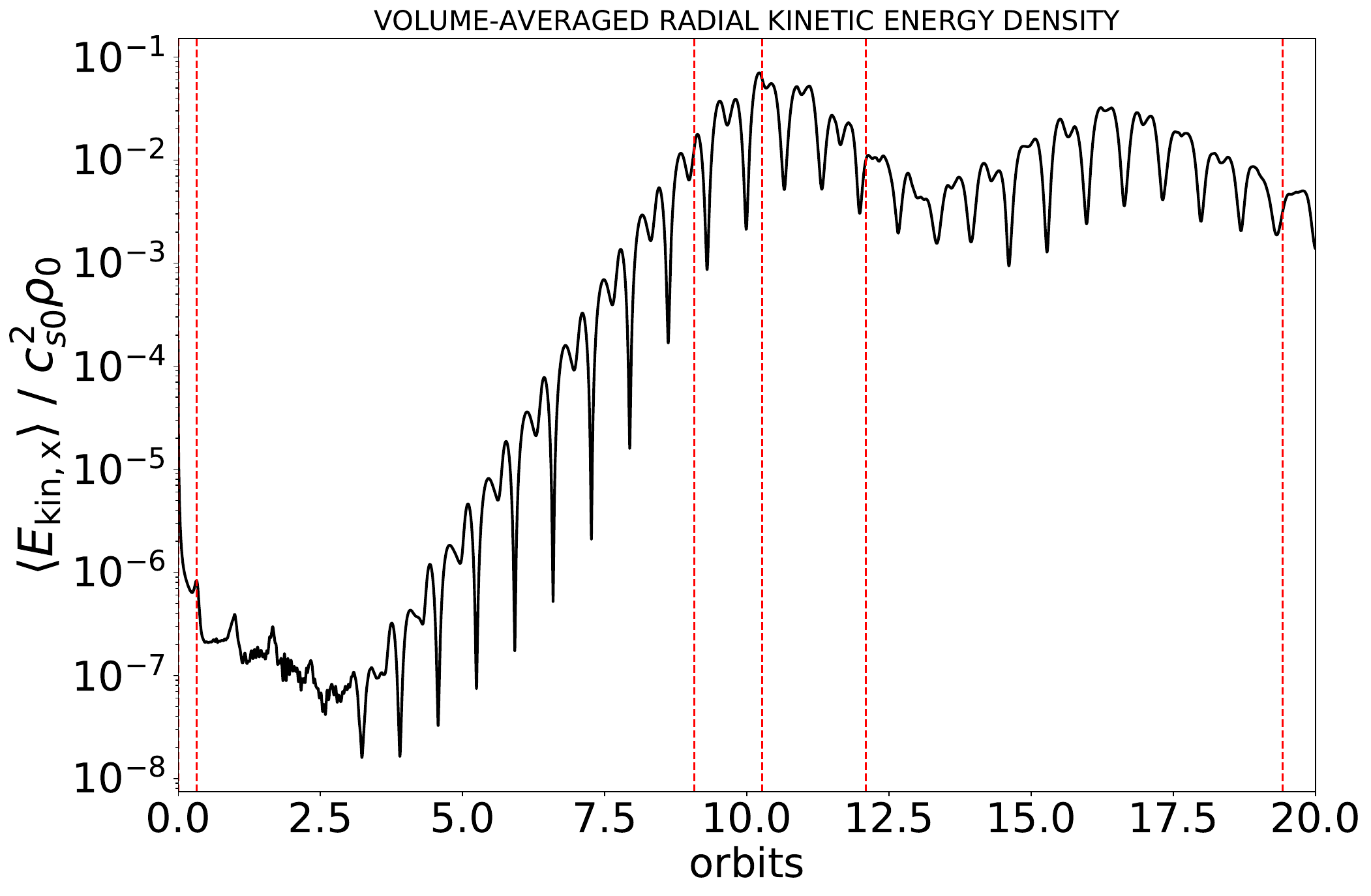}
\caption{Top: time-evolution (over first 20 orbits) of dynamical scale-height $H(t)$ from fiducial freely oscillating 3D simulation. Bottom: time-evolution of volume-averaged radial kinetic energy density. Vertical dashed lines correspond to times of snapshots shown in Fig.~\ref{FIGURE_FiducialSimulationFlowField} (except the top-left one, which is taken at initialization).}
\label{FIGURE_FiducialSimTimeEvolutionDynamicalHAndEkinx}
\end{figure}

\subsubsection{Flow field}
In Fig.~\ref{FIGURE_FiducialSimulationFlowField} we plot the density in the $xz$-plane from snapshots at different times to illustrate the evolution of our initially out-of-equilibrium disc. After initialization (top-left panel of Fig.~\ref{FIGURE_FiducialSimulationFlowField}) the disc rapidly collapses towards the mid-plane, reaching a thickness of around $0.3H_0$ after around 0.33 orbits (top-middle panel), before expanding again due to pressure support. For the first six orbits the disc continues oscillating in this manner. We observe two shocks propagating towards the disc mid-plane from the reflective vertical boundaries shortly after the disc reaches its maximum thickness (we check the effects of more realistic outflow boundary conditions below). The bounce amplitude decreases over the course of each bounce, with the dissipation dominated by shocks as discussed below. 

However, around orbit 7 the disc begins to exhibit a noticeable warp or bending wave (top-right panel), which develops into a standing wave as the disc continues to bounce. This wave grows as a result of the parametric instability described in Section~\ref{SECTION_ParametricInstabilityTheory}, where it corresponds to the $n=1$ mode. As the amplitude of the warp grows we see that the counter-propagating radial fluid motions, which are a component of the bending wave, eventually become so strong that they compress the disc in the radial direction by a factor of nearly 3.5 around orbit 10 (bottom left-hand panel), corresponding to nonlinear saturation of the parametric instability (see below). After nonlinear saturation the disc continues to oscillate in the vertical direction (albeit with reduced amplitude) while maintaining a time-dependent warped structure in the form of a standing wave. The mid-plane and atmosphere of the disc during this period are characterized by shocks and waves (bottom-middle and bottom-right panels in Fig.~\ref{FIGURE_FiducialSimulationFlowField}).

To investigate the structures we observe in the 2D flow field in greater detail, in Fig.~\ref{FIGURE_FiducialFreeBounceSimRadialdiscstructure} we show radial slices through $z = 2H_0$ and $z = 4H_0$ taken from a snapshot shortly after nonlinear saturation (bottom-middle panel of Fig.~\ref{FIGURE_FiducialSimulationFlowField}). At $z = 2H_0$ (bottom two panels) we find four discontinuities in the density at $x/H_0 = -2.64, 0.23, 2.03$ and $3.51$, respectively. These coincide with discontinuities in the radial velocity. The profiles resemble N-waves resulting from wave steepening and are likely shocks.\footnote{In a frame moving with the shock, the normal velocity must decrease from supersonic to subsonic as it passes through the shock. But if the shock is moving in the computational frame, the fluid velocities may be subsonic on both sides.} In the disc atmosphere (top two panels), we find clear evidence of strong shocks at the radial locations $x = -3.68H_0$ and $x=1.46H_0$, respectively. Here the motion is supersonic with the radial velocity at the discontinuities reaching around $2 c_\text{s}$. (Note that we have repeated this analysis in one of our quasi-2D simulations at a much higher resolution of $512/H_0$. The wave- and shock-fronts are sharper at higher resolution, but the disc structure is the same.)

\subsubsection{Time-series}
\label{RESULTS_FiducialSimulationTimeSeries}
To quantify the behaviour we observe in the flow field, we next turn to the top panel of Fig.~\ref{FIGURE_FiducialSimTimeEvolutionDynamicalHAndEkinx} where show the time-series of the dynamical scale-height (see Equation \ref{EQUN_DynamicalScaleheight}). After initialization the disc rapidly collapses, reaching a thickness of around $0.3H_0$ half-way through the first bounce. Pressure support then causes the disc to expand. The oscillation or bounce period $\Delta T$ is around 0.66 orbits or around 1.5 bounces per orbit, which is consistent with the theoretical prediction (Fig.~\ref{FIGURE_FreeBounceHmaxHminPrediction}) when $H_\text{max} = 2$. The oscillation amplitude (cf. Equation \ref{EQUN_OscillationAmplitude}) decays gradually over the first few orbits of the simulation from an initial amplitude of $\Delta H \sim 1.67H_0$ at initialization to around $1.55 H_0$ at orbit 8. Thereafter the amplitude of the oscillation drops rapidly over several orbits, reaching a minimum of $\sim 0.55 H_0$ around orbit 10. As we will describe shortly, this occurs during the nonlinear saturation of the parametric instability during which a warp (bending mode) is seeded in the disc, which feeds off the bouncing motion. Surprisingly, during the nonlinear phase of the instability (orbit 10 to 20) we observe periods during which the oscillation is `re-invigorated' or increased over several orbits. The amplitude of the oscillation is smaller in each successful re-invigoration phase, indicating that the bouncing motion is decaying with time, as expected since there is nothing to drive the oscillation in this simulation. In Appendix~\ref{APPENDIX_Coupling_Model} we provide a mathematical model of coupled and damped oscillators that can explain the observed re-invigoration.

Next we turn to the time-evolution of the volume-averaged radial kinetic energy density, shown for this simulation in the bottom panel of Fig.~\ref{FIGURE_FiducialSimTimeEvolutionDynamicalHAndEkinx}, which is a good diagnostic for tracking the development of the parametric instability of the background vertically oscillating flow. On short (orbital) timescales this oscillates with the oscillation of the disc scale-height $H(t)$ (top panel). More markedly, however, it exhibits exponential growth between around orbits 4 and 10, before saturating at around $7\times10^{-2}$ (in units of $c_\text{s}^2 \rho_0$). Note that by the time the parametric instability starts to grow, the maximum thickness has been damped from the initial value of $H(0) = 1.97H_0$ to $H_\text{max} \sim 1.8H_0$. The growth rate of the amplitude (i.e.\ half the growth rate of the radial kinetic energy) during the linear phase is around $0.169\Omega^{-1}$, in very good agreement with value predicted from theory of $0.17\Omega^{-1}$ for an oscillation with $H _\text{max}= 1.8H_0$ (see right-hand panel of Figure \ref{FIGURE_FreeBounceGrowthRates}). After nonlinear saturation, the radial kinetic energy density also exhibits phases of decay and re-invigoration, similar to (but out of phase with) the dynamical scale-height. This predator--prey type behaviour can be understood as follows. First, the vertical oscillation (bouncing) excites a warp/bending mode by means of parametric instability, leading to a decrease in the oscillation amplitude, but an increase in radial kinetic energy. The warp/bending mode then excites vertical oscillations due to the combination of tilt and shear, as explained in the last paragraph of Section \ref{SECTION_INTRO_VerticalOscillations}. This cyclic behaviour is damped, however, because of the dissipation associated with shocks.

Finally, we have also looked at the time-evolution of the various Reynolds stress tensor components (not shown). During most of the linear phase the Reynolds stress is zero. Before and after nonlinear saturation, however, the stress becomes highly oscillatory, peaking in amplitude at 0.04 (in units of $c_\text{s}^2 \rho_0$) around nonlinear saturation of the instability at orbit 10. Overall, the stress oscillates around zero indicating that the instability does not lead to a net transport of angular momentum; averaged from orbit 5 to orbit 20 we find $\langle \langle R_{xy} \rangle \rangle_t \sim (-5\times 10^{-5}) \pm (7\times10^{-3})$. The $xy$-component of the stress dominates, but we also observe oscillations of non-negligible amplitude in the $xz$- and $yz$-components of the stress.

\begin{figure}
\centering
\includegraphics[scale=0.3]{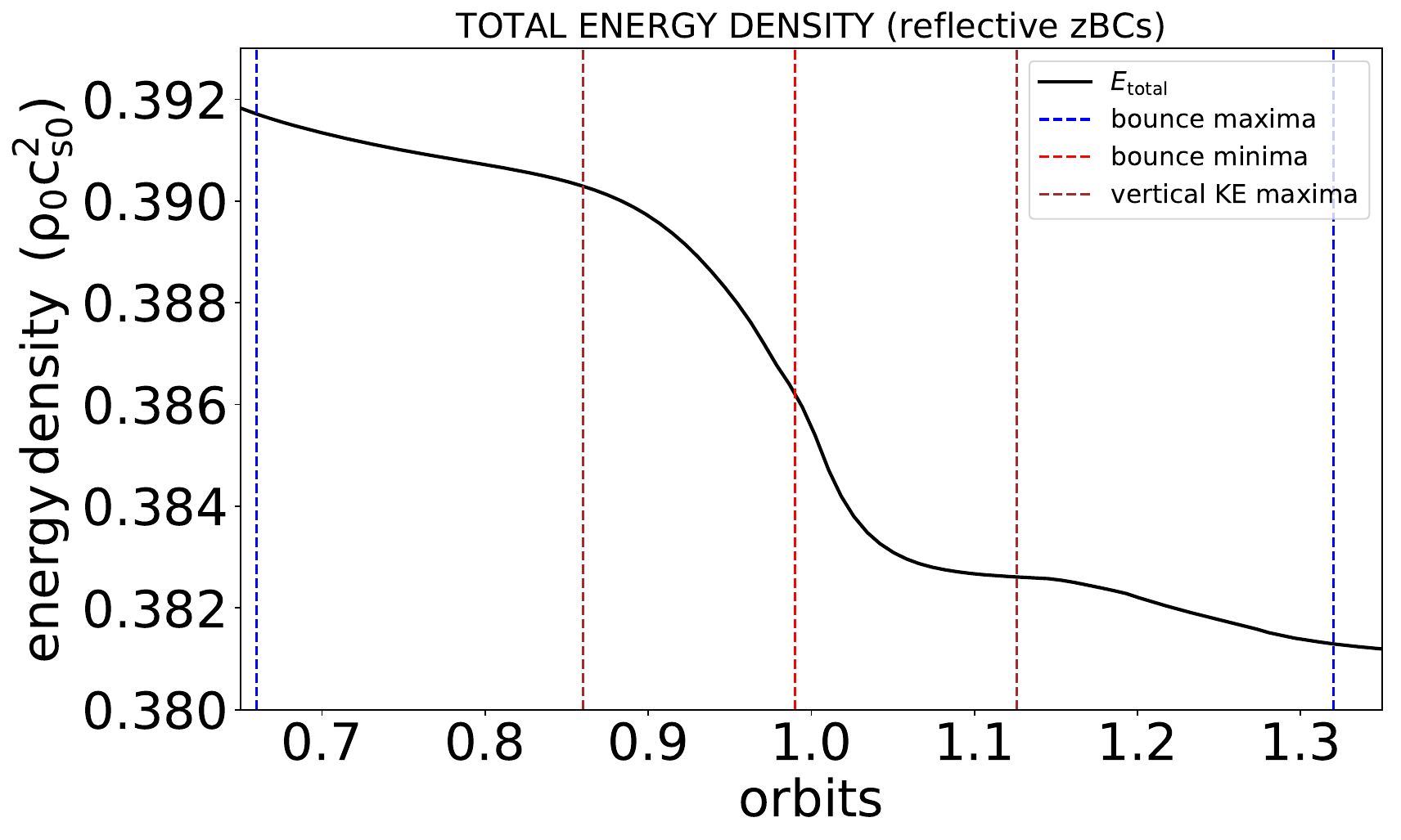}
\includegraphics[scale=0.3]{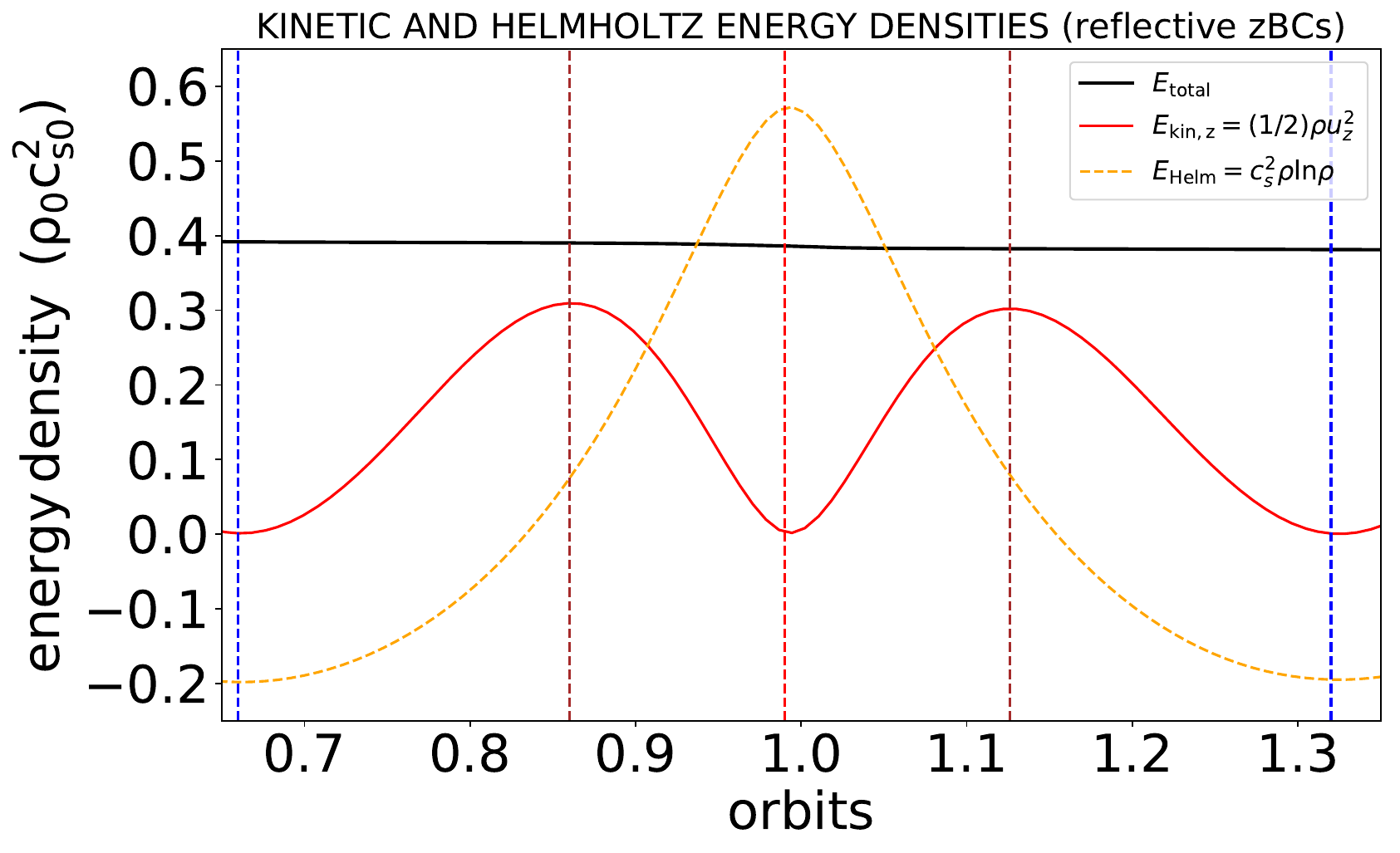}
\caption{Time-evolution of various energy components (in units of $c_\text{s}^2 \rho_0$) during the second oscillation (orbit 0.66-1.33) from fiducial 3D simulation of free oscillations. Top panel: time-evolution of total kinetic energy (black curve). Bottom panel: time-evolution of Helmholtz free energy (orange, dashed curve) and vertical kinetic energy density (red curve). Vertical blue and red dashed lines mark the locations of bounce maxima and minima, respectively. Vertical brown dashed lines mark times at which the vertical kinetic energy density reaches a maximum.}
\label{FIGURE_FiducialFreeBounceEnergiesFirstFewBoucnes}
\end{figure}

\subsubsection{Energetics: first few bounces}
\label{SECTION_EnergeticsStronglyBouncingCase}
To determine the mechanisms damping the oscillations before the parametric instability sets in, we conduct an energy analysis of the first few bounces (an oscillation or bounce being defined as the interval between successive peaks in the dynamical scale-height $H$). During this interval the disc undergoes six full oscillations. The bounce amplitude 
decreases by $6.8\%$ from $\Delta H/H_0 \sim 1.673$ 
to $\sim 1.559$, 
while the total energy decreases by around $12.5\%$ 
. The main energy terms of interest at this stage are the vertical kinetic energy density $E_{\text{kin},z}$ (which tracks the vertical oscillatory motion of the disc as it collapses and expands periodically), the Helmholtz free energy $c_\text{s}^2 \rho \ln{\rho}$ (which takes into account losses/gains in total energy due to compression/expansion), and the total energy itself $E_\text{tot}$ (see Equation \ref{EQUN_TotalEnergy}), so we will focus on these terms.

During the very first half oscillation (as the disc collapses from its initial thickness of $H \sim 1.97H_0$ at orbit 0 down to $H \sim 0.297H_0$ at orbit 0.33, the total energy actually \textit{increases} very slightly over the first 0.16 orbits. Total energy can only increase due to a positive Reynolds stress 
extracting energy from the radial shear, and indeed during these very early stages of the simulation we observe a sharp spike in the Reynolds stress (not shown) soon after initialization, reaching $\sim3\times 10^{-6}$. This probably results from the transient amplification of non-axisymmetric perturbations present in the initial noise. The stress then drops rapidly, reaching around $\sim 5\times 10^{-7}$ at orbit 0.16. Over the course of the next 4 orbits, the Reynolds stress oscillates but does not exceed $\sim 5\times 10^{-8}$. Therefore the stress plays a negligible role in the energy budget during this interval, so we will restrict our subsequent energy budget analysis to the second and third oscillations, in order to avoid the transient behaviour immediately after initialization.

The second oscillation/bounce (orbit 0.66--1.33) begins with the disc at a maximum thickness of $H \sim 1.95H_0$ at orbit~0.66 and ends with the disc at maximum thickness of $H \sim 1.93H_0$ one cycle later. (Subsequent oscillations exhibit similar behaviour.) We plot the total energy over the course of this oscillation in the top panel of Fig.~\ref{FIGURE_FiducialFreeBounceEnergiesFirstFewBoucnes}. Other energy components are plotted in the bottom panel. The overall behaviour of the key energy terms over the course of the bounce is as follows: (i) the total energy decreases monotonically by around $2.5\%$ over the course of the entire oscillation (orbit 0.66--1.32), from $\sim 0.391$ to $\sim 0.381$ (in units of $c_\text{s}^2 \rho_0$); 
(ii) the vertical kinetic energy density (red curve, bottom panel) increases from zero to a maximum value and decreases back to zero again during each half cycle; (iii) the Helmholtz free energy (dashed orange curve, bottom panel) increases as the disc collapses, peaks when the disc is most compressed, and decreases again as the disc expands.

\begin{figure*}
\includegraphics[scale=0.33]{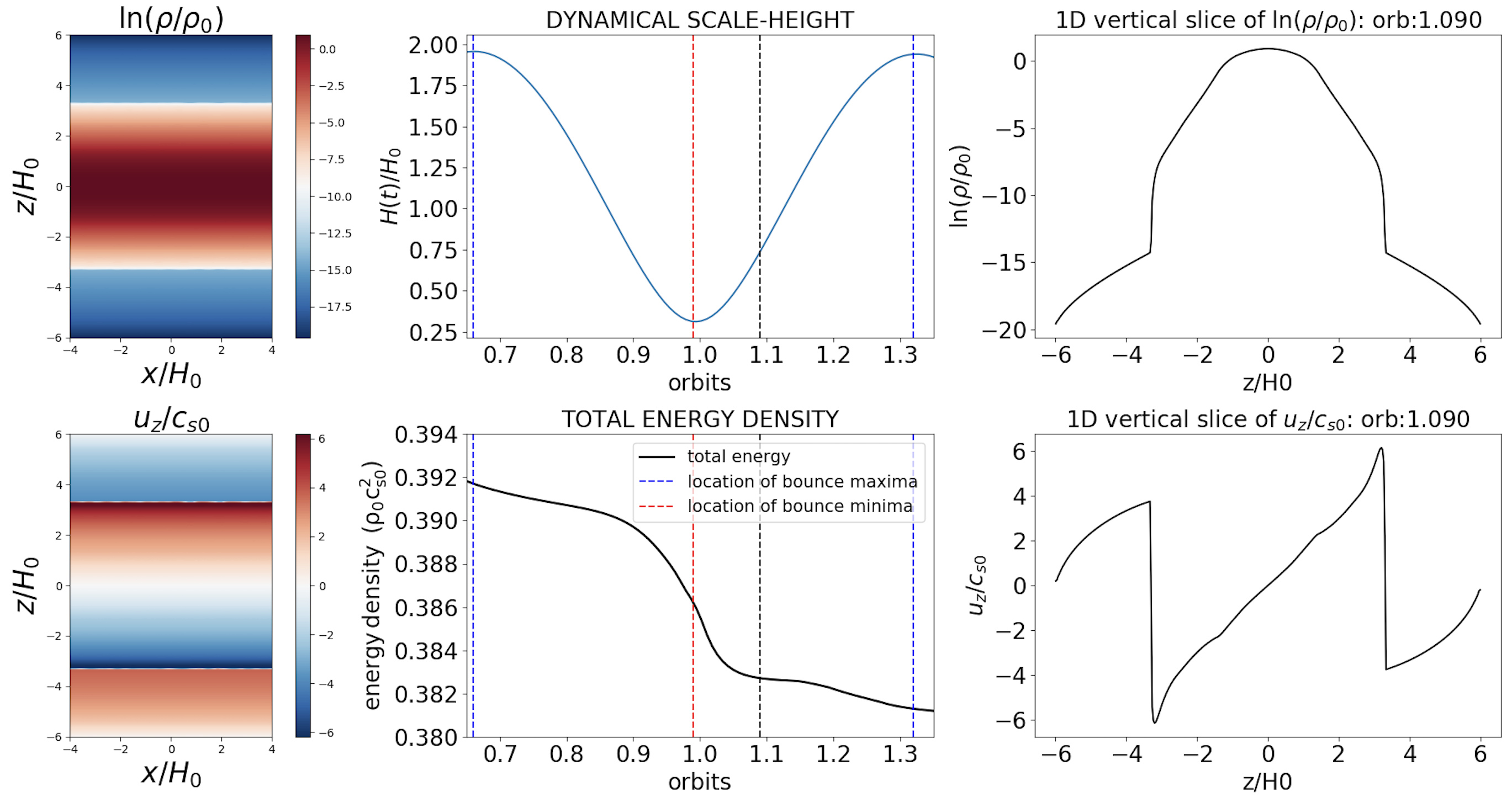}
\caption{Flow structure in free vertical oscillation simulation with reflective $z$-boundary conditions. Left column: 2D snapshots in $xz$-plane of density logarithm (top) and vertical velocity (bottom) taken as the disc is expanding during the second oscillation (orbit 1.09). Middle column: time-series of dynamical scale-height (top) and total energy density (bottom) during the second bounce. The dashed black line shows the time at which the snapshots were taken. Right panel: 1D vertical slices (through $x/H_0=0$) of density logarithm (top) and vertical velocity (bottom).}
\label{FIGURE_FreeBounceOutflowBCEnergyBudget}
\end{figure*}

Although the total energy decreases monotonically during the course of the bounce, it does not do so at a constant rate: in the top panel of Fig.~\ref{FIGURE_FiducialFreeBounceEnergiesFirstFewBoucnes} we have identified four different stages. These four stages can be understood by combining information about the flow field (density $\rho$ and vertical velocity $u_z$) and the total energy. In Fig.~\ref{FIGURE_FreeBounceOutflowBCEnergyBudget} we show 2D (in the $xz$-plane; left panel) and 1D vertical slices (through $x=0$; right-panel) of the density logarithm (top) and $u_z$ (bottom), taken at orbit 1.09 (as the disc is expanding). The middle panels show the evolution of the disc thickness (top) and total energy density (bottom) over the course of the entire bounce (orbit 0.66--1.33). During the initial stage of disc contraction (orbit 0.66--0.875), a shock propagates down from the vertical boundaries through the atmosphere. This is associated with a mild decrease in total energy due to shock dissipation. When the shock hits the bulk of the disc (around orbit~0.875) the dissipation rate increases as it passes through denser gas. The disc starts to expand after orbit~1, during which the total energy is nearly constant. The rate of energy dissipation then increases again after orbit~1.15. This coincides with time at which highly supersonic fluid (Mach number $\text{M}\sim8$) hits the rigid outer boundary, which launches a new shock from the boundary. Overall, the total energy density decreases by around $3\%$ during the course of the bounce, all of which is due to shock dissipation and ultimately related to the interaction of the bouncing motion with the unphysical vertical boundaries.

\subsubsection{Energetics: nonlinear saturation}
\label{SECTION_EnergeticsStronglyBouncingCase}

\begin{figure}
\centering
\includegraphics[scale=0.3]{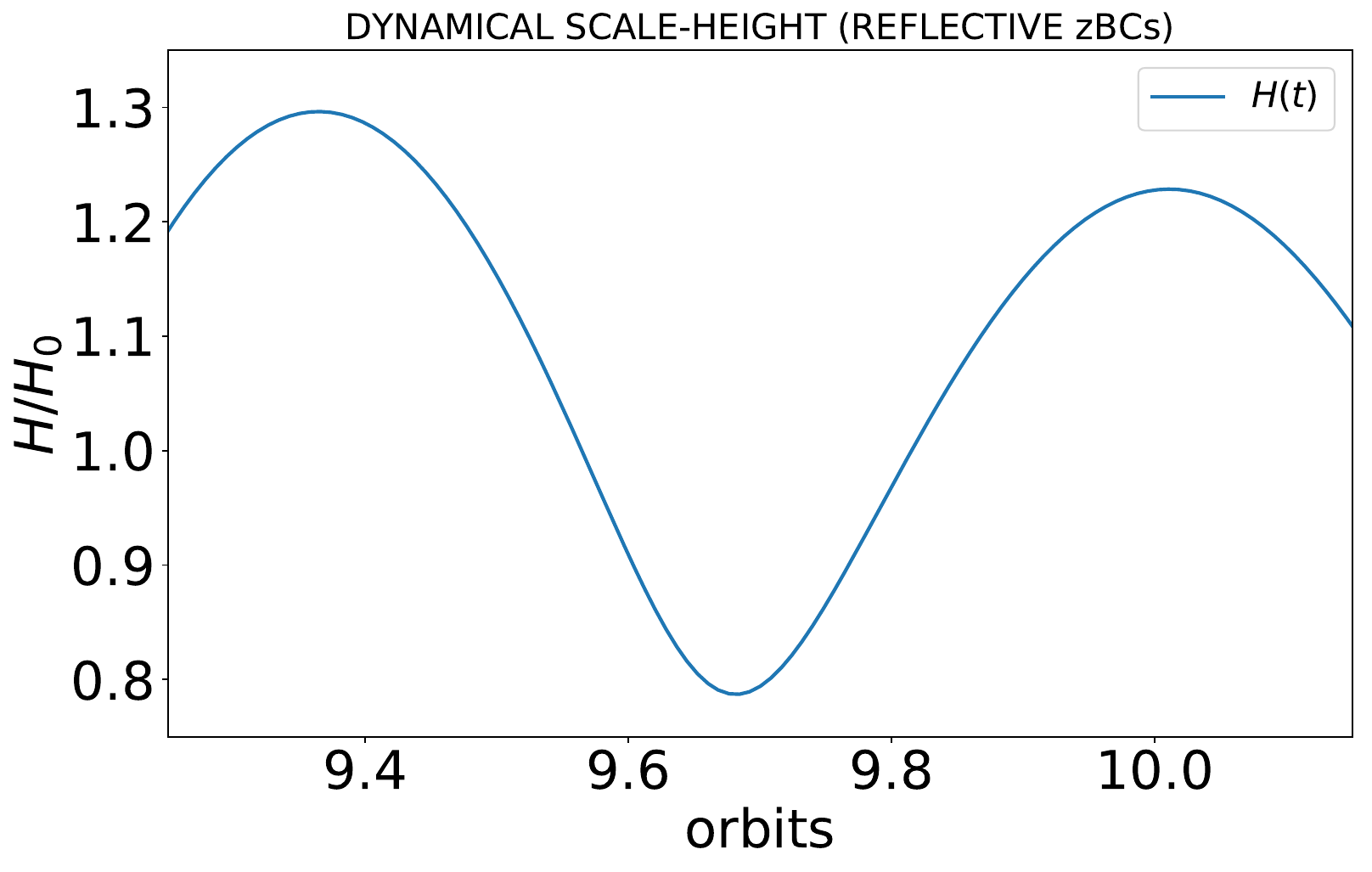}
\includegraphics[scale=0.3]{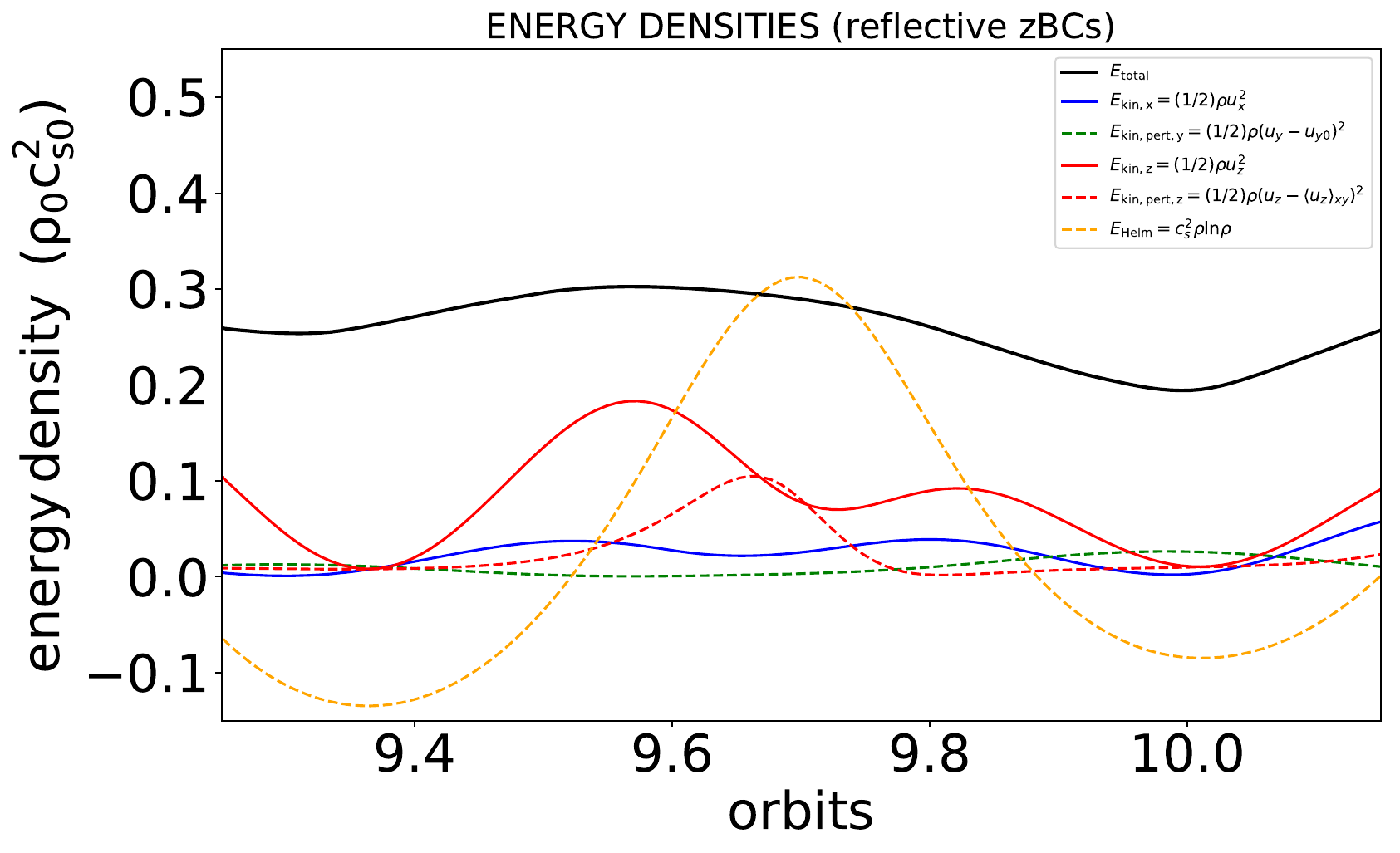}
\caption{Top: time-evolution of dynamical scale-height during an oscillation over the course of nonlinear saturation of the parametric instability (orbit 9.35-10.01) Bottom: corresponding evolution of total energy density (black) and various components (see legend).}
\label{FIGURE_EnergeticsNonlinearSaturationFreeBounceSim}
\end{figure}

Next we analyse a bounce during nonlinear saturation of the parametric instability. We focus on a vertical oscillation between around orbit~9.35 and orbit~10.01 (see Fig.~\ref{FIGURE_EnergeticsNonlinearSaturationFreeBounceSim}). At this point the disc is noticeably warped and strong compression and rarefaction can be seen in the disc due to radial motion induced by the instability. The dynamics is complicated: in addition to the periodic vertical oscillation of the disc and the behaviour and dissipation associated with that motion (described in the previous sub-section), the warped/corrugated disc is now also oscillating as a standing wave and this corrugation drives radial flows. 

The bounce amplitude decreases by around $22\%$ during this cycle, while the total energy decreases by around $25\%$, a marked increase in dissipation compared to when the disc was undergoing pure vertical oscillations (orbit~0 to orbit~4). Interestingly, the total energy (Fig.~\ref{FIGURE_EnergeticsNonlinearSaturationFreeBounceSim}, bottom panel, black curve) does not decrease monotonically over the course of the oscillation, but actually \textit{increases} as the disc compresses (roughly orbit 9.35 to 9.68). Since the only source term in the total energy equation is flux of energy through radial boundaries due to the Reynolds stress, this must be due to a non-zero Reynolds stress. Indeed, we find the Reynolds stress (not shown) is positive over this short time interval. The implication is that the unstable bending wave gains energy from the Keplerian shear (via the Reynolds stress) during this half cycle over which the total energy is increasing. [A related phenomenon can occur in the vertical shear instability, in which an axisymmetric inertial wave gains energy both from the vertical shear and from the Keplerian shear (Ogilvie, Latter \& Lesur, in preparation).]

The vertical component of the kinetic energy density (solid red curve) is non-zero even when the disc is maximally compressed, on account of vertical motion associated with the corrugation rather than the background oscillation of the disc. This can be seen more clearly by inspecting the perturbed part of the vertical kinetic energy density, defined as $E_{\text{kin},z,\text{pert}} \equiv \langle (1/2)\rho (u_z-\langle u_z \rangle_{xy})^2 \rangle$ (dashed red line), which peaks at around orbit 9.68. At this point in the oscillation cycle, the disc has actually reached its minimum thickness and has not begun to expand yet, so there is no vertical motion associated with expansion/contraction of the disc itself. Instead vertical motion during this phase is driven by the corrugation as the disc evolves from planar (at orbit 9.60) to warped (easily visible at orbit 9.68). Finally, we also show the (perturbed) azimuthal kinetic energy (green dashed curve). This is out of phase with the radial kinetic energy (solid red curve) due to epicyclic motions (conservation of angular momentum).

\subsection{Quasi-2D simulations}
\label{SECTION_FreeBounceQuasi2DSimulations}
To determine whether non-axisymmetry is important to the problem, we have run a series of quasi-2D simulations. The reduced dimensionality also enabled us to probe much higher resolutions, from $32/H_0$ (the same resolution used in our fiducial 3D simulation described in the previous section) up to $512/H_0$. The box size across all simulations was fixed at $[L_x, L_y, L_z] = [8,0.008,12]H_0$. Furthermore, we employed reflective boundary conditions in the vertical direction, and included a small explicit viscosity ($\text{Re} = 4687$), as in our fiducial 3D simulation.

\begin{figure}
\centering
\includegraphics[scale=0.23]{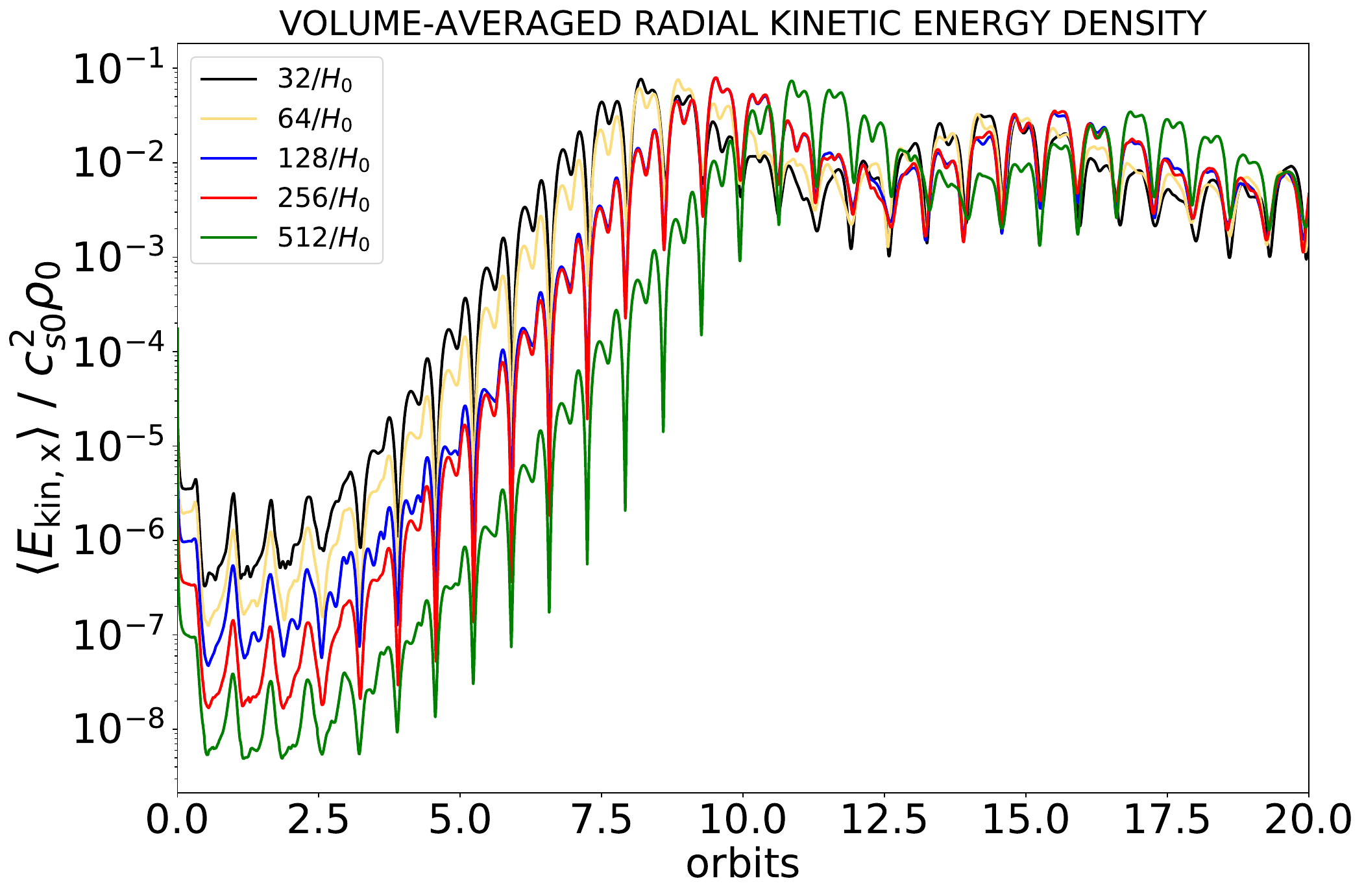}
\includegraphics[scale=0.23]{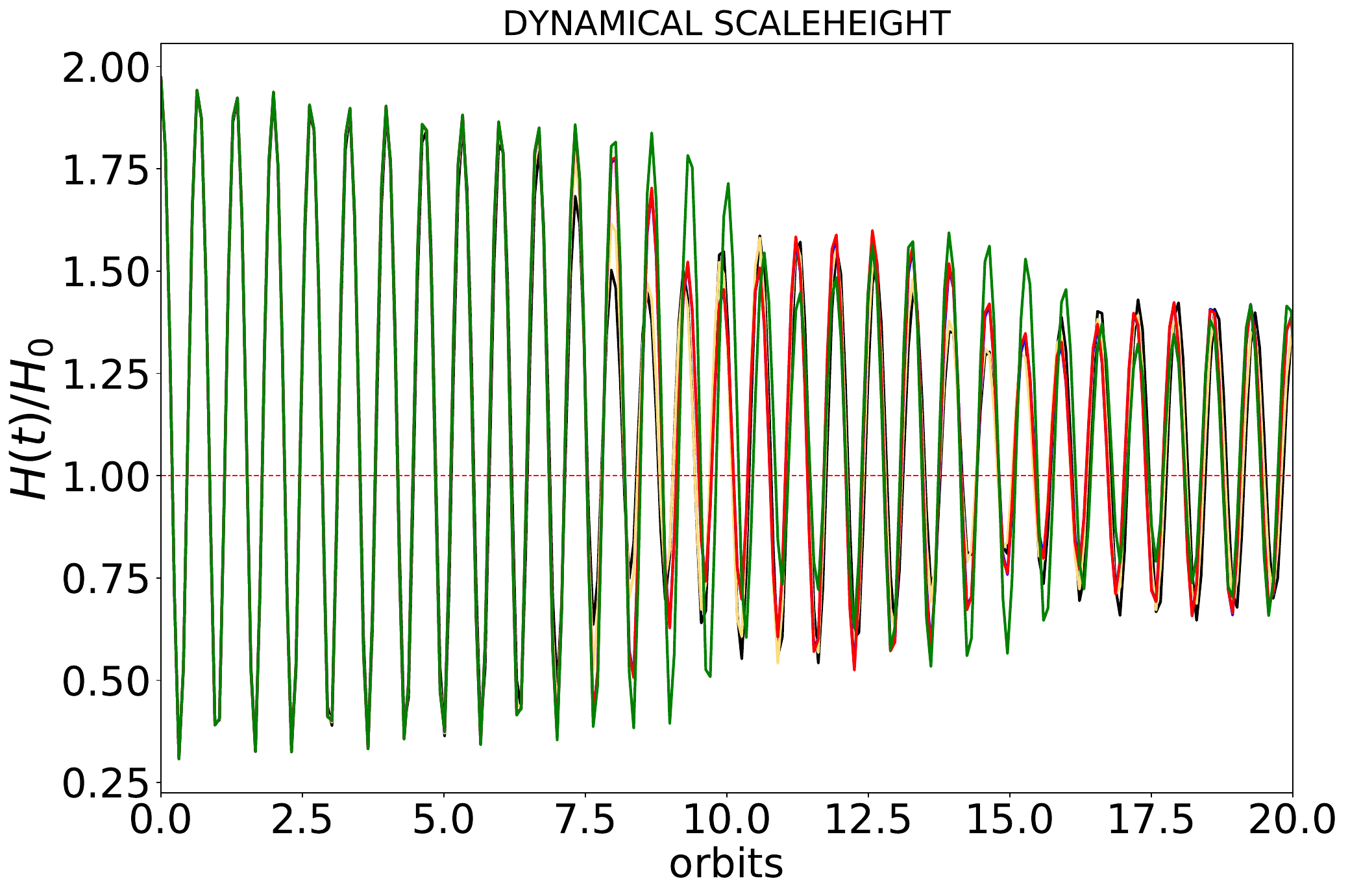}
\caption{Top: time-evolution of dynamical scale-height from fiducial quasi-2D simulations at different resolutions of free vertical oscillations. Bottom: time-evolution of volume-averaged radial kinetic energy density from the simulations. Black: $32/H_0$, yellow $64/H_0$, blue $128/H_0$, red $256/H_0$, green $512/H_0$.}
\label{FIGURE_Quasi2DSimsTimeEvolutionDynamicalHAndEkinx}
\end{figure}

\begin{figure}
\centering
\includegraphics[scale=0.25]{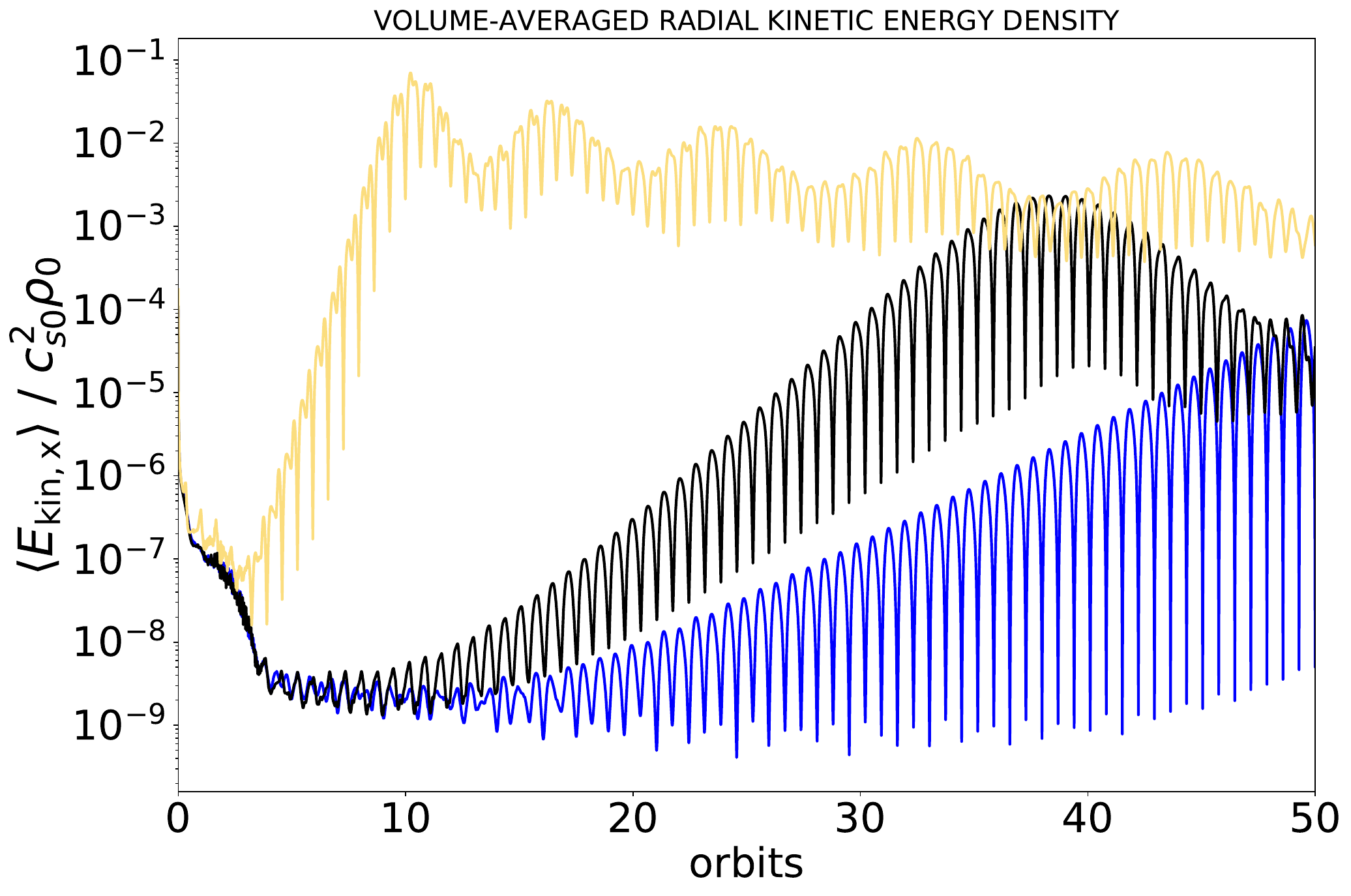}
\includegraphics[scale=0.25]{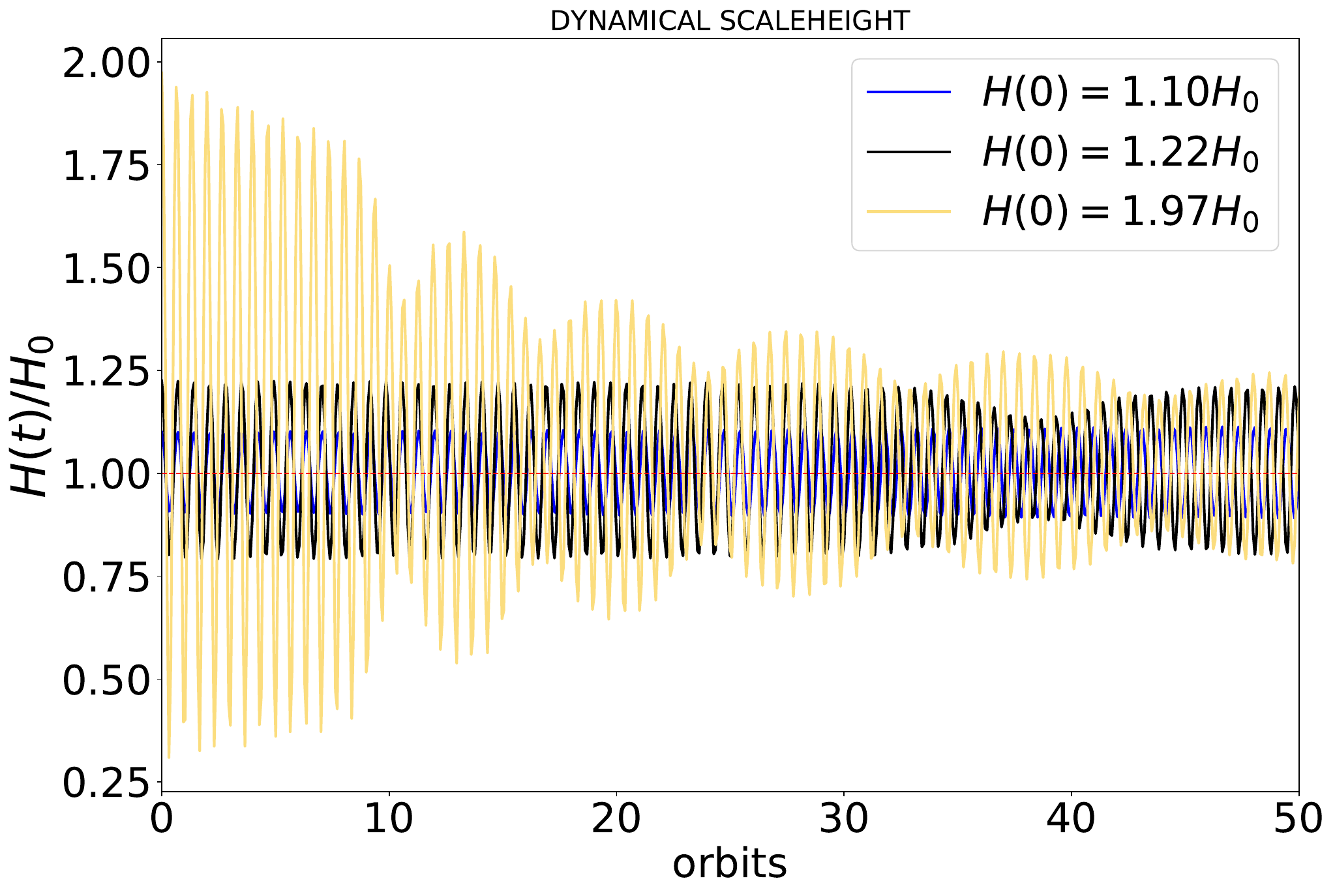}
\caption{Comparison of 3D simulations of free vertical oscillations with different initial oscillation amplitudes $H(0)$. Top panel: time-evolution of radial kinetic energy density. Bottom panel: time evolution of disc thickness (dynamical scale-height) $H(t)$. Yellow curves: large oscillation amplitude [$H(0)\sim1.97H_0$]. Black curves: intermediate oscillation amplitude [$H(0)\sim1.22H_0$]. Blue curves: small oscillation amplitude [$H(0)\sim1.10H_0$].}
\label{FIGURE_FreeBounceAmplitudeComparison}
\end{figure}

\begin{figure}
\centering
\includegraphics[scale=0.25]{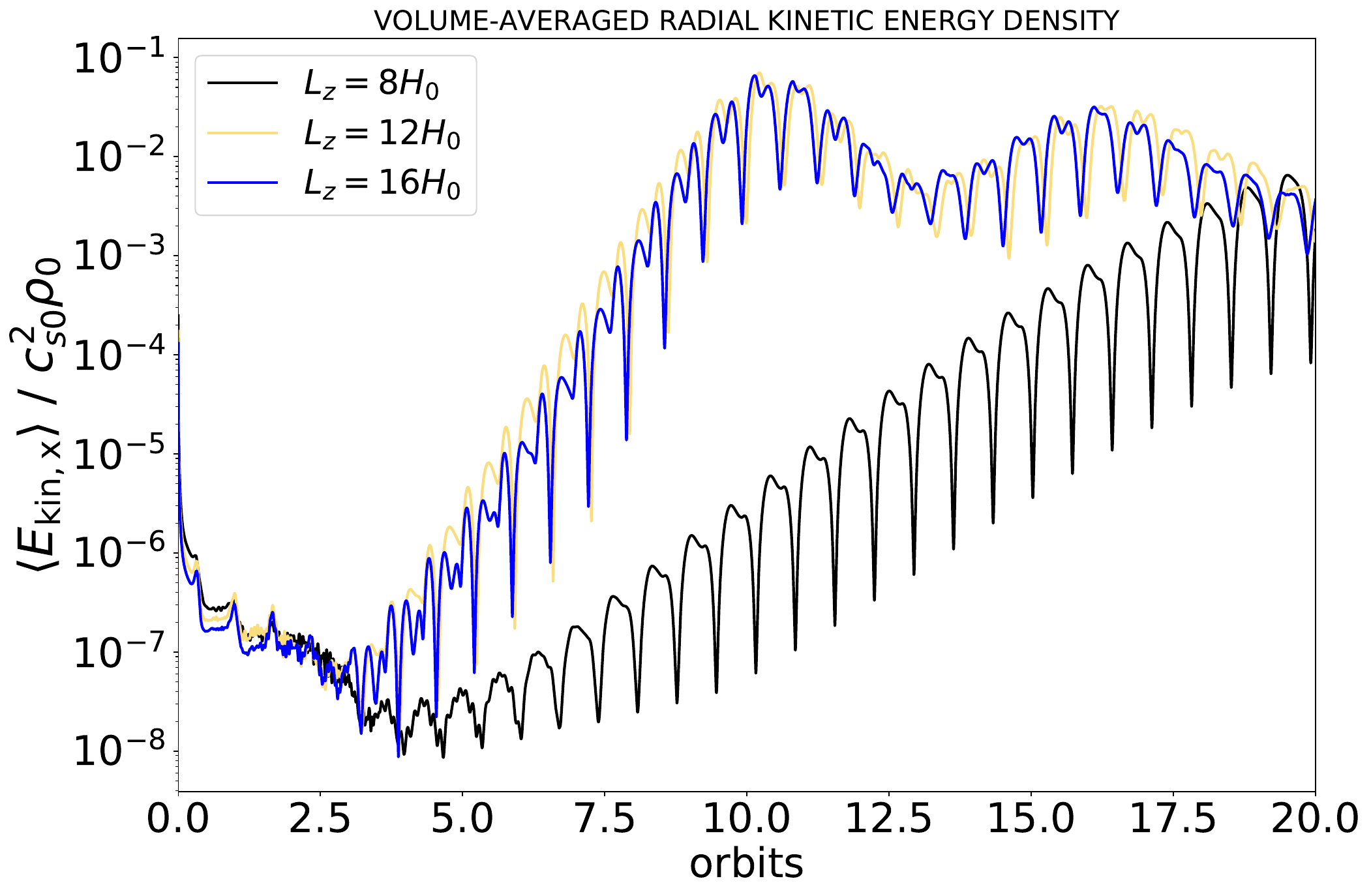}
\includegraphics[scale=0.25]{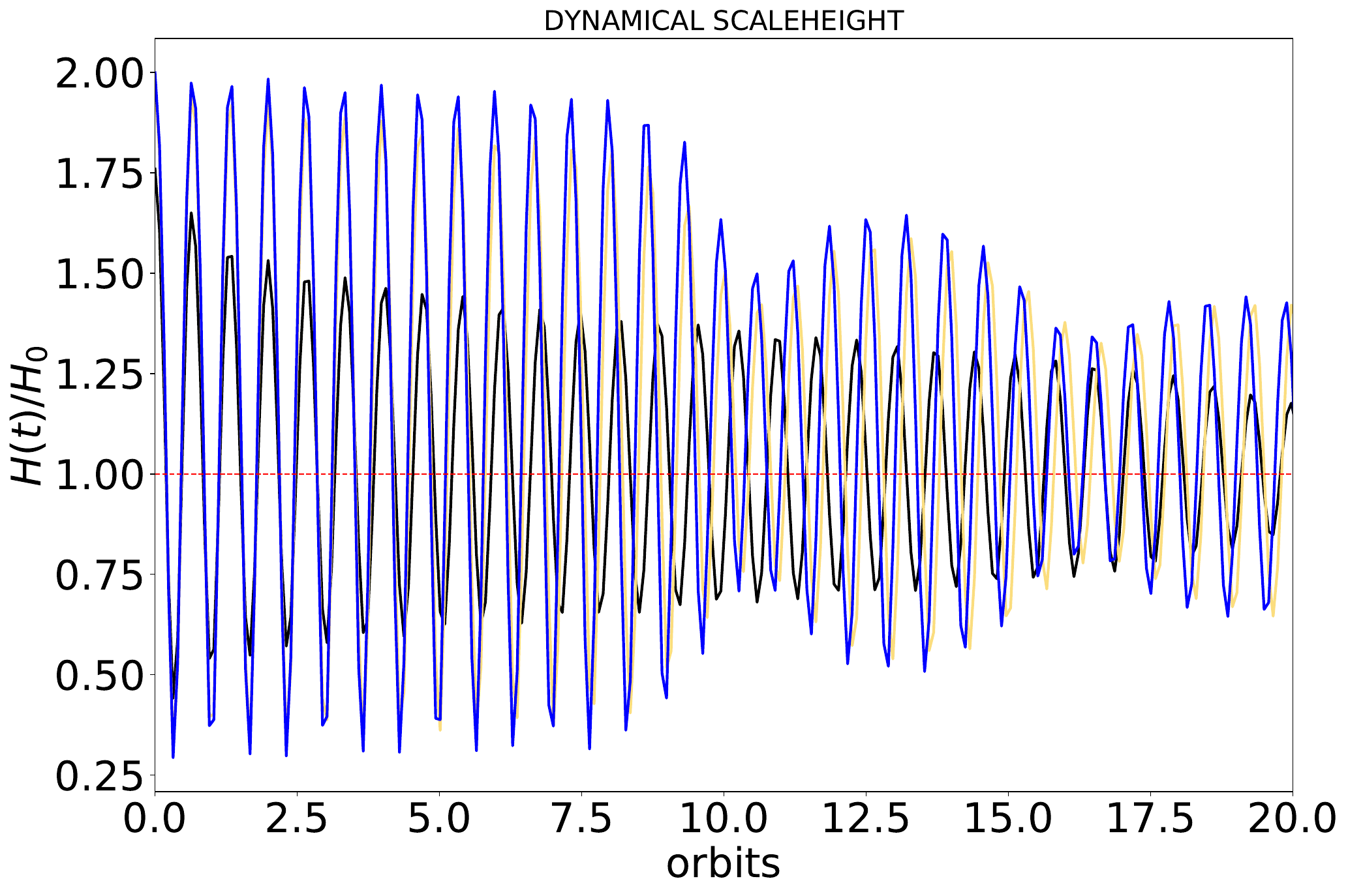}
\caption{Effect of vertical box size on 3D simulations of free vertical oscillations. Black: $L_z = 8H_0$. Yellow: $L_z = 12H_0$ (fiducial simulation). Blue: $L_z = 16H_0$. All simulations were of size $L_x = 8H_0$ and $L_y = 4H_0$ in the radial and azimuthal directions, respectively, and employed reflective boundary conditions in the vertical direction.}
\label{FIGURE_FreeBounceVerticalBoxSizeComparison}
\end{figure}

In Fig.~\ref{FIGURE_Quasi2DSimsTimeEvolutionDynamicalHAndEkinx} we show the time-evolution of the dynamical scale-height $H(t)$ (top) and volume-averaged radial kinetic energy density $E_{\text{kin},x}$ (bottom). Reassuringly, we find almost no difference in the bounce period and amplitude between our lowest resolution run (32 cells per equilibrium scale-height $H_0$) and highest resolution run ($512/H_0$). During the first half-cycle, the minimum disc thickness is around $H \sim 0.26$. This corresponds to around 133 cells across the disc at the highest resolution, so the disc remains well resolved even when it is most compressed. Parametric instability sets at slightly later times as the resolution is increased, on account of the lower level of initial noise in higher-resolution simulations.\footnote{Note that the rms velocity amplitude of the noise is similar at all resolutions, but as the resolution is increased the power is spread over a wider range of wavenumbers and therefore there is less power at the scales needed to seed the instability.}  The growth rate of the instability is very similar regardless of the resolution, e.g. $s/\Omega \sim 0.177$ and $0.175$ at resolutions of $32/H_0$ and $512/H_0$, respectively. The level of nonlinear saturation and the behaviour after nonlinear saturation are very similar at all resolutions.

We compared the flow field (density in the $xz$-plane) at different resolutions in these quasi-2D simulations (not shown). Regardless of the resolution, after nonlinear saturation the disc exhibits corrugation/warping and radially propagating waves and shocks. At higher resolutions we can see finer structure where the shocks meet (at $x \sim -3H_0$ and  $x \sim 1.5H_0$), but the overall structure of the flow is the same.

In conclusion, our high-resolution simulations indicate that the resolution of $32/H_0$ used in our fiducial 3D simulation is sufficient to capture all the salient physics associated with the vertical oscillations, including damping of the oscillations by shocks, the development of parametric instability, and the subsequent corrugation of the disc. The fact that the quasi-2D simulations reproduce all the physics of the fiducial fully 3D simulation confirms that the problem is axisymmetric (which we also determined directly through visual inspection of density snapshots in the $yz$-plane in our fiducial 3D simulation).

\subsection{Effect of numerical and physical parameters on the bouncing motion}
\label{SECTION_FreeBounceParameterStudy}

\subsubsection{Effect of initial bounce amplitude}
In Fig.~\ref{FIGURE_FreeBounceAmplitudeComparison} we compare 
three different initial disc thicknesses: $H(0) = 1.10, 1.22$, and $1.97$, which we refer to as `small', 
`intermediate' 
and `large'-amplitude oscillations. We carried out the small-amplitude run in a slightly larger box ($L_x = 9H_0$ compared to $L_x=8H_0$) so that the box size fits comfortably within the peak of the $n=1$ mode predicted by theory (see left-hand panel of Fig.~\ref{FIGURE_FreeBounceGrowthRates}). As predicted by the theory (see Fig.~\ref{FIGURE_FreeBounceHmaxHminPrediction}), the initial oscillation amplitude determines the subsequent oscillation period. We find that the disc oscillates $\sim 1.41$ times per orbit in the small-amplitude case, exactly as predicted by theory in the limit of small oscillations. For the intermediate-amplitude case the disc oscillates $\sim 1.43$ times per orbit, while in the large-amplitude case the disc oscillates $\sim 1.51$ times per orbit.

All three simulations exhibit parametric instability, with growth rates $s/\Omega \sim 0.024$, $0.044$ and $0.169$, respectively, in good agreement with the theoretically predicted values (see Table \ref{TABLE_3DSimsFreelyBouncing_ComparisonBounceAmplitude}). The onset of nonlinear saturation is delayed as the initial bounce amplitude is decreased, and does not occur until 100 orbits after initialization in the small amplitude case ($H(0) \sim 1.10H_0$). For this very weakly oscillating case, the radial kinetic energy density saturates at around  $10^{-3}$ (in units of $c_\text{s}^2 \rho_0$), about 10 times smaller than the saturation amplitude in the large-amplitude oscillation simulation ($H(0) \sim 1.97H_0$).

\subsubsection{Effect of vertical box size}
Since we employ reflective boundary conditions, the periodic expansion of the disc results in reflection of material at the vertical boundaries. Ideally we want the box to be as tall as possible to avoid significant boundary effects such as this from interfering with the dynamics closer to the bulk of the disc, which lies within $|z|< 2H_0$. Thus we have repeated our fiducial simulation in shorter ($L_z = 8H_0$) and taller ($L_z = 16H_0$) boxes (keeping the resolution the same). In Fig.~\ref{FIGURE_FreeBounceVerticalBoxSizeComparison} we show the results of this study with the radial kinetic energy density in the top panel and the dynamical scale-height $H(t)$ in the lower panel. Note that in the shortest box a significant fraction of the disc atmosphere is cut off by the vertical boundaries and the initial dynamical scale-height is therefore $H(0) \sim 1.67H_0$, compared to $H(0) \sim 1.97$ and $\sim 1.99$ in the $L_z=12H_0$ and $16H_0$ boxes, respectively. We see that the results are converged for boxes of size $L_z > 8H_0$, with the growth rate of the parametric instability and evolution of the bounce amplitude nearly identical in the $L_z = 12 H_0$ (fiducial run) and $L_z = 16H_0$ boxes. In the shortest box $L_z = 8H_0$ (Fig. \ref{FIGURE_FreeBounceVerticalBoxSizeComparison}, black curve), however, first the oscillation amplitude is smaller to begin with because a significant fraction of the initial disc profile is cut-off by the boundaries, and second the oscillation amplitude is strongly damped even before the parametric instability starts to grow because the boundaries are lower. Thus the growth rate of the instability is significantly reduced in the shortest box.

\subsubsection{Effect of vertical boundary conditions}
To compare vertical boundary condition effects (zBCs), we have repeated our fiducial simulation (Section \ref{SECTION_FreeBounceFiducialSimulation}) using outflow rather than reflective boundary conditions in the vertical direction. The dynamical scale-height and radial kinetic energy density are shown in Fig.~\ref{FIGURE_FreeBouncezBCComparison}.

Overall the solution in the simulation with outflow boundary conditions is very similar to that with reflective boundary conditions (see Fig.~\ref{FIGURE_FreeBouncezBCComparison}), although we do find small qualitative differences between the two. The oscillation is damped somewhat faster in the outflow case during the linear phase, although by the end of the simulation the oscillation amplitude is very similar in the two cases. The linear growth rate is somewhat larger in the reflective case ($0.169\Omega$, compared to $0.154\Omega$ in the outflow case), which means the reflective zBC simulation reaches nonlinear saturation earlier. This results in a slight phase difference (of a few orbits) in the oscillation of the disc between the two simulations by orbit 20.

\begin{figure}
\centering
\includegraphics[scale=0.3]{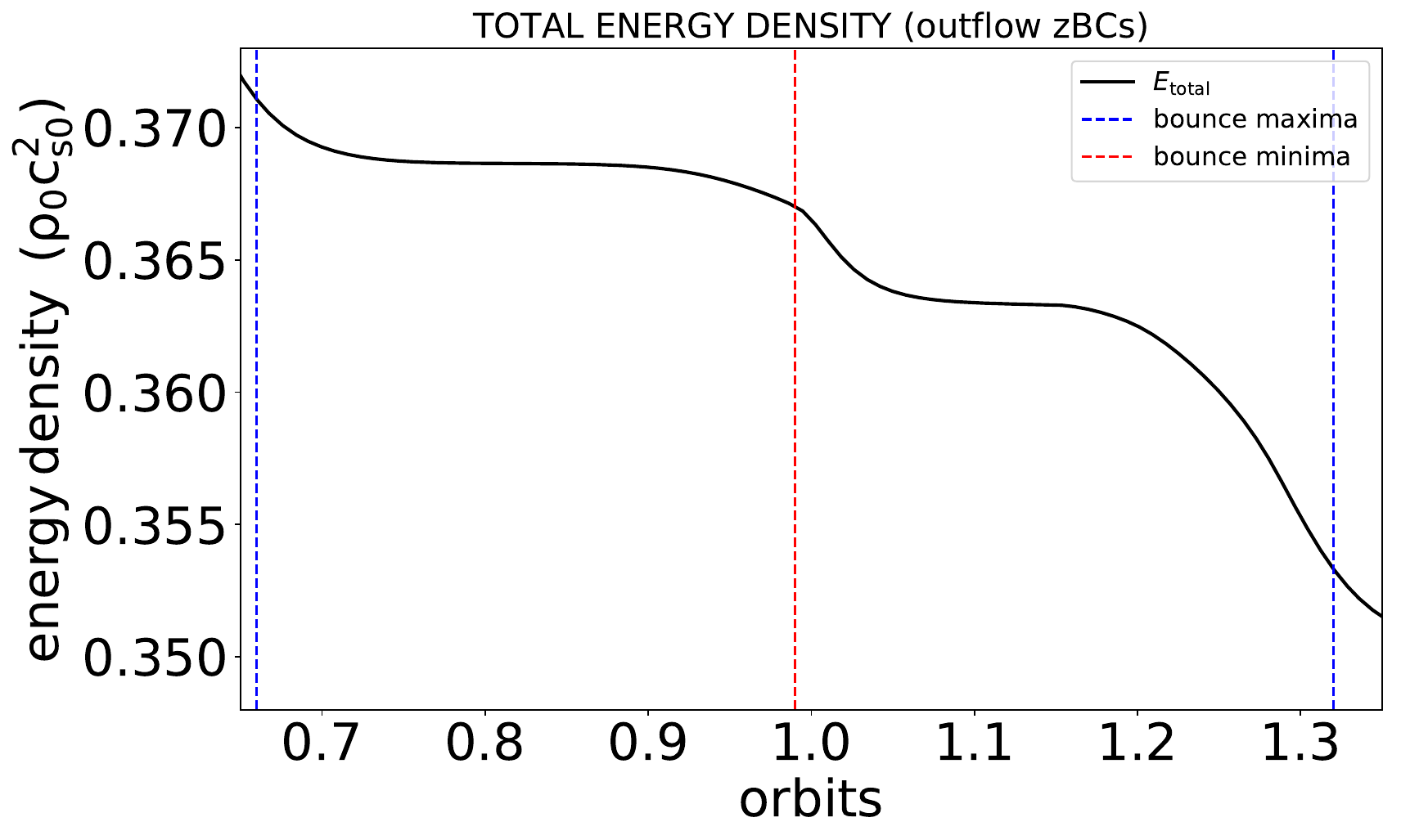}
\caption{Time-evolution of total energy density (in units of $c_\text{s}^2 \rho_0$) during the second oscillation (orbit 0.66-1.33) from the simulation with outflow vertical boundary conditions.}
\label{FIGURE_FiducialFreeBounceEnergiesFirstFewBoucnesOUTFLOW}
\end{figure}

Although the overall result over 20 orbits is very similar in the two cases, we find more marked differences between the two runs when we focus on the first few bounces (before parametric instability sets in). We have carried out an energy budget analysis of the outflow simulation during the first four orbits (during which the disc bounces around 6 times), similar to the one we presented for the fiducial simulation with reflective zBC (see Section~\ref{SECTION_EnergeticsStronglyBouncingCase}). This interval is before the onset of parametric instability, so the disc is undergoing only vertical oscillations. The total energy in the outflow run drops by around $25\%$ between orbit 0 and orbit 4 (6 oscillations) compared to a drop of around $13\%$ in the reflective case.

In the outflow zBC simulation we did not use a mass source term (to avoid interfering with the vertically oscillating flow), and thus mass and energy were able to leave the box through the vertical boundaries. Indeed, over the entire duration of the simulation (50 orbits) around $2.2\%$ of the initial mass in the box was lost through the vertical boundaries due to the periodic expansion of the disc, although most of this mass loss occurs during the first 10 orbits (linear phase), with over half of the total mass loss ($1.2\%$ of the initial mass) taking place during the first four orbits alone. The rapid drop in mass loss after orbit 10 coincides with the onset of nonlinear saturation of the parametric instability, which significantly damps the oscillations, and thus the resultant mass loss through the vertical boundaries.

The behaviour of the total energy over the second bounce (orbit 0.66 to orbit 1.33) in the outflow simulation is shown in Fig.~\ref{FIGURE_FiducialFreeBounceEnergiesFirstFewBoucnesOUTFLOW}. The total energy is constant during the initial stages of disc contraction since there are no shocks from the boundaries to dissipate the energy. A shock eventually does form as the disc contracts supersonically, which results in a rapid drop in the total energy during the final moments of contraction and the subsequent initial moments of expansion (orbits 0.907--1.05). When this shock (moving at Mach number $\text{M}\sim8$) hits the outer boundary, there is another sharp drop in the total energy, this time because mass and energy are advected across the vertical boundary. Overall, the total energy density decreases by around $5\%$ during the course of the oscillation (most of that due to loss of energy through the vertical boundaries).

\begin{figure}
\centering
\includegraphics[scale=0.24]{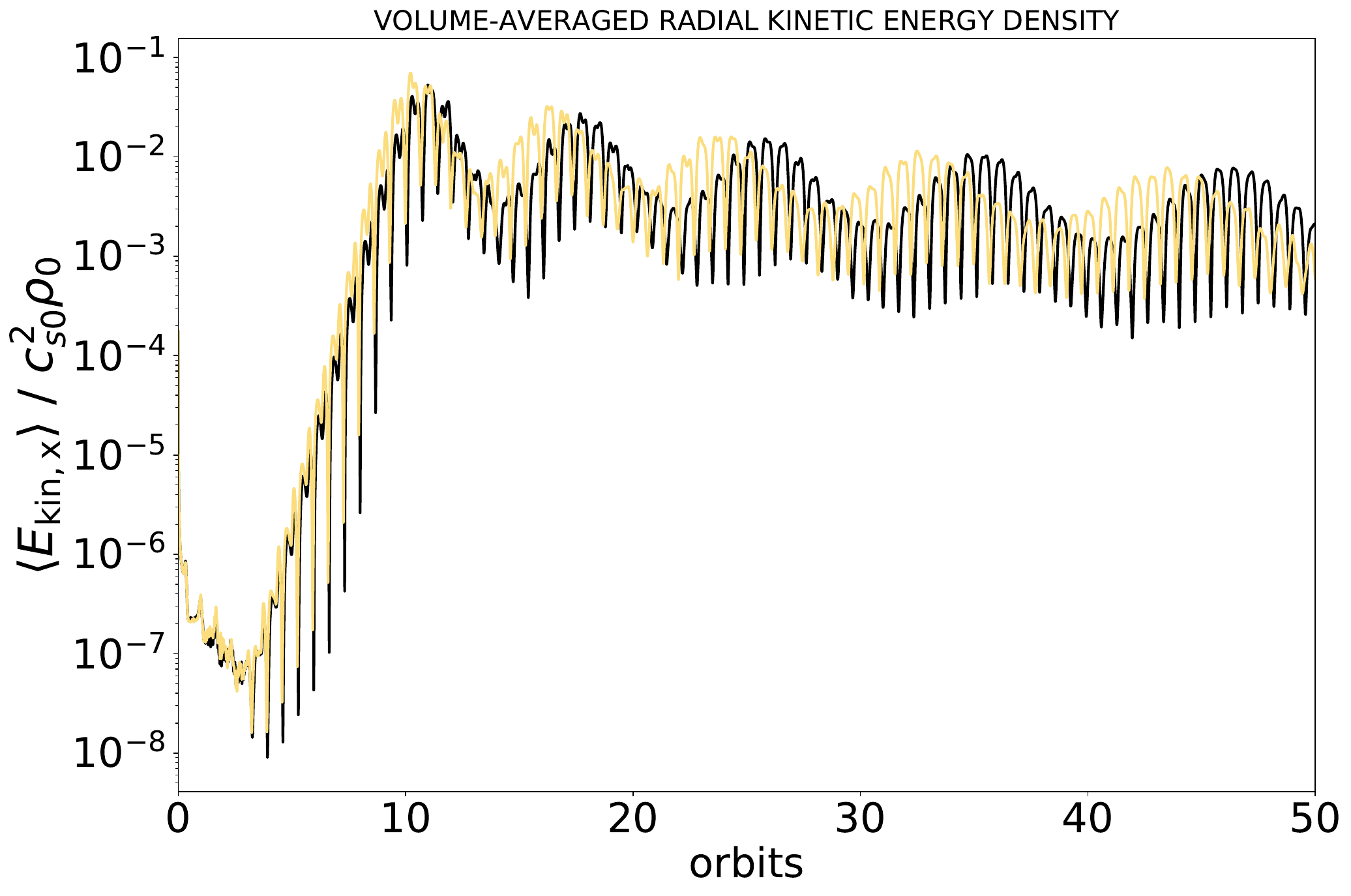}
\includegraphics[scale=0.24]{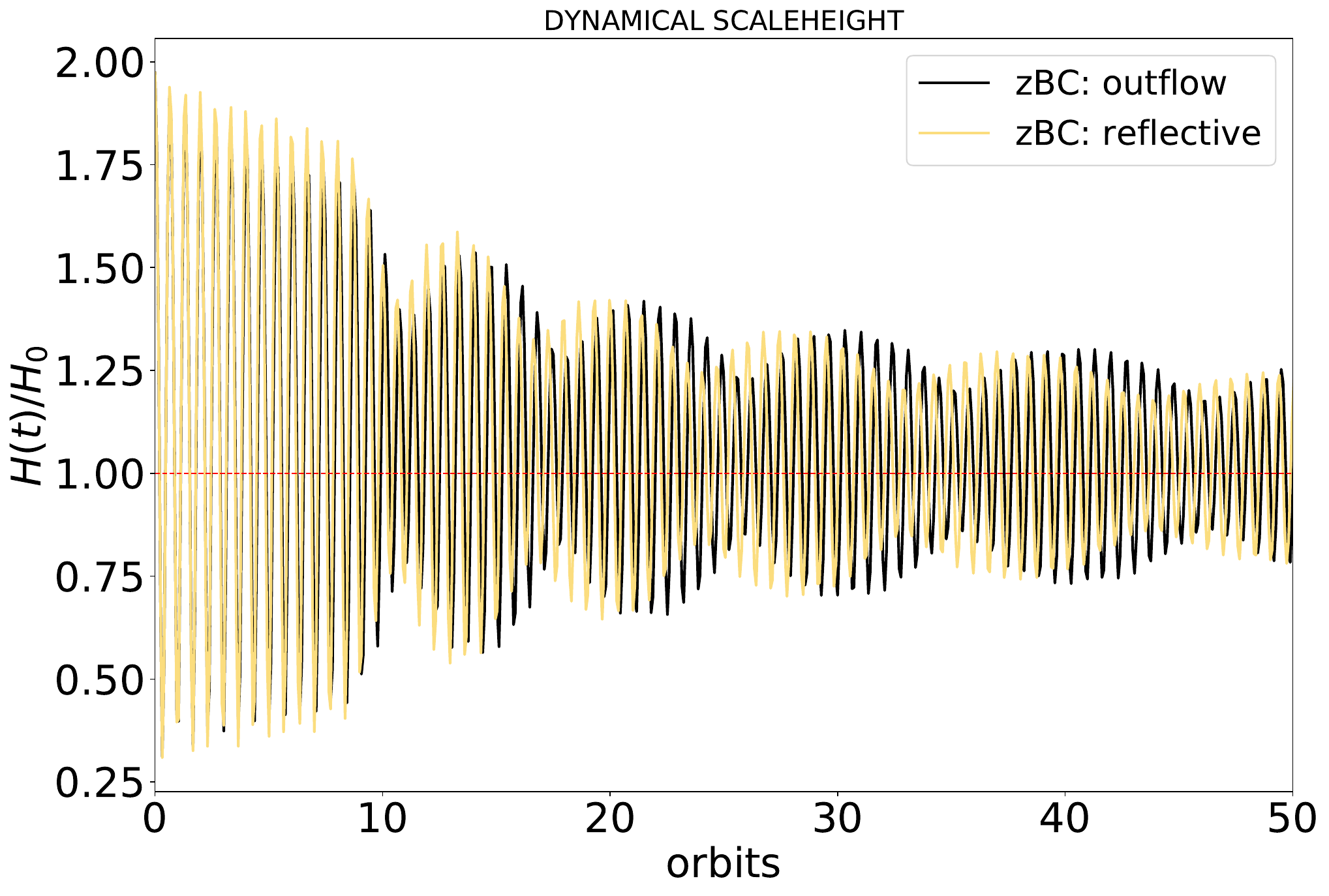}
\caption{Comparison of dynamical scale-height (top panel) and radial kinetic energy density (bottom panel) from 3D simulations of free vertical oscillations with different vertical boundary conditions. Yellow curves: reflective boundary conditions. Black curves: outflow boundary conditions. Both simulations were run in a box of size $[L_x,L_y,L_z]=[8,4,12]H_0$ at a resolution of $32/H_0$.}
\label{FIGURE_FreeBouncezBCComparison}
\end{figure}

\subsubsection{Effect of explicit viscosity}
In our fiducial 3D run and in most of our quasi-2D runs we included a small explicit viscosity, corresponding to a Reynolds number of $\text{Re} = 4687$. To ensure the resolution we used is large enough to capture the viscous length-scale in our non-ideal simulations, and if so, to determine the effect of the explicit viscosity on the dissipation of oscillations, we have tested the effects of explicit and numerical viscosity. We do so primarily using quasi-2D simulations since these enable us to achieve much higher resolutions.\footnote{We have also repeated our fiducial 3D run (with $\text{Re} = 4687$) at twice the resolution ($64/H_0$) and found that the results are in very good agreement with the lower resolution run.} At $32/H_0$ we find that the oscillation amplitude and growth rates are nearly identical whether we include explicit viscosity or not. However, the viscous scale [which is given by $l_{\nu} \sim 2\pi/\sqrt{\text{Re}}$; see discussion in \cite{heldmamatsashvili2022}] in the non-ideal run is poorly resolved at $\text{Re}=4687$ ($\sim 3$ cells per viscous length scale). Thus we have repeated our ideal and non-ideal runs at $256/H_0$. In this case the viscous length-scale is well resolved ($\sim 25$ cells). Again we find very little difference between the evolution of the dynamical scale-height, the parametric instability growth rate ($\sim 0.17\Omega$ in both cases), or any of the dynamics during the nonlinear phase between the ideal and non-ideal runs. Thus we conclude that the explicit viscosity we have included in our simulations plays a negligible role in dissipating the vertical oscillations.

\section{Simulations of forced oscillations}
\label{RESULTS_ForcedBounceSimulations}
In the previous section we analysed the behaviour of a disc undergoing free oscillations in the vertical direction in order to isolate the physical and numerical mechanisms that dissipate the oscillation. We now turn to the case in which the vertical motion is forced. This can occur when the vertical component of gravity varies along a streamline, for example in an eccentric disc, or in a disc that is tidally distorted by a binary companion.
In a warped disc, the forcing arises due to the combination of tilt and shear even in the absence of an external perturber \citep{ogilvie2013local,fairbairn2021non}.

In order to drive vertical oscillations, we introduce a time-dependent gravitational acceleration in the vertical direction of the form $g_{\text{eff},z} = -\Omega^2 z (1+a\cos{\omega t})$ (see Section~\ref{METHODS_GoverningEquations}). This allows us to set the forcing frequency $\omega$ (in units of the angular frequency of the disc $\Omega$) and the forcing amplitude $a$ (in units of $\Omega^2 H_0$). Thus we have two parameters which we can use to control the forcing. We first discuss astrophysically relevant forcing frequencies and forcing amplitudes in Section \ref{SECTION_ForcedBounceAstrophysicalApplications}. Then, in Section \ref{SECTION_ForcedBounceTheory}, we extend our 1D model for vertical disc oscillations to include forcing, and calculate the expected parametric growth rates for six select forcing frequencies. We also calculate the amplitude of oscillations of the disc scale-height $H(t)$ for each forcing frequency (see Fig.~\ref{FIGURE_ForcedBounceInitialScaleheight}), which we subsequently use to determine the appropriate initial condition for each simulation. Finally we discuss fully 3D simulations with forcing in Section~\ref{SECTION_ForcedBounceSimulations}.

\subsection{Astrophysical applications: forcing frequencies and amplitudes}
\label{SECTION_ForcedBounceAstrophysicalApplications}
\subsubsection{Forcing frequencies}
Let us first consider which forcing frequencies are relevant for each type of distorted disc (eccentric, tidally distorted, and warped). 

In an eccentric disc, the distance to the central star varies along each eccentric streamline. The vertical component of gravity due to a central point source is strongest at pericentre and weakest at apocentre, so the forcing frequency experienced by the gas in an eccentric disc 
is simply equal to the orbital frequency ($\omega = \Omega$), if we neglect any precession of the eccentricity. A sinusoidal variation of the vertical gravity is a good model for small eccentricities.

\begin{figure}
\centering
\includegraphics[scale=0.24]{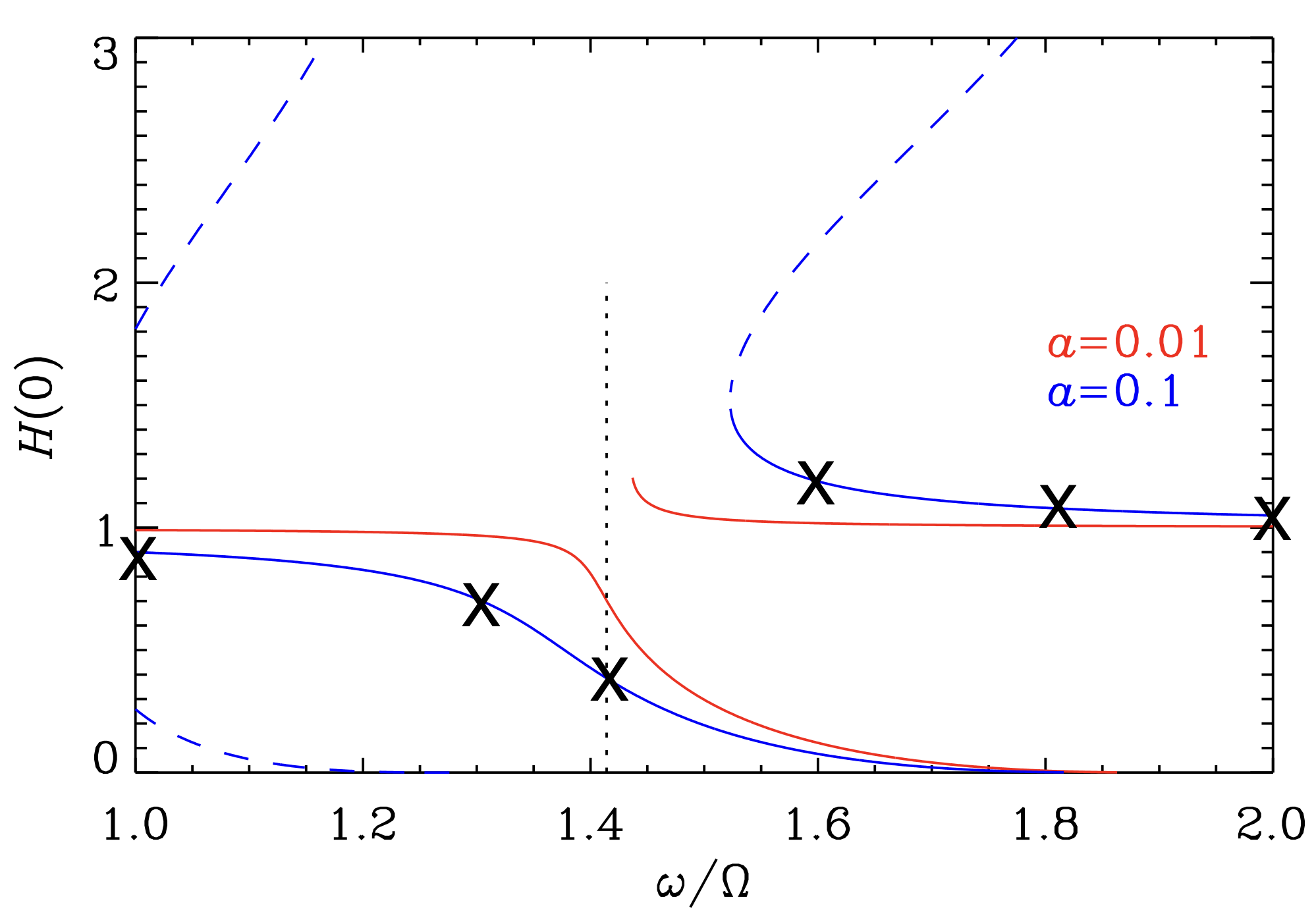}
\caption{Resonant response plot of dimensionless scale-height $H$ (in units of equilibrium scale-height $H_0$) at $t = 0$ as a function of the forcing frequency $\omega$ (in units of disc angular frequency $\Omega$) at two different forcing amplitudes $a$ (units of $H_0 \Omega^2$). The gas is isothermal ($\gamma = 1$). The dotted vertical line shows the location of the vertical resonance at $\omega = \sqrt{\gamma+1} = \sqrt{2} \sim 1.41$ (for an isothermal gas). The crosses correspond to simulations initialized with the corresponding initial disc thickness and forcing frequency. (For each value of~$a$, the right-hand branch folds over near the resonance and connects with an unstable upper branch, marked by the dashed line.)}
\label{FIGURE_ForcedBounceInitialScaleheight}
\end{figure}

In a tidally distorted disc (i.e. a circumstellar disc orbited by a binary companion, or a circumbinary disc), the distance -- and thus the vertical component of gravity -- also varies along any particular orbit, but the forcing frequency is a function of cylindrical radius $r$. Consider  a circumstellar disc in a circular binary system: in the frame rotating with the binary angular frequency $\Omega_\text{b}$, the disc can have a stationary, non-circular shape as a result of the tidal force due to the companion (see Fig.~\ref{FIGURE_VerticalOscillationsSchematic}). We can decompose the non-axisymmetric part of the gravitational potential into Fourier modes with azimuthal mode number $m$ \citep{lubow1981vertically, ogilvie2002tidally, ogilvie2002non}. At any given cylindrical radius $r$ in the disc, the angular frequency of the disc relative to the tidal deformation is $\Omega(r) - \Omega_\text{b}$, where $\Omega (r) \equiv (GM/r^3)^{1/2}$ is the Keplerian angular frequency of the disc (ignoring effects due to the companion). The forcing frequency (in units of the disc angular frequency) associated with mode number $m$ is given by $\omega/\Omega = m(1-\Omega_\text{b}/\Omega)$. \cite{lubow1981vertically} showed that vertical resonances occur where $m(1-\Omega_\text{b}/\Omega) = \sqrt{\gamma+1}$ and the vertical oscillation mode is excited. In practice $m = 2$ is the most important mode, both because of the need to fit the resonance inside the disc and because $m = 2$ is the strongest component of the tidal deformation. Note that the forcing frequency \textit{decreases} as we move radially outwards from the primary. Closest to the central star, the forcing frequency (for the $m=2$ mode) is $\omega \sim 2\Omega$ (since $\Omega \gg \Omega_\text{b}$), although here the deformation of the orbit is smallest and so the forcing actually vanishes. Farthest from the central star (approaching the 2:1 resonance, which actually lies outside the disc), $\Omega \gtrsim 2\Omega_\text{b}$ and so $\omega \gtrsim \Omega$. The resonance should lie somewhere in between these two locations, but within the disc.\footnote{See Fig.~3 of \citet{lubow1981vertically}, which compares the radial locations of the vertical resonance, the disc tidal truncation radius (outer edge), and the Lindblad radius, for secondary-to-primary mass ratios $q \gg 1$ to $q \ll 1$. For $q \lesssim 1$ the vertical resonance lies inside the disc, while the Lindblad resonance lies outside the disc.}

Finally, a warped disc can be thought of as a series of nested circular orbits, each with a different tilt, so the vertical component of gravity does not vary along any given streamline. However, the combination of tilt and shear experienced by the gas along any given orbit results in periodic vertical compression and rarefaction as if the vertical component of gravity were periodic \citep{ogilvie2013local, ogilvie2022hydrodynamics}. The forcing frequency is equal to twice the orbital frequency ($\omega = 2\Omega$), if we neglect any precession of the warp.

\subsubsection{Forcing amplitudes}

In the case of an eccentric disc, the forcing amplitude $a$ can be directly related to the eccentricity $e$ through $a=3e$. The factor of $3$ comes from the fact that the vertical gravity scales with $r^{-3}$.

Next, we obtain an estimate of the appropriate forcing amplitude $a$ for a tidally distorted disc. The gravitational (plus indirect) potential due to a primary star of mass $M_1$ and secondary star of mass $M_2$ in a circular binary of separation $r_\text{b}$ is given by \citep{lubow1981vertically}
\begin{align}
  \Phi(r,\phi',z) = & \Phi_1(r,z) + \Phi_2(r,\phi',z) \nonumber \\
  = & - \frac{GM_1}{(r^2 + z^2)^{1/2}} - \frac{GM_2}{(r^2-2 r r_\text{b} \cos{\phi'} + r_\text{b}^2 + z^2)^{1/2}} \nonumber \\
  & \qquad + \frac{M_2}{M} \Omega_\text{b}^2 r_\text{b} r \cos{\phi'},
\end{align}
where $\Omega_\text{b} = (GM/r_\text{b}^3)^{1/2}$ is the binary angular frequency, $M=M_1+M_2$, and the relative azimuthal coordinate is $\phi' = \phi -\Omega_\text{b} t$. Expanding the non-axisymmetric part for $r \ll r_\text{b}$ (at $z=0$), we obtain $\Phi_2 \approx -(3/4)G M_2 (r^2/r_\text{b}^3) \cos{2\phi'}$. This gives a horizontal force that distorts the orbital motion from a circular form, causing the vertical gravity due to $M_1$ to vary around the non-circular orbit. (There is also a weaker component, beyond the quadrupolar approximation, that gives a direct contribution to the vertical gravity from $M_2$. Both effects scale linearly with the mass ratio $q=M_2/M_1$.) 

\begin{figure*}
\centering
\includegraphics[scale=0.45]{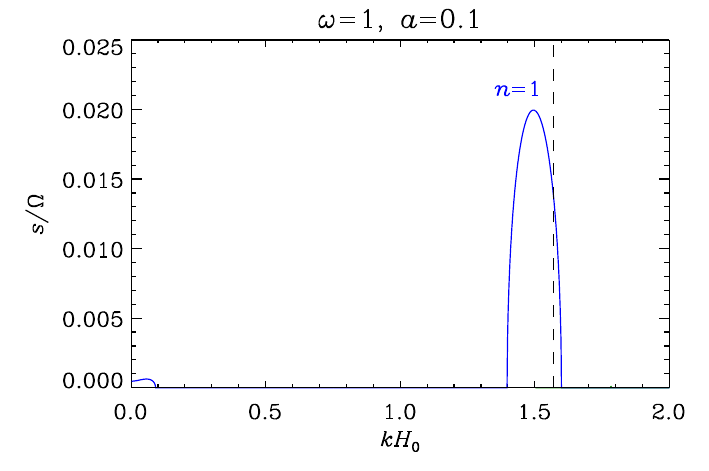}
\includegraphics[scale=0.45]{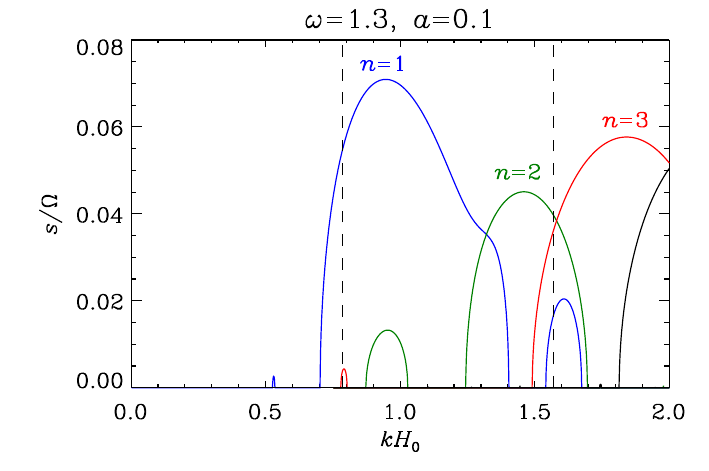}
\includegraphics[scale=0.45]{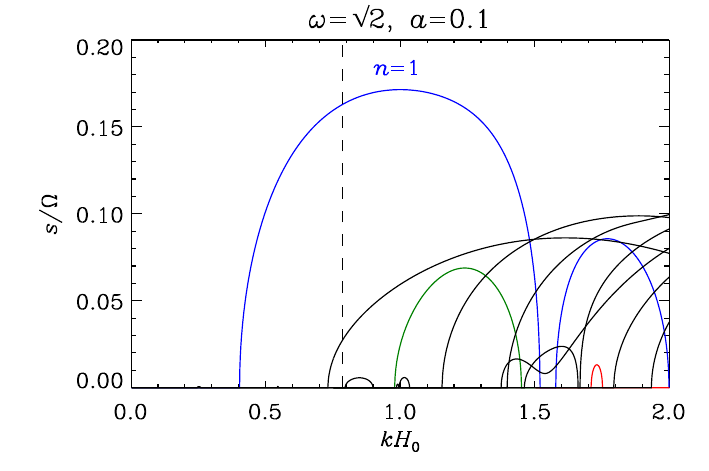}
\\
\includegraphics[scale=0.45]{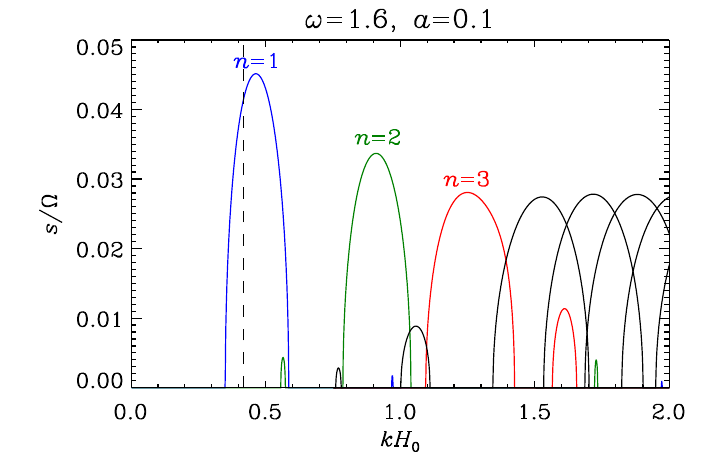}
\includegraphics[scale=0.45]{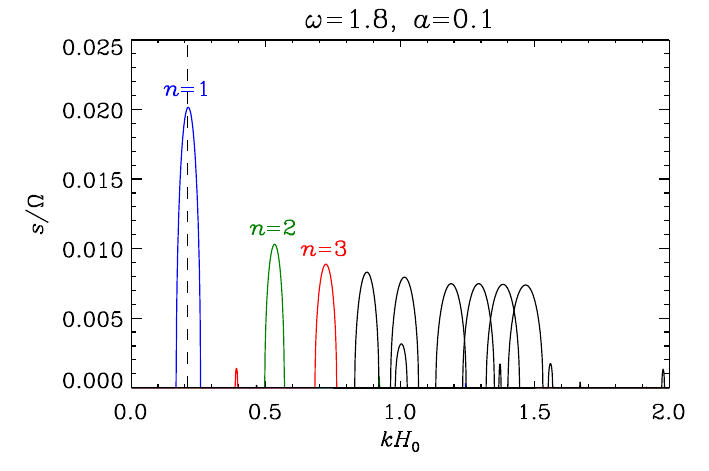}
\includegraphics[scale=0.45]{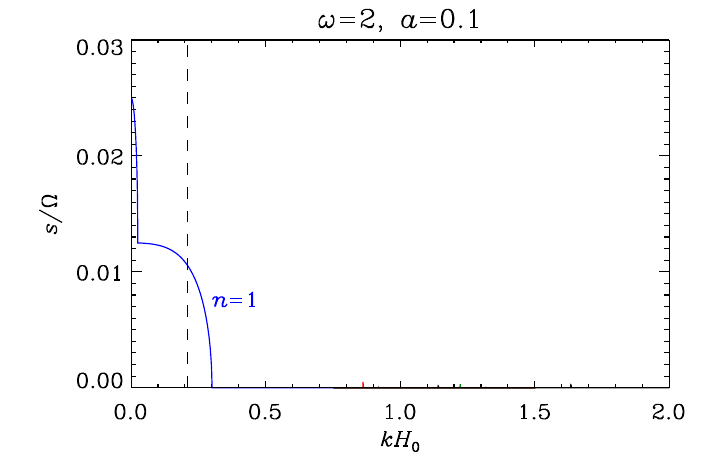}
\caption{Theoretical calculation of growth rate (in units of $\Omega$) of the parametric instability for forced vertical oscillations at different forcing frequencies: $\omega / \Omega = 1.0, 1.3, \sqrt{2}$ (top row), and $\omega / \Omega = 1.6, 1.8, 2.0$ (bottom row). The $x$-axis is the radial wavenumber in units of $H_0^{-1}$. Different colored solid lines correspond to growth rates for different mode numbers $n$. All calculations are for a forcing amplitude of $a = 0.1 H_0 \Omega^2$. Vertical dashed lines mark $2\pi/L_x$, where $L_x$ is the radial box size of a simulation carried out at the corresponding forcing frequency.}
\label{FIGURE_ForcedBounceLinearTheoryGrowthRates}
\end{figure*}

The radial displacement of the tidal deformation can be calculated (neglecting pressure) from
\begin{equation}
  (\Omega^2-\omega^2)\xi_r=-\frac{\partial\Phi_2}{\partial r}-\frac{2m}{r}\frac{\Omega}{\omega}\Phi_2,
\end{equation}
where $m=2$ and $\omega=2(\Omega-\Omega_\text{b})$. This gives
\begin{equation}
  \frac{\xi_r}{r}=-\frac{3q}{2} f(\tilde r,q)\cos{2\phi'},
\end{equation}
where
\begin{equation}
  f(\tilde{r},q)=\frac{\tilde{r}^3\left[1+\frac{1}{1-(1+q)^{1/2}\tilde{r}^{3/2}}\right]}{4[1-(1+q)^{1/2}\tilde{r}^{3/2}]^2-1},
\end{equation}
with $\tilde{r}\equiv r/r_\text{b}$.
Since the vertical gravity due to $M_1$ is $GM_1z/r^3\propto r^{-3}$, the amplitude of its fractional variation (assumed small) is $3$ times the amplitude of $\xi_r/r$, i.e.
\begin{equation}
  a=\frac{9q}{2}f(\tilde r,q).
\end{equation}
For $q=1$ and $\tilde{r}=r/r_\text{b}$ between $0.2$ and $0.3$, $a$ is approximately between $0.04$ and $0.2$, so $a \sim 0.1$ (in units of $\Omega^2 H_0$) is a plausible forcing amplitude.

\subsection{Solutions from theory}
\label{SECTION_ForcedBounceTheory}
In Section \ref{RESULTS_Theory} we presented a 1D theory of vertical oscillations, resulting in an evolution equation for the disc thickness or dimensionless scale-height $H(t)$ (Equation \ref{EQUATION_1DTheoryDynamicalHEvolution3}). We now modify this equation to include the forcing term $a \cos{\omega t}$:
\begin{equation}
    \ddot H = -(1 + a \cos\omega t )H + \frac{1}{H^\gamma}.
\end{equation}
(From here onwards, we will consider only the case $\gamma = 1$ since our simulations are isothermal.) To determine appropriate initial conditions for the simulations we solve this equation for a given forcing amplitude $a$ and frequency $\omega$. The response curve for the scale-height $H(0)$ ($t=0$ being the time when forcing is maximal) is shown in Fig.~\ref{FIGURE_ForcedBounceInitialScaleheight} for forcing frequencies between $\omega/\Omega = 1.0$ and $\omega/\Omega = 2.0$. We plot the response curve for two cases: small-amplitude forcing ($a=0.01$, blue curve) and large-amplitude forcing ($a=0.1$, red curve). Linear resonance \citep{lubow1981vertically} occurs at $\omega = \sqrt{2} \approx 1.41$ (indicated by the dashed vertical line). Below the resonant frequency, only one (stable) solution branch exists. At frequencies higher than the resonant one, multiple solutions may exist: there is a weakly oscillating upper branch, and a strongly oscillating lower branch, and hysteresis may be possible. (There are also unstable solutions denoted by the dashed lines.) The nonlinearity of the resonance can be seen in the fact that the endpoint of the upper branch (which in fact folds over and becomes unstable at this point) moves to the right as the forcing amplitude is increased.

\begin{figure*}
\centering
\includegraphics[scale=0.16]{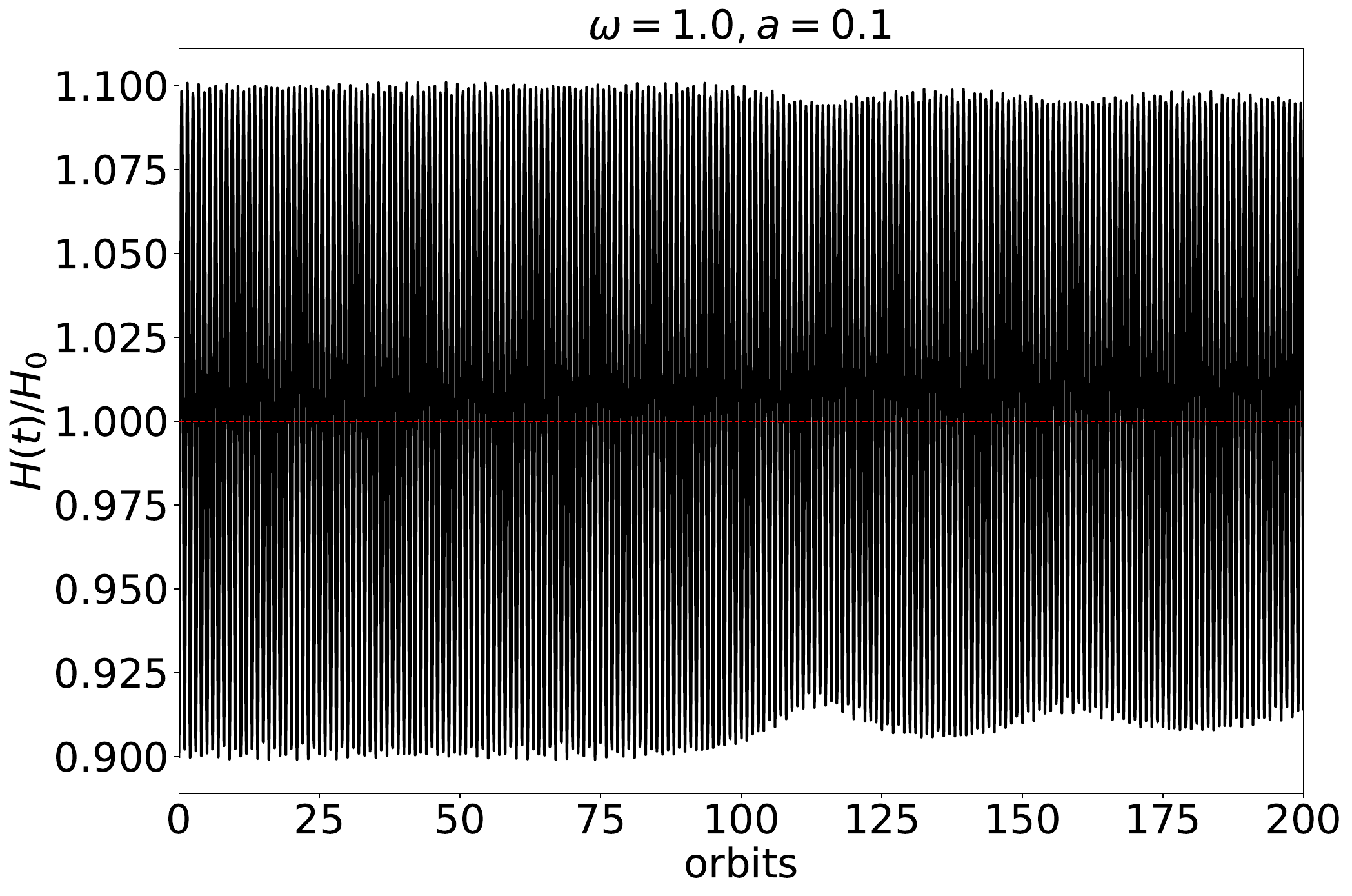}
\includegraphics[scale=0.16]{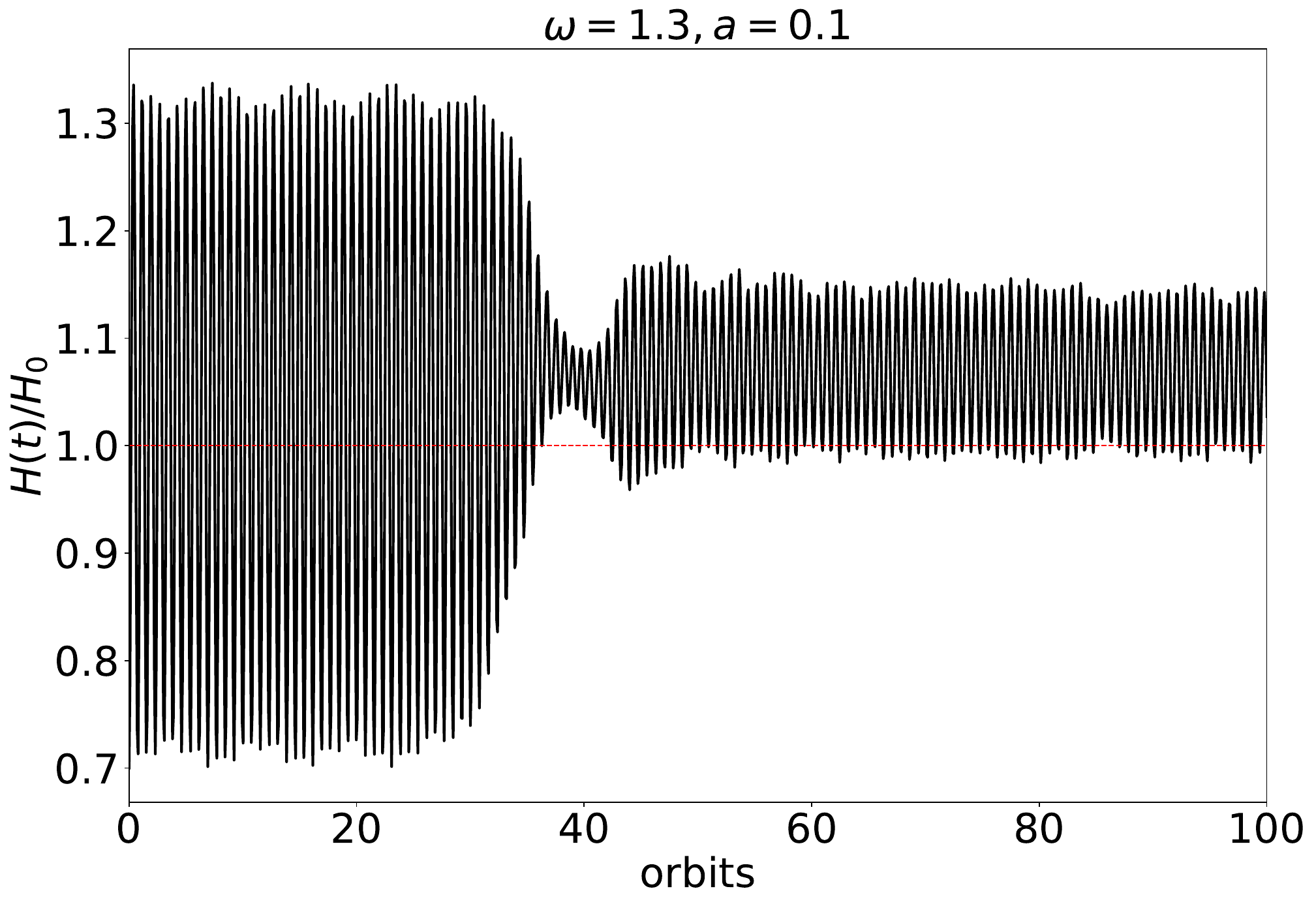}
\includegraphics[scale=0.16]{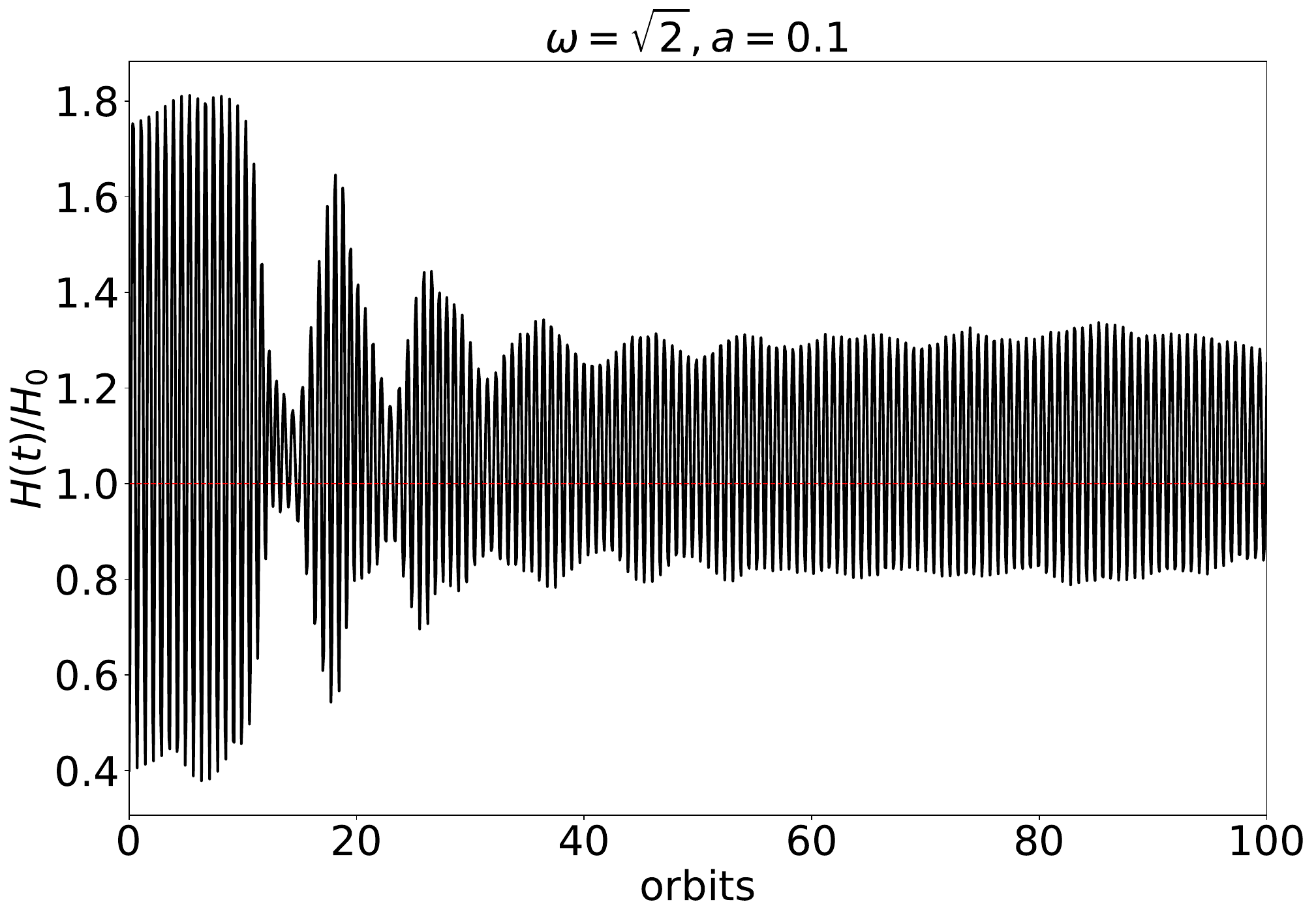}
\includegraphics[scale=0.16]{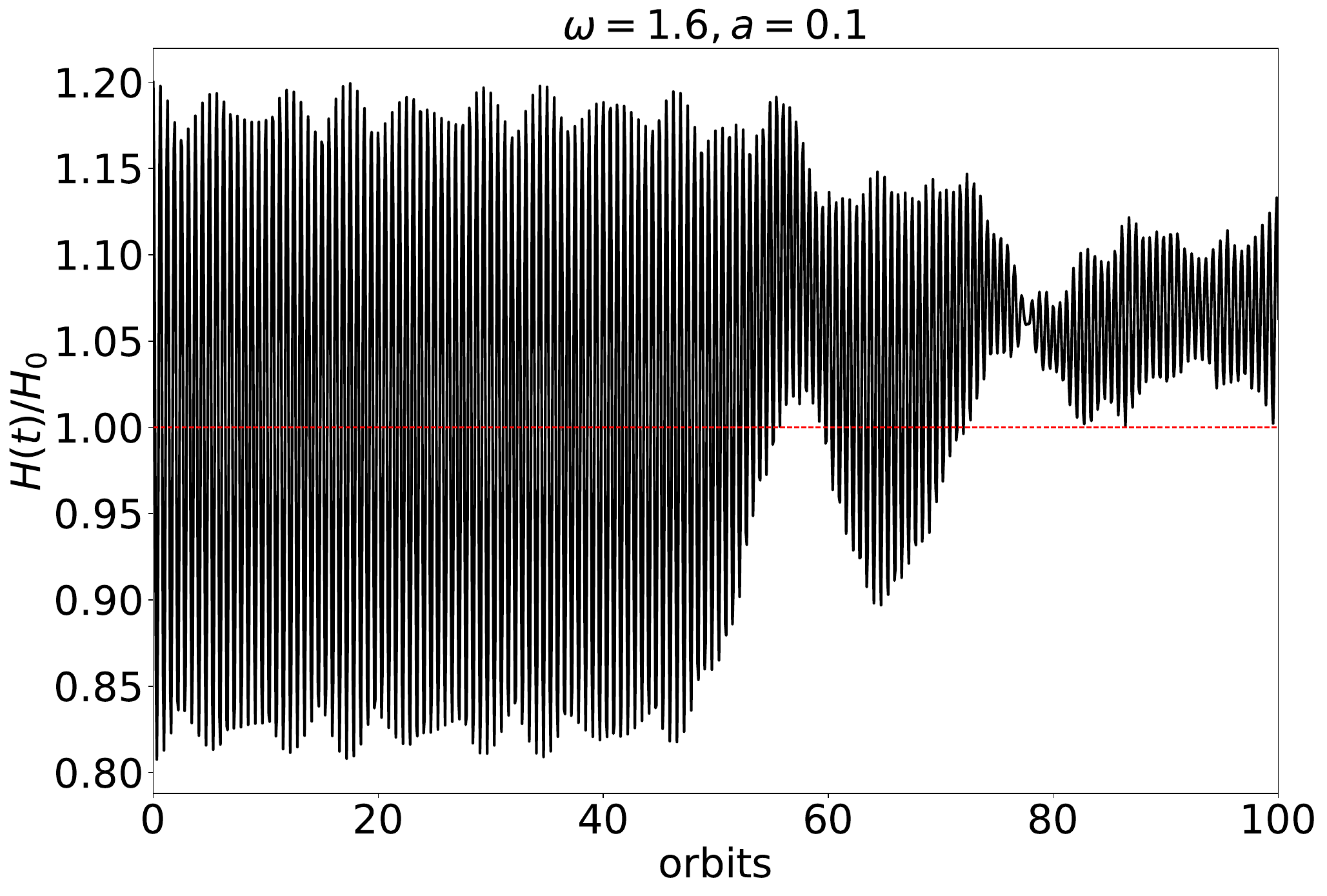}
\includegraphics[scale=0.16]{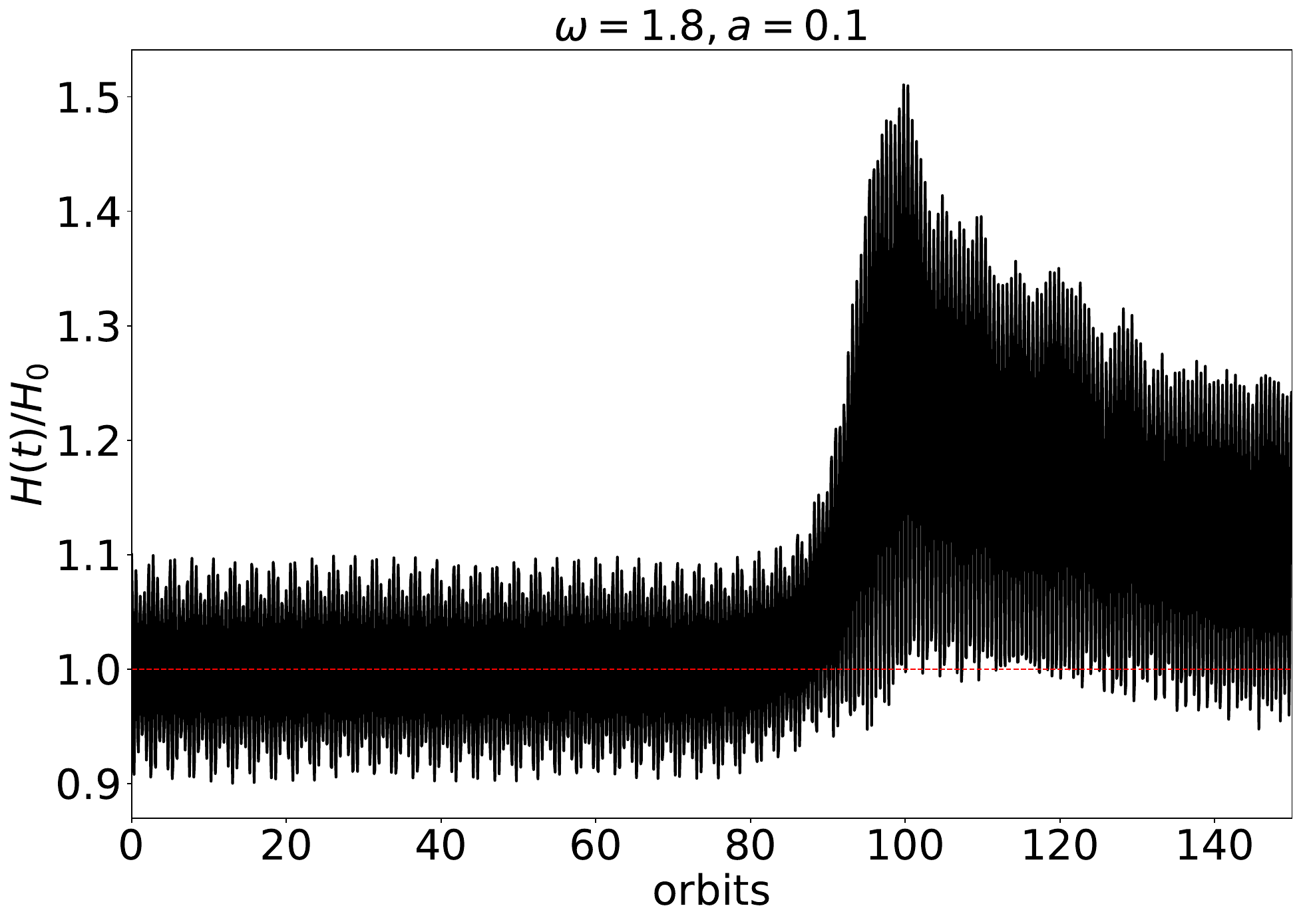}
\includegraphics[scale=0.16]{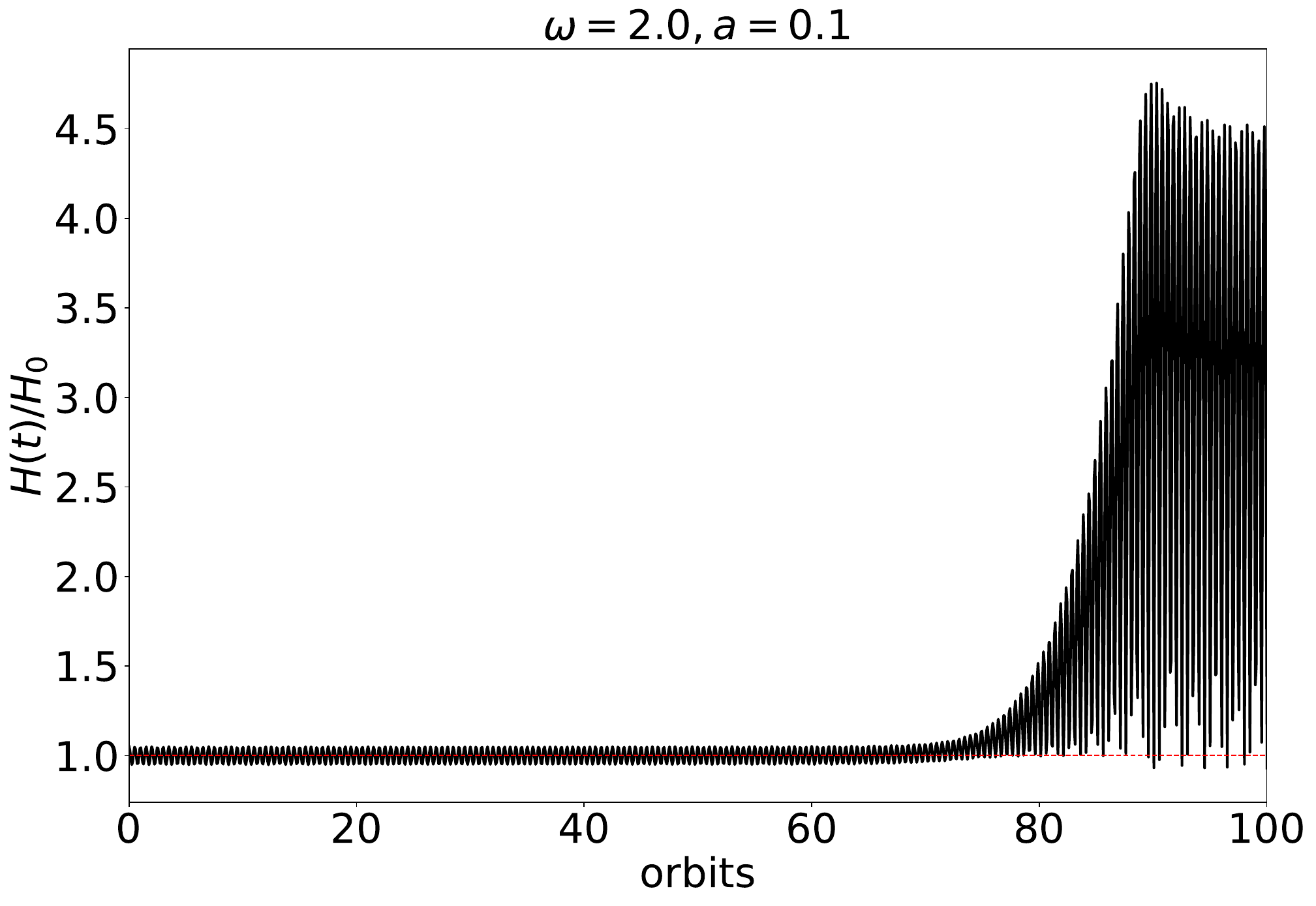}
\caption{Time-evolution of dynamical scale-height $H$ (see Equation \ref{EQUN_DynamicalScaleheight}) from simulations with different forcing frequencies $\omega$ (in units of disc angular frequency $\Omega$). Top row (left-to-right): $\omega/\Omega = 1.0, 1.3, \sqrt{2}$. Bottom row: $\omega/\Omega = 1.6, 1.8, 2.0$. (Note that the vertical axis range is different in each plot.) All simulations were run at a constant forcing amplitude of $a = 0.1 H_0\Omega^2$.}
\label{FIGURE_ForcedBounceDynamicalHForcingFrequencies  }
\end{figure*}

Next we analyse the stability of the forced bouncing solutions to radially dependent perturbations, in the same manner as we did for the freely bouncing case (see Section \ref{SECTION_ParametricInstabilityTheory}). In Fig.~\ref{FIGURE_ForcedBounceLinearTheoryGrowthRates} we plot the growth rates for six different forcing frequencies (at a constant forcing amplitude of $a=0.1 H_0\Omega^2$). We show the growth rate as a function of radial wavenumber for the first ten modes (highlighting $n=1$ [blue], $n=2$ [green] and $n=3$ [red]). At forcing frequencies below resonance ($\omega/\Omega = \sqrt{2}$), increasing the forcing frequency results in an increase in the growth rate of the fastest growing mode, while the peak simultaneously broadens and moves to the left (i.e. to longer wavelengths). On the other side of the resonance, the growth rate of the fastest growing mode \textit{decreases} with increasing forcing frequency (as does the oscillation amplitude on the upper branch), and the peak narrows. At $\omega/\Omega = 2$ we find a very narrow peak in the limit of very large wavelength, sitting on top of a broader peak. We will analyse this behaviour in more detail when we discuss the corresponding simulation (Section \ref{SECTION_ForcedBounceSimulation_Omega2pt00_a0pt1}).

\subsection{Simulations at different forcing frequencies}
\label{SECTION_ForcedBounceSimulations}
Having presented the theory of forced (driven) vertical oscillations, we turn to the simulations. We discuss, in turn, six runs with forcing frequencies ranging from $\omega/\Omega = 1$  (one oscillation or bounce per orbit), which is relevant to an eccentric disc, or the outer part of a tidally distorted single disc, to $\omega/\Omega = 2$ (two bounces per orbit), which is relevant to a warped disc, or the inner part of a tidally distorted single disc). Forcing frequencies between these two limits correspond to different radial locations in a tidally distorted disc. We used a forcing amplitude of $a = 0.1  H_0\Omega^2$ for all runs. Note that we adjust the radial box size in each simulation to ensure that we can capture the $n=1$ mode of the parametric instability for that forcing frequency (see Fig.~\ref{FIGURE_ForcedBounceLinearTheoryGrowthRates}). We initialize each run with a disc thickness $H(0)$ that is equal to (or very close to) the initial thickness predicted from theory for that forcing frequency and amplitude (blue curve in Fig.~\ref{FIGURE_ForcedBounceInitialScaleheight}), because otherwise an additional free oscillation with a different period is superimposed on the forced oscillation. In the azimuthal and vertical directions we keep the box size fixed at $L_y = 4H_0$ and $L_z = 12H_0$, respectively. Finally, we use a resolution of $32/H_0$ for all runs, and run each simulation for 100--200 orbits ($628$--$1257\,\Omega^{-1}$).

\subsubsection{The case $\omega/\Omega = 1$}
\label{SECTION_ForcedBounceSimulation_Omega1pt00_a0pt1}
The lowest forcing frequency we investigated is the case where the forcing frequency is equal to the local orbital frequency, i.e. $\omega/\Omega = 1$, which is far from resonance. This could be the case for an eccentric disc around a single star \citep[see, e.g.,][]{wienkers2018non}. The growth rates of the parametric instability for this forcing frequency and amplitude are shown in the top-left panel of Fig.~\ref{FIGURE_ForcedBounceLinearTheoryGrowthRates}. After initialization the disc oscillates once per orbit, and its thickness varies between $H \sim 1.1H_0$ and $\sim 0.9H_0$. We observe very little change in the amplitude of the bounce over the first 100 orbits. The radial kinetic energy density grows exponentially after a few orbits, but with a relatively weak growth rate of $s/\Omega = 0.012$ (the predicted value from theory for a wavelength equal to the radial box size $L_x=4H_0$ is $s/\Omega = 0.014$). Nonlinear saturation of the parametric instability does not occur until around orbit 110 at $\langle E_\text{kin,x} \rangle \sim 7\times10^{-4}$. Around this time we can begin to see the disc developing a warp; however, even in the nonlinear phase the behaviour is quite gentle compared to our fiducial freely bouncing simulation, consisting only of small oscillations together with periodic corrugation even after 200 orbits. 

\subsubsection{The case $\omega/\Omega = 1.3$}
\label{SECTION_ForcedBounceSimulation_Omega1pt30_a0pt1}
Next, we consider a slightly larger forcing frequency of $\omega/\Omega = 1.3$, corresponding to a radius larger than the radius at which linear resonance occurs in a tidally distorted single disc in a binary. During the linear phase, the disc thickness $H(t)$ varies between $0.7H_0$ and $1.35H_0$. In addition the scale-height exhibits a mild beating on a timescale of around 8--9 orbits. This arises due to the mismatch between the forcing frequency at which the oscillation is driven ($\omega/\Omega = 1.3$), and the natural or resonant frequency ($\sqrt{2} \approx 1.41\Omega$).\footnote{The expected beat frequency is $(1.41-1.3)\Omega=0.11\Omega$, corresponding to one beat every 9 orbits, as seen in the simulation.}

The oscillating disc undergoes parametric instability with a measured growth rate of $0.041 \Omega$ in the linear phase, compared to the predicted value of $\sim 0.055 \Omega$ at $kH = 2\pi/8 \approx 0.78$, and nonlinear saturation occurs around orbit~35. Around this time, the oscillation is considerably damped by shocks resulting from radially converging flows around $x = \pm 2H_0$ due to the pressure misalignment between neighbouring fluid columns in the corrugated disc. Note that after nonlinear saturation the dynamical scale-height oscillates around a slightly larger value ($1.06H_0$) than before nonlinear saturation, which is due to this diagnostic picking up the standing-wave corrugation of the disc. During the nonlinear phase we observe many of the features that we observed in our fiducial unforced simulation (Section \ref{SECTION_FreeBounceFiducialSimulation}): the disc continues to expand and contract in the vertical direction (due to the forcing) and simultaneously oscillates as a standing wave. On top of these large-scale motions we observe many radially propagating waves and shocks.

\subsubsection{The case $\omega/\Omega = \sqrt{2}$ (resonance)}
\label{SECTION_ForcedBounceSimulation_Omega1pt41_a0pt1}
Next we consider a disc oscillating at the resonant frequency $\omega / \Omega = \sqrt{2} \approx 1.41$. The resonant forcing causes the disc to oscillate with a large amplitude $\Delta H \sim 1.3H_0$, most closely resembling that in our fiducial unforced simulation (Section \ref{SECTION_FreeBounceFiducialSimulation}). Between initialization and around orbit 5 the time-averaged disc thickness appears to increase, possibly due to a very long beat phenomenon in this case (as the natural frequency of nonlinear oscillations differs from the forcing frequency by an amount that depends on the amplitude). The linear phase growth rate measured between orbit 5 and orbit 10 is $0.15 \Omega$, compared to the value of $0.162 \Omega$ predicted from theory at $k H_0 = 2\pi/8 \approx 0.78$. Around orbit~7 we see a corrugation begin to develop in the disc, and nonlinear saturation sets in around orbit~10, significantly earlier than at other forcing frequencies, owing to the larger growth rate of the instability at resonance. After nonlinear saturation, the disc exhibits strong shocks and radial compression, which significantly damp the oscillation amplitude (orbit~10 to orbit~15). 

Nevertheless, we observe at least five phases during which the oscillation is re-invigorated after nonlinear saturation, most noticeably between orbits 15 and 18, similar to what we observed in our fiducial freely bouncing simulation (Section \ref{RESULTS_FiducialSimulationTimeSeries}). During each of these phases we see the disc become increasingly corrugated until the resultant radial shocks in the disc atmosphere are so strong that the subsequent bouncing is damped; this can be clearly seen from a snapshot (not shown) taken at orbit~19.257 which is taken half a cycle after the bounce amplitude peaks during the first of these re-invigoration phases. The disc is highly corrugated, and two intersecting waves near the mid-plane of the disc at $x \sim -1H_0$ and $\sim +3H_0$ have steepened into shocks (with Mach number $M \sim 4$ in the disc atmosphere). The peak amplitude in the vertical oscillation during successive phases of 
re-invigoration is always weaker than the last. From orbit 50 onwards, the disc settles down into a quasi-steady equilibrium consisting of a complicated flow field with many intersecting waves and shocks.

\subsubsection{The case $\omega/\Omega = 1.6$}
\label{SECTION_ForcedBounceSimulation_Omega1pt6_a0pt1}
Next we turn to a simulation with a forcing frequency of $\omega/\Omega = 1.6$, i.e. a forcing frequency that is \textit{greater} than the resonant frequency. In a tidally distorted single disc in a binary this is expected to occur at a radius \textit{inside} the resonant radius. The disc undergoes parametric instability with a measured growth rate in the linear phase of $0.033 \Omega$, compared to the theoretical prediction of $\sim 0.041\Omega$ for the $n = 1$ mode at this wavelength. Like at forcing frequencies less than the resonant frequency ($\omega/\Omega = \sqrt{2}$), we observe a beating of the dynamical scale-height, but the envelope is more complicated than in the $\omega/\Omega = 1.3$ simulation. We measure an average beat period (interval between successive peaks in the envelope) of around 5.8 orbits, compared to the expected value of around $1/|1.6-\sqrt{2}| \sim 5.4$ orbits for this forcing frequency. Nonlinear saturation sets in at around orbit~55, after which the time-averaged dynamical scale-height increases to $\langle H \rangle_t \sim 1.08H_0$. During the nonlinear phase the behaviour of the dynamical scale-height is complicated, exhibiting alternating phases during which the amplitude of the oscillation increases (for example, between orbits 55 and 65 the oscillation amplitude increases from around $\Delta H \sim 0.1H_0$ to $\sim 0.25 H_0$ ), and then decreases again (orbits 65 to 75).

\subsubsection{The case $\omega/\Omega = 1.8$}
\label{SECTION_ForcedBounceSimulation_Omega1pt80_a0pt1}
At a forcing frequency of $\omega = 1.8 \Omega_0$ the disc first oscillates with an amplitude of around $\Delta H \sim 0.2 H_0$ and a beat of $\sim2.6$ orbits (compared to the expected beat of $1/(1.8-\sqrt{2}) \sim 2.6$ orbits). We measured a growth rate of $\sim 0.019 \Omega_0$, in excellent agreement with that predicted by theory for this radial box size ($L_x = 30H_0$, i.e. $kH_0 \sim 0.21$) of $\sim 0.20 \Omega_0$ (see middle panel of Fig.~\ref{FIGURE_ForcedBounceLinearTheoryGrowthRates}). The disc begins to corrugate around orbit 80 -- this is development of the bending mode, as seen in our other simulations.

\subsubsection{The case $\omega/\Omega = 2$}
\label{SECTION_ForcedBounceSimulation_Omega2pt00_a0pt1}
Finally we consider the case where the disc is forced to oscillate twice per orbit, relevant to a warped disc around a single star. The shape of the dispersion relation for the $n=1$ mode (we remind the reader that here $n$ is the vertical mode number) is markedly different in this case compared to the other forcing frequencies (see bottom-right panel of Fig.~\ref{FIGURE_ForcedBounceLinearTheoryGrowthRates}). Only the $n=1$ mode grows, and only at long ($\gtrsim 21H_0$) wavelengths. Thus we need a large radial domain to capture this mode: we chose a domain of size $L_x = 30H_0$ in the radial direction, corresponding to a wavenumber of $kH_0 \sim 0.2$. Note that this box size is rather extreme,\footnote{ Suppose $H_0/R \sim 0.01$, which is reasonable for X-ray binaries. Then a box of radial size $L_x = 30H_0$ would cover $0.3R$. Boxes of this size are of questionable applicability, unless disc is extremely thin. What would be excited in this regime is a global warp, which is outside the regime of applicability of the local model.} but we include this run for completeness.

The simulation at this forcing frequency differs markedly from the earlier cases. The disc initially exhibits small-amplitude oscillations about the equilibrium value ($\Delta H \sim 0.1H_0$), in keeping with the theoretical prediction at this forcing frequency (Fig.~\ref{FIGURE_ForcedBounceInitialScaleheight}). The radial kinetic energy density grows exponentially, but with a small growth rate of around $s \sim 0.007\Omega$ (the theoretical growth rate at this box size / radial wavenumber is $s_{\text{th}} \sim 0.01\Omega$). Unlike at smaller forcing frequencies, the disc does not begin to develop a warp as the mode grows. Instead, starting at around orbit~40, the entire disc begins to be oscillate uniformly about the mid-plane ($z=0$) twice per orbit even as the disc itself continues to expand and contract at the same rate. The amplitude of this oscillation continues to grow until, at around orbit~85, the disc hits the reflective vertical boundaries at $z = \pm 6H_0$ and is subsequently disrupted.\footnote{Note that, because the forcing is quadratic in $z$, increasing the box size in the vertical direction would not prevent the disc from eventually hitting the vertical boundaries.} 

What we are seeing is the excitation of an $n=1$ mode with radial wavenumber $k=0$. The up-and-down motion (about $z=0$) of the entire disc is how a tilted disc would appear to an observer moving along an untilted circular orbit, and can be related to the tilt instability at the 3:1 resonance\footnote{In relating the forcing frequency to the radial location in a tidally distorted disc, we have generally assumed that the $m=2$ component of the tidal potential is being used. The $3:1$ tilt instability uses the weaker $m=3$ component but is located in the outer part of the disc in binaries of sufficiently small mass ratio \citep{lubow1992tidally}.} in a tidally distorted disc in binary \citep{lubow1992tidally}. This can understood quantitatively by deriving an evolution equation (we omit the details of this calculation) for the height of the centre of mass of the disc (defined as $Z(t)=(1/M)\iint\rho z\,dx\, dz$, where $M=\iint\rho\,dx\,dz$ is the total mass) with periodic forcing
\begin{equation}
  \ddot Z+\Omega^2(1+a\cos\omega t)Z=0,
\end{equation}
which is a form of the Mathieu equation. In particular, parametric resonance occurs for $\omega$ sufficiently close to $2\Omega$. The expected growth rate is
\begin{equation}
  \frac{\Omega}{2}\left[\left(\frac{a}{2}\right)^2-\left(\frac{\omega}{\Omega}-2\right)^2\right]^{1/2},
\end{equation}
where the quantity inside the square root is positive, i.e.\ where the detuning is sufficiently small, i.e.\ for $\omega/\Omega$ between $2\pm(a/2)$. The maximum growth rate of $a\Omega/4$ is expected at $\omega=2\Omega$. For a forcing amplitude of $a=0.1$ this works out to be $\sim 0.025\Omega$, corresponding to the spike near $k=0$ in Fig.~\ref{FIGURE_ForcedBounceLinearTheoryGrowthRates}, and in good agreement with the value of $\sim 0.022\Omega$ measured directly in the simulation from the time-series of vertical kinetic energy density (not shown).

If the case $\omega=2\Omega$ is identified with the forcing of vertical oscillations in a warped disc, then to say that such oscillations can parametrically excite a large-scale warp illustrates the fact that there are nonlinear couplings between such a warp and vertical oscillations that can exchange energy is either direction. This is consistent with the picture derived in the oscillating torus model of \citet{fairbairn2021non}.

\section{Conclusions}
\label{CONCLUSIONS}
We have studied vertical oscillations in accretion discs, involving periodic expansion and contraction of the gas perpendicular to the plane of the disc. These breathing or bouncing motions are expected to occur in all types of distorted accretion discs (warped, tidally distorted and eccentric discs) on account of the lack of vertical hydrostatic equilibrium in those objects. 
In theory, periodic nonlinear oscillations are exact solutions of the equations of ideal gas dynamics in a local model of the disc. To investigate the dynamics of these oscillations we carried out both quasi-2D and fully 3D local (shearing-box) hydrodynamic simulations with an isothermal equation of state. To complement our numerical simulations, we have also presented a one-dimensional model of vertical oscillations and analysed their stability.

To isolate the mechanisms damping these vertical oscillations, we first considered the case of a freely oscillating disc (i.e.\ in the absence of any forcing). When the oscillation amplitude is sufficiently large, the disc collapses and expands supersonically, with pressure providing an impulsive restoring force. The resultant shocks dominate the damping of the vertical oscillations, at least in the very early stages of the simulation when the disc is undergoing only vertical oscillations, but their existence and properties are dependent on the artificial boundary conditions at the top and bottom of the box.

A significant new result is that the vertical oscillations can excite a corrugation or bending mode by means of the parametric instability. This instability, which is an example of the parametric instability of inertial waves in distorted discs, further dampens the vertical oscillations as their kinetic energy is converted into that of the bending mode. The disc develops a radially periodic corrugation or warp that oscillates like a standing wave. The resultant radial pressure gradients in the corrugated disc lead to radially converging flows. After nonlinear saturation the mid-plane of the corrugated disc consists of many intersecting waves which steepen into shocks in the atmosphere ($z \gtrsim 2H_0$). The dissipation due to these shocks ultimately dampens the vertical oscillations in the absence of forcing. However, on short time-intervals ($\sim 2-5$ orbits) the oscillations can actually be re-invigorated by means of a three-mode coupling between the vertical oscillation and the oppositely directed traveling waves that form the warp or standing wave. These results are robust to changes in resolution, vertical box size, vertical boundary conditions, and numerical and physical (explicit) viscosity.

We then investigated the case in which the oscillations were driven by means of periodic variation in the vertical component of gravity for various forcing frequencies $\omega$, ranging from $\omega=\Omega$ (relevant to an eccentric disc) to $\omega= 2\Omega$ (relevant to a warped disc). Various values between these two limits correspond to the forcing frequency at different radial locations in a tidally distorted disc. Here, too, a disc that is initially oscillating only in the vertical direction becomes unstable, leading to the development of a bending mode or warp. However, for forcing frequencies greater than the resonant frequency ($\sqrt{2}\Omega$) we observe, in addition to the bending mode, vertical oscillations in the centre of mass of the disc. Viewed in the plane perpendicular to the disc in a local (shearing-box) frame this is manifested as uniform vertical oscillation of the entire disc, which becomes more extreme as the forcing frequency approaches twice the orbital frequency. These oscillations are related to the tilt instability expected to occur at the 3:1 resonance in tidally distorted binary discs \citep{lubow1992tidally}.

\section*{Acknowledgements}
This research was funded by STFC through grant ST/X001113/1.
The authors would like to thank Hongping Deng, Henrik Latter, Callum Fairbairn, and Jim Stone for helpful discussions. LEH thanks Masaru Shibata for providing access to the Cobra and Raven clusters at the Max Planck Computing and Data Facility (MPCDF) in Garching, Germany, and to the Yamazaki workstations at the Max Planck Institute for Gravitational Physics (Albert Einstein Institute) in Potsdam, Germany.

\section*{Data availability}
The data underlying this article will be shared on reasonable request to the corresponding author.




\bibliographystyle{mnras}
\bibliography{BouncingDisksBib} 




\appendix

\section{Model of coupled and damped oscillators}
\label{APPENDIX_Coupling_Model}

\begin{figure}
\centering
\includegraphics[scale=0.45]{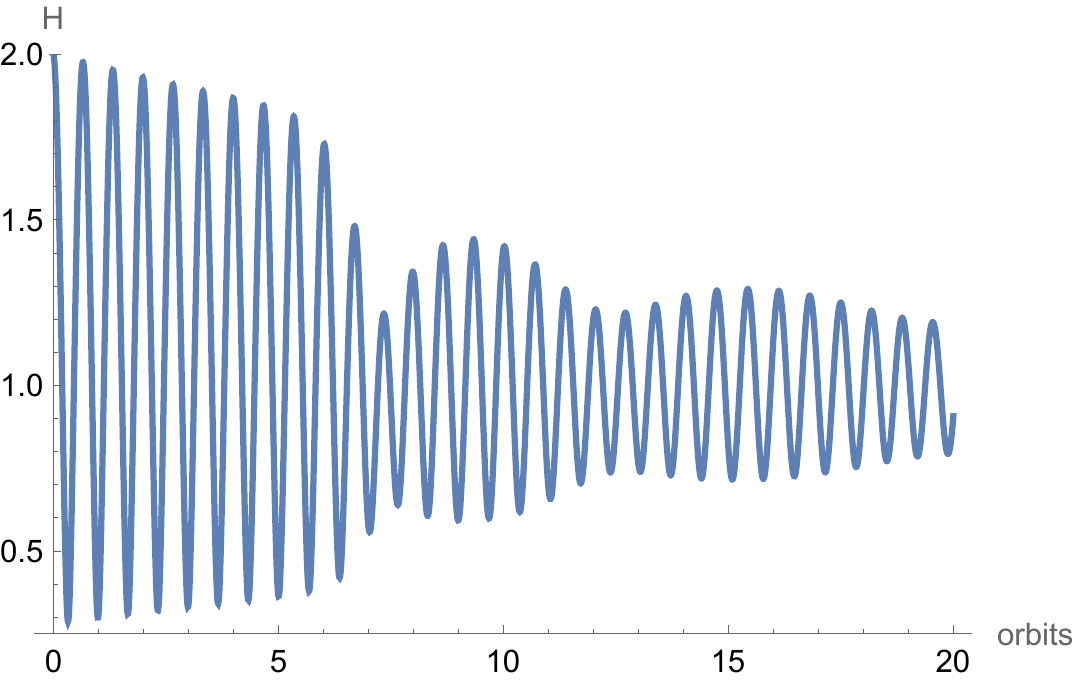}
\caption{Evolution of the dynamical scale-height according to the model of coupled and damped oscillators (\ref{coupling1})--(\ref{coupling3}), for an unforced oscillation ($a=0$) with parameters $k=2\pi/8$, $\alpha_1=0.01$, $\beta_1=1$, $\alpha_2=\alpha_3=0.5$, $\beta_2=\beta_3=2$ and initial conditions $H=2$, $\dot H=0$, $Y=10^{-3}$, $\dot Y=0$, $Z=0$ and $\dot Z=0$.}
\label{FIGURE_Coupling_Model}
\end{figure}

To try to explain the behaviour seen in several of the simulations, in which there are phases of re-invigoration of the bouncing motion, we consider the following model:
\begin{align}
  &\ddot H=-gH+\frac{1}{H}-\frac{k}{H^2}\,\text{Re}\left(Y^*Z\right)-\alpha_1|\dot H|^{\beta_1}\dot H,\label{coupling1}\\
  &\ddot Y=-(1+k^2)Y+\frac{kZ}{H}-\alpha_2|\dot Y|^{\beta_2}\dot Y,\label{coupling2}\\
  &\ddot Z=-gZ+\frac{kX}{H}-\alpha_3|\dot Z|^{\beta_3}\dot Z,\label{coupling3}
\end{align}
where $g=1+a\cos\omega t$ is the strength of vertical gravity. This model is based on equations (\ref{floquet1})--(\ref{floquet3}) in the case $n=1$, with $Y=\ii X$, but also includes the conservative nonlinear feedback of the daughter modes on the parent oscillation; the term involving $Y^*Z$ in the first equation can be derived from a Lagrangian formulation of the three-mode system. It also includes nonlinear damping terms (the final term in each equation) to model the dissipative effects of shocks. We have not made an extensive exploration of the model parameters $\alpha_i$ and $\beta_i$, but in Fig.~\ref{FIGURE_Coupling_Model} present an illustrative solution for $H(t)$ in a setup corresponding to the fiducial model of Section~\ref{SECTION_FreeBounceFiducialSimulation}; this can be compared with Fig.~\ref{FIGURE_FiducialSimTimeEvolutionDynamicalHAndEkinx} and shows phases in which the vertical oscillation alternately loses and gains energy from the daughter modes, combined with an overall damping.

\section{Tables of Simulations}
\label{APPENDIX_TablesOfSimulations}

\begin{table*}
\centering
\caption{3D simulations of a freely bouncing disc: comparison of initial disc thickness $H(0)$ in units of the equilibrium disc thickness $H_0$. The first column gives the box size in the $x$-, $y$-, and $z$-directions, respectively (in units of $H_0$). The second column gives the resolution in cells per $H_0$. All simulations were run with an explicit viscosity $\nu$ with corresponding Reynolds number $\text{Re}\equiv H_0^2 \Omega / \nu \approx 4687$. The fourth column gives the vertical boundary conditions (zBCs). The bounce period (in orbits) measured in the simulation is denoted by $\Delta T$, the growth rate of the parametric instability measured in the simulation by $s/\Omega$ (rounded to 3 decimals), and the growth rate predicted from theory by $s_{\text{th}}/\Omega$. (Note that the theoretical calculation does not take into account dissipation.) All simulations were run for 50 orbits. Note that for the $H(0)/H_0 = 1.97$ run, the maximum disk thickness at onset of parametric instability is actually closer to $H(0)/H_0 \sim 1.8$, so the theoretical growth rate was calculated for $H(0)/H_0 = 1.8$.}
\label{TABLE_3DSimsFreelyBouncing_ComparisonBounceAmplitude}
 	\begin{tabular}{lcccccccr}
		\hline
		Run	& Box Size & Resolution & Re & zBC & $H(0)/H_0$ & $\Delta T$ /orb & $s/\Omega$& $s_{\text{th}}/\Omega$\\ 
        \hline
        FreeBounceRes32Lz12H0-SmallBounce & [9,4,12] & $32/H_0$ & $4687$ & reflective & 1.10 & 0.708 &  0.024 & 0.023\\
        FreeBounceRes32Lz12H0-IntBounce & [8,4,12] & $32/H_0$ & $4687$ & reflective & 1.22 & 0.701 &  0.044 & 0.045\\
        FreeBounceRes32Lz12H0-LargeBounce & [8,4,12] & $32/H_0$ & $4687$ & reflective & 1.97 & 0.66 & 0.169 & 0.17\\
        \hline
	\end{tabular}
\end{table*}

\begin{table*}
\centering
\caption{3D simulations of a freely bouncing disc: vertical box size and resolution comparisons. Simulations were run for 50 orbits. Note that simulation FreeBounceRes32Lz12H0 (2nd row) is the same as our fiducial simulation (FreeBounceRes32Lz12H0-LargeBounce) in Table \ref{TABLE_3DSimsFreelyBouncing_ComparisonBounceAmplitude}). Note that the theoretical growth rate for the $L_z = 8H_0$ run assumes $H \sim 1.4H_0$ which is close to the disc thickness in this run at the onset of parametric instability.}
\label{TABLE_3DSimsFreelyBouncing_ComparisonBoxSizeResolution}
 	\begin{tabular}{lcccccccr}
		\hline
		Run	& Box Size & Resolution & Re & zBC & $H(0)/H_0$ & $\Delta T$/orb &$s/\Omega$& $s_{\text{th}}/\Omega$\\ 
        \hline
        FreeBounceRes32Lz8H0 & [8,4,8] & $32/H_0$ & $4687$ & reflective & 1.76 & 0.68 & 0.075 & 0.086 \\
        FreeBounceRes32Lz12H0 & [8,4,12] & $32/H_0$ & $4687$ & reflective & 1.97 & 0.66 & 0.169 & 0.17\\
        FreeBounceRes32Lz16H0 & [8,4,16] & $32/H_0$ & $4687$ & reflective & 1.99 & 0.66 & 0.168 & 0.17\\
        \hline
        FreeBounceRes64Lz12H0 & [8,4,12] & $64/H_0$ & $4687$ & reflective & 1.97 & 0.66 & 0.157 & 0.17 \\
        \hline
	\end{tabular}
\end{table*}

\begin{table*}
\centering
\caption{3D simulations of a freely bouncing disc: vertical boundary condition comparison. Simulations were run for 50 orbits. Note that simulation FreeBounceRes32Lz12H0-zBCRef (1st row) is the same as our fiducial simulation (FreeBounceRes32Lz12H0-LargeBounce) in Table \ref{TABLE_3DSimsFreelyBouncing_ComparisonBounceAmplitude}).}
\label{TABLE_3DSimsFreelyBouncing_ComparisonVerticalBCs}
 	\begin{tabular}{lcccccccr}
		\hline
		Run	& Box Size & Resolution & Re & zBC & $H(0)/H_0$ & $\Delta T$/orb & $s/\Omega$& $s_{\text{th}}/\Omega$ \\ 
        \hline
        FreeBounceRes32Lz12H0-zBCRef & [8,4,12] & $32/H_0$ & $4687$ & reflective & 1.97 & 0.66 & 0.169 & 0.17\\
        FreeBounceRes32Lz12H0-zBCOut & [8,4,12] & $32/H_0$ & $4687$ & outflow & 1.97 & 0.66 & 0.154 & 0.17 \\
        \hline
	\end{tabular}
\end{table*}

\begin{table*}
\centering
\caption{Quasi-2D simulations of a freely bouncing disc at different resolutions. Runs without explicit viscosity are labeled as "ideal". All simulations were run for 20 orbits, and used reflective boundary conditions in the vertical direction.}
\label{TABLE_3DSimsFreelyBouncing_ComparisonExplicitNumericalViscosity}
	\begin{tabular}{lcccccccr}
		\hline
		Run	& Box Size & Resolution & Re  & $H(0)/H_0$ & $\Delta T$/orb & $s/\Omega$&$s_{\text{th}}/\Omega$\\ 
        \hline
  FreeBounce2DRes32  & [8,0.008,12] & $32/H_0$ & $4687$ & 1.97 & 0.66 & 0.177 & 0.17\\
  FreeBounce2DRes64  & [8,0.008,12] & $64/H_0$ & $4687$ & 1.97 & 0.66 & 0.170 & 0.17\\
  FreeBounce2DRes128  & [8,0.008,12] & $128/H_0$ & $4687$ & 1.97 & 0.66 & 0.165 & 0.17 &\\
  FreeBounce2DRes256  & [8,0.008,12] & $256/H_0$ & $4687$ & 1.97 & 0.66 & 0.178 & 0.17 &\\
  FreeBounce2DRes512  & [8,0.008,12] & $512/H_0$ & $4687$ & 1.97 & 0.66 & 0.175 & 0.17 &\\
  \hline
FreeBounce2DRes256Ideal  & [8,0.008,12] & $256/H_0$ & ideal & 1.97  & 0.66 & 0.171 & 0.17 &\\
		\hline
	\end{tabular}
\end{table*}

\begin{table*}
\centering
\caption{3D forced bounce simulations, in which bouncing motion was driven using a time-dependent vertical gravitational acceleration of the form $g_{\text{eff,z}} = -\Omega^2 z (1 + a \cos{\omega t})$, where $a$ is the forcing amplitude (in units of $H_0 \Omega^2$), and $\omega$ is the forcing frequency (in units of the local orbital angular frequency $\Omega$). Note that $\Delta T$ is the interval (in orbits) between bounces measured directly from the simulation -- it should correspond to the inverse of the forcing frequency. All simulations were run for 100-200 orbits, at fixed Reynolds number of $\text{Re}=4687$, and used reflective boundary conditions in the vertical direction.}
\label{TABLE_3DSimsFreelyBouncing_ComparisonExplicitNumericalViscosity}
	\begin{tabular}{lcccccccr}
		\hline
		Run	& Box Size & Resolution & $H(0)/H_0$ & $\omega/\Omega$ & $a/(H_0 \Omega^2)$ & $\Delta T$/orb &$s/\Omega$ & $s_{\text{th}}/\Omega$ \\ 
		\hline
		ForcedBounce1.0    & [4,4,12] & $32/H_0$ & $0.9$ & $1.0$ & 0.1 & 1.0 & 0.012 &0.014 \\
        ForcedBounce1.3   & [8,4,12] & $32/H_0$ & $0.7$ & $1.3$ & 0.1 & 0.778 & 0.041 & 0.055 \\
        \hline
		ForcedBounce1.41    & [8,4,12] & $32/H_0$ & $0.4$ & $\sqrt{2}$ & 0.1 & 0.708 & 0.150 &0.162 \\
        \hline
		ForcedBounce1.6   & [15,4,12] & $32/H_0$ & $1.2$ & $1.6$ & 0.1 & 0.622 & 0.033 & 0.041\\
        ForcedBounce1.8   & [30,4,12] & $32/H_0$ & $1.1$ & $1.8$ & 0.1 & 0.554 & 0.019 & 0.020\\
		ForcedBounce2.0  & [30,4,12] & $32/H_0$ & 1.05 & 2.0 & 0.1 & 0.5 & 0.007 & 0.011\\
        \hline
	\end{tabular}
\end{table*}

\bsp	
\label{lastpage}
\end{document}